\documentclass[fleqn,usenatbib]{mnras}

\usepackage{newtxtext,newtxmath}

\usepackage[T1]{fontenc}

\DeclareRobustCommand{\VAN}[3]{#2}
\let\VANthebibliography\thebibliography
\def\thebibliography{\DeclareRobustCommand{\VAN}[3]{##3}\VANthebibliography}

\usepackage{graphicx}	
\usepackage{amsmath}	
\usepackage{makecell}

\title[ICL fraction and cluster properties]{The dependence of the intracluster light fraction on galaxy cluster properties}

\author[L. Canepa et al.]{
Louisa Canepa,$^{1}$\thanks{E-mail: l.canepa@unsw.edu.au}
Sarah Brough,$^{1}$
Mireia Montes,$^{2}$
Nina Hatch$^{3}$
\\
$^{1}$School of Physics, University of New South Wales, NSW 2052, Australia\\
$^{2}$Institute of Space Sciences (ICE, CSIC), Campus UAB, Carrer de Can Magrans, s/n, 08193 08193, Cerdanyola del Valles, Spain\\
$^{3}$School of Physics and Astronomy, University of Nottingham, University Park, Nottingham NG7 2RD, UK
}

\date{Accepted XXX. Received YYY; in original form ZZZ}

\pubyear{\the\year{}}

\begin{document}
\label{firstpage}
\pagerange{\pageref{firstpage}--\pageref{lastpage}}
\maketitle

\begin{abstract}
We use machine learning to measure the intracluster light (ICL) fractions of 177 galaxy groups and clusters identified from Hyper Suprime-Cam Subaru Strategic Program imaging to explore how the ICL varies with the properties of its host cluster. We study the variation in ICL fraction with host cluster redshift, halo mass, and magnitude gap to investigate how the ICL develops over time, in various cluster environments, and with cluster relaxation. We find that there is a decreasing correlation with redshift (Spearman correlation $r_S=-0.604$, p-value $=9\times10^{-10}$), however this can be plausibly accounted for by including the effects of cosmological surface brightness dimming and the passive aging of stellar populations. There is a weak negative correlation with halo mass ($r_S=-0.330$, p-value $=8\times 10^{-5}$) where ICL fractions are higher in lower halo mass groups than higher halo mass clusters. We also find that there is a marginal positive correlation with magnitude gap ($r_S=0.226$, p-value = 0.01), indicating that relaxed clusters are more likely to host higher ICL fractions. These results are consistent with a scenario where the dominant formation mechanism of the ICL is galaxy-galaxy interactions such as tidal stripping, and demonstrates the capability of the method to easily construct large samples and study large-scale trends in the ICL fraction.
\end{abstract}




\section{Introduction}
The intracluster light (ICL) or intragroup light (IGL) is a diffuse, low surface brightness stellar component that is visible throughout galaxy clusters and groups (see \citealp{mihos_deep_2019}; \citealp{contini_origin_2021}; \citealp{montes_faint_2022}; \citealp{arnaboldi_kinematics_2022} for reviews). For simplicity, throughout the paper we use ICL to refer to both intracluster and intragroup light. The ICL can form through several different mechanisms, with the mechanisms widely considered as the ones that contribute the most stars being either violent relaxation after major mergers with the brightest cluster galaxy (BCG; e.g. \citealp{murante_importance_2007, lidman_importance_2013}), or tidal stripping of surviving satellite galaxies (e.g. \citealp{contini_formation_2014, montes_intracluster_2014, demaio_lost_2018, montes_intracluster_2018}). Simulations have also found that in-situ formation (e.g. \citealp{puchwein_intracluster_2010, barfety_assessment_2022, ahvazi_star_2024}) and complete disruption of dwarf satellite galaxies (e.g. \citealp{purcell_shredded_2007}) could also be important contributors. Because the ICL forms mainly due to the interactions and mergers that have occurred throughout the accretion history of the cluster (e.g. \citealp{rudick_formation_2006}), it forms an important record of the assembly history of the cluster and holds valuable information about how galaxies interact and evolve within the cluster environment. Studying the ICL is particularly important for understanding the evolution of the BCG, around which the ICL is concentrated (e.g. \citealp{puchwein_intracluster_2010}; \citealp{lidman_importance_2013}; \citealp{groenewald_close_2017}). Despite this, the ICL remains difficult to measure and characterise accurately due to its very low surface brightness.

The ICL fraction, defined as the fraction of total stellar light in the cluster that belongs to the ICL component, is a commonly used metric to quantify the amount of ICL present in a cluster. It is a useful way of studying the ICL as it holds information about the efficiency of the interactions that have shaped the cluster, and allows for easier comparison of the quantity of light captured in the ICL component among a variety of different clusters. This reveals how different cluster properties and accretion histories can facilitate a more or less efficient production of the ICL.

However, a key issue with quantifying the amount of ICL in clusters is the ambiguous definition of the ICL in contrast to the BCG in both observations and simulations. Both the ICL and the BCG share similar formation and growth pathways (e.g. \citealp{conroy_hierarchical_2007, pillepich_first_2018, chun_formation_2023}), and there is a significant region of overlap, making it difficult to define the separation between the two. There are a variety of different methods currently used in both observations and simulations (see \citealp{brough_preparing_2024} for a summary and comparison between a number of different ICL definition methods). In observations, the methods for defining the ICL range from using a surface brightness cut with the threshold placed anywhere between 25 mag/arcsec$^2$ (e.g. \citealp{burke_coevolution_2015, furnell_growth_2021}) and 26.5 mag/arcsec$^2$ (e.g. \citealp{feldmeier_deep_2004, montes_buildup_2021}) in various observational bands, applying a set physical aperture (e.g. \citealp{zhang_dark_2024, golden-marx_hierarchical_2025}), modeling and subtracting cluster galaxies using 2D parametric (e.g. \citealp{gonzalez_intracluster_2005, morishita_characterizing_2017}) or non-parametric models (e.g. \citealp{jimenez-teja_disentangling_2016}), modeling the ICL component with an assumed profile (e.g. \citealp{iodice_fornax_2016, spavone_galaxy_2024, garate-nunez_correcting_2025}), or using wavelet-based algorithms (e.g. \citealp{da_rocha_intragroup_2005, ellien_dawis_2021}). Simulations, which can take advantage of information not available in observations, choose to use methods such as 3D apertures either based on a physical aperture (e.g. \citealp{sommer-larsen_simulating_2005, pillepich_first_2018, contreras-santos_characterising_2024}) or defined as fractions of the cluster radius (e.g. \citealp{montenegro-taborda_stellar_2025, kimmig_intra-cluster_2025}), or kinematic separation of the BCG and ICL (e.g. \citealp{murante_importance_2007, dolag_dynamical_2010, remus_outer_2017}). Each of these methods come with different assumptions and systematics, and measurements are also complicated by the dependence of the ICL fraction on photometric band and limiting depth used for the observation (e.g. \citealp{burke_coevolution_2015, montes_intracluster_2018, montes_faint_2022}). It is therefore challenging to compare between different samples. In addition, many of these methods require significant time and manual tweaking, making it also challenging to construct large homogeneous samples of ICL fractions and study trends on a large scale. Large samples are necessary for drawing robust conclusions about how the ICL fraction changes with redshift and in different cluster environments. 

The trends that are revealed by our current samples of simulated and observed ICL fractions show some discrepancies. Most simulations predict that the ICL fraction will grow over time, although there is no consensus about the rate at which this growth is expected to happen. \citet{tang_investigation_2018} finds a stronger evolution with redshift as compared to \citet{rudick_quantity_2011} and \citet{canas_stellar_2020}. \citet{contini_connection_2024} finds that the ICL fraction grows at different rates during different redshift intervals, with the growth occurring fastest between $z=0.5$ and $z=0$.

Observations reveal similar discrepancies. For example, \citet{burke_coevolution_2015}, studying clusters with $0.19\leq z \leq 0.44$, and \citet{furnell_growth_2021}, studying clusters with  $0.08 \leq z \leq 0.50$, find a very strong evolution with redshift in their ICL fractions. In contrast, \citet{krick_diffuse_2007} ($0.05 \leq z \leq 0.31$) and \citet{montes_intracluster_2018} ($0.31 \leq z \leq 0.55$) find only a mild evolution across redshift, whereas \citet{jimenez-teja_unveiling_2018} find no evolution with redshift in their sample of clusters ($0.19 \leq z \leq 0.55$).

The evolution of the ICL fraction with halo mass gives us information about the efficiencies of the ICL production mechanisms in different environments. In simulations, there is no consensus of the expected relationship of ICL fraction with halo mass. Some simulations find an increasing relationship of ICL with increasing halo mass (e.g. \citealp{purcell_shredded_2007}; \citealp{contini_connection_2024}), others find no trend (e.g. \citealp{proctor_identifying_2024}, \citealp{contreras-santos_characterising_2024}), whereas others find a decreasing trend with halo mass (e.g. \citealp{cui_characterizing_2014}; \citealp{tang_investigation_2018}). \citet{montenegro-taborda_stellar_2025}, studying simulated clusters, found that the identified relationship between ICL fraction and halo mass is highly dependent on the method used to measure the ICL fraction, hinting that the use of different measurement methods could be probing different parts of the ICL component. Measurements of the ICL fraction from observational methods currently indicate no trend with halo mass, but with a very large scatter (e.g. \citealp{burke_coevolution_2015, kluge_photometric_2021, ragusa_does_2023, zhang_dark_2024}). 

Finally, we are also interested in the relationship between the dynamical state of the cluster and its ICL fraction. Simulations have generally found that more relaxed clusters have larger ICL fractions (e.g. \citealp{canas_stellar_2020}; \citealp{contreras-santos_characterising_2024}; \citealp{contini_connection_2024}; \citealp{montenegro-taborda_stellar_2025}; \citealp{kimmig_intra-cluster_2025}). Several observational studies have also found hints of this same trend, using various observational proxies for cluster relaxation (e.g. \citealp{ragusa_does_2023}; \citealp{golden-marx_hierarchical_2025}). Cluster dynamical state has also been found to impact ICL colours, revealed through comparison of ICL fractions measured in different observational bands \citep{jimenez-teja_unveiling_2018}. 

The wide variety of different measurement methods, photometric bands, and limiting observational depth that these studies are subject to means that small observational samples are not necessarily comparable. This makes it difficult to know how much of the scatter observed in these relationships is due to intrinsic scatter in the ICL fraction due to its production being tied to stochastic processes, and how much is due to different systematics in measurement method. It is therefore important to study these questions with a large homogeneous sample which allows us to disentangle these two elements. 

The main challenge for the creation of a large homogeneous sample of ICL fractions is the difficulty in scaling most current measurement methods, as many rely on manual involvement and tuning on an individual cluster basis for their accuracy (e.g. \citealp{brough_preparing_2024}). With current and upcoming imaging surveys such as the Hyper Suprime-Cam Subaru Strategic Program (HSC-SSP; \citealp{miyazaki_hyper_2018}), \emph{Euclid} \citep{mellier_euclid_2025}, and the Legacy Survey of Space and Time (LSST; \citealp{ivezic_lsst_2019}) which reach the depth needed to study the ICL and other low surface brightness features (e.g. \citealp{montes_intracluster_2019, kluge_euclid_2025}), we have too many images to rely on human involvement for measurement. For example, LSST will image the entire Southern sky to the depth needed to study the ICL in an estimated 100,000 clusters and 1 million groups up to $z\sim1.2$ \citep{brough_vera_2020}. 

Machine learning is a method that is well suited to dealing with these types of data-intensive problems, and has proven capable of matching human performance in a variety of astronomy applications (see \citealp{huertas-company_dawes_2023} for a review). In particular, it has seen successful use in image classification (e.g. \citealp{pearson_identifying_2019, hayat_self-supervised_2021, desmons_detecting_2024, huertas-company_galaxy_2024}), and image regression (e.g. \citealp{hayat_self-supervised_2021, canepa_measuring_2025}). \citet{canepa_measuring_2025} presented a machine learning based method which outputs accurate ICL fractions as benchmarked on measurements made with the manual surface brightness cut method, but without any human involvement necessary. This allows us to efficiently process a much larger sample than previously possible, using a consistent method, and begin to study the ICL at a large scale to reveal how its formation and evolution differs across a variety of clusters. 

In this work, we study how the ICL fraction varies with several cluster properties in 177 galaxy groups and clusters. 101 of these clusters were among the model training data presented in \citet{canepa_measuring_2025}, and 76 additional galaxy groups are added here. Section \ref{sec:method} describes our data sources and methods for obtaining the cluster parameters and using the machine learning method to measure the ICL fractions in our groups and clusters. Section \ref{sec:results} presents results for the ICL fraction dependence on cluster redshift, halo mass, and dynamical state. In Section \ref{sec:discussion} we discuss these results and make comparisons with other observational and simulation results from the literature. Finally, a summary of our work and conclusions are presented in Section \ref{sec:summary}. Throughout the paper, we assume a standard cosmological model with parameters $H_0=70$ km s$^{-1}$ Mpc$^{-1}$, $\Omega_m=0.3$, and $\Omega_\Lambda=0.7$.

\section{Method}
\label{sec:method}
\subsection{Data sources}
\subsubsection{Photometry}
We perform our measurements and analysis using images from the HSC-SSP Public Data Release 2 (PDR2; \citealp{aihara_second_2019}). Public Data Release 3 (PDR3; \citealp{aihara_third_2022}) is currently available, however our machine learning model was trained using images from PDR2. Using PDR3 would necessitate retraining the model from scratch due to differences in sky conditions, observing strategy, and data treatment pipeline that the model may not be able to account for. The main difference that could affect our measured ICL fractions arises from the sky subtraction pipelines, which changed slightly from PDR2 to PDR3 mainly to address issues in compute time and faint galaxy photometry \citep{aihara_third_2022}. \citet{garate-nunez_correcting_2025} and \citet{martinez-lombilla_galaxy_2023} measured ICL fractions for the same group, the former with PDR3 imaging and the latter with PDR2 imaging. They found compatible ICL fractions for this group, with the differences in calculated fractions being attributed to differences in measurement method as opposed to differences in the processing pipeline, although they note that some minor impact could be due to using different data releases. \citet{garate-nunez_correcting_2025} recovers the same trends observed by \citet{martinez-lombilla_galaxy_2023} despite the different data releases used and the differences in their measurement methods. From this, we conclude that our choice of PDR2 over PDR3 is unlikely to affect our conclusions, although it is possible that there could be small differences in ICL fractions measured as a result of the different pipelines. PDR2 has also been extensively tested and shown to be well suited for low surface-brightness studies due to its sky subtraction implementation (e.g. \citealp{huang_weak_2020}; \citealp{li_reaching_2022}; \citealp{martinez-lombilla_galaxy_2023}; \citealp{desmons_galaxy_2023}). For these reasons, we choose to continue with this data release in this work. HSC-SSP is a three layered multi-band (\textit{grizy}) imaging survey, carried out using the Hyper Suprime-Cam instrument on the 8.2m Subaru telescope. Each layer has a different limiting depth: Wide ($m_r\sim26.4$ mag), Deep ($m_r\sim27.4$ mag), and Ultradeep ($m_r\sim 28.0$ mag). In this work, we make use of both the Deep and Ultradeep layers, as these layers are imaged to a depth sufficient to study low surface brightness features such as ICL.

\subsubsection{Group and cluster catalogues}
We make use of the CAMIRA catalogue \citep{oguri_cluster_2014}, an optically-selected catalogue of clusters run on all HSC-SSP images. The algorithm is a red-sequence cluster finder, using the expectation that all massive clusters will show a ``red sequence" of galaxies (e.g. \citealp{gladders_new_2000}) to identify likely red sequence galaxies with a stellar population synthesis model and group them into clusters. The algorithm is fully described in \citet{oguri_cluster_2014}. The CAMIRA catalogue consists of 248 clusters within the Deep/Ultradeep layers of the HSC survey, where the central coordinates of each cluster are the coordinates of the BCG identified by the algorithm.

We also use the Galaxy and Mass Assembly (GAMA) survey \citep{driver_galaxy_2011} Group Catalogue (G$^3$Cv10; \citealp{robotham_galaxy_2011}) to identify additional galaxy groups. GAMA is a spectroscopic survey carried out using the optical AAOmega multi-object spectrograph on the 3.9m Anglo-Australian Telescope (AAT). The survey has a depth of $m_r\sim 19.8$ mag. Groups are identified with a friends-of-friends based grouping algorithm, described in detail in \citet{robotham_galaxy_2011}. We find 111 GAMA groups that overlap with the PDR2 HSC-SSP Deep/Ultradeep imaging area with $\ge 5$ members. The central coordinates of these clusters are the coordinates of the iterative central galaxy, which is not necessarily also the BCG, however, these are equivalent for 93 of these groups ($84$\%).

We perform the same sample cuts as in \citet{canepa_measuring_2025} (restricting our sample to clusters that have a redshift of less than 0.5, and visually inspecting the images for significant contamination due to bright stars or bad photometry). This results in 101 clusters in the CAMIRA sample, and 88 in the GAMA sample. We also remove 12 GAMA groups that match clusters within the CAMIRA catalogue (i.e. the BCG coordinates are within 1 arcsec of each other), giving us a final sample of 177 groups and clusters. 

\subsubsection{Group and cluster parameters}
\label{sec:group_and_cluster_parameters}
\begin{figure*}
    \centering
    \includegraphics[width=\textwidth]{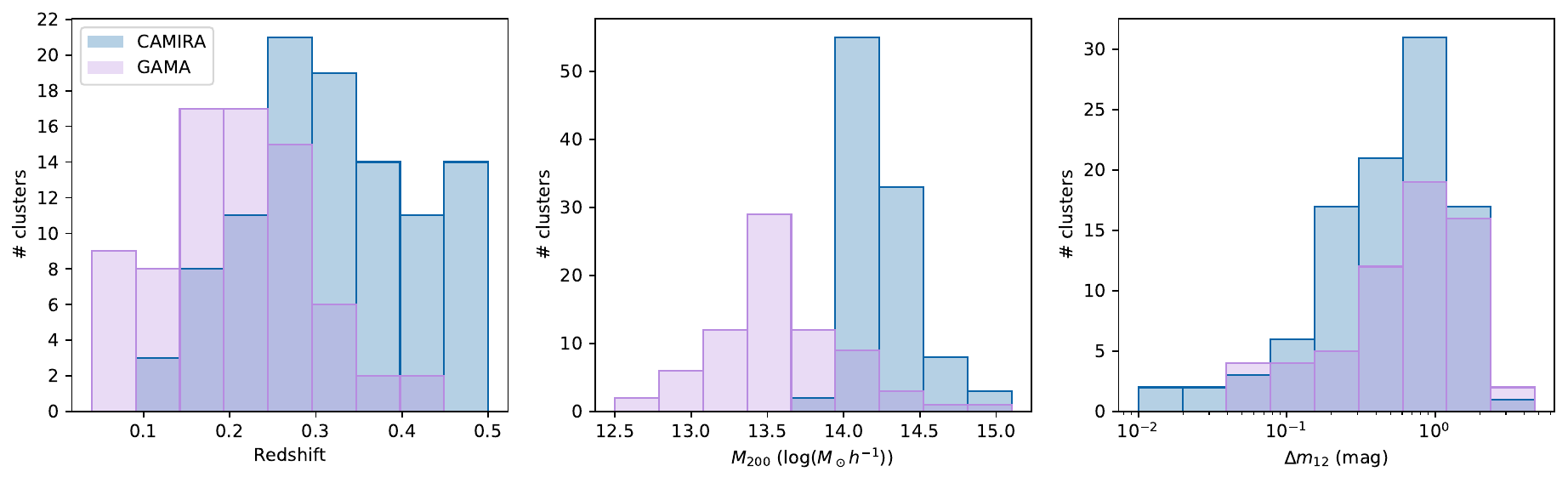}
    \caption{Distribution of redshift (left), halo mass (centre), and magnitude gap (right) for CAMIRA cluster and GAMA group sample.}
    \label{fig:params}
\end{figure*}

The CAMIRA catalogue provides BCG spectroscopic redshifts where available from other catalogues, as well as photometric cluster redshifts calculated by the algorithm based on the most likely redshift for a particular likely red sequence of galaxies. We take the spectroscopic redshifts to be the cluster redshifts where available, and for the 25 clusters where this is not available we use the photometric cluster redshift. The $1\sigma$ accuracy of the CAMIRA photometric redshift $z_\textrm{phot}$ is $0.0081(1+z_\textrm{phot})$ \citep{oguri_optically-selected_2018}. For the GAMA sample, we take the redshifts of the GAMA groups to be the median spectroscopic redshift of the group as given by the G$^3$Cv10 catalogue. The distribution of redshifts across the CAMIRA and GAMA samples is shown in Figure \ref{fig:params}, with a minimum redshift of 0.04 and a maximum redshift of 0.5.

We estimate the CAMIRA cluster halo mass using a scaling relation between CAMIRA richness (the number of red member galaxies with stellar masses $M_\star \gtrsim 10^{10.2}M_\odot$ within an aperture of radius $R\sim1h^{-1}$ Mpc) and cluster halo mass $M_{200}$ (the mass within a spherical boundary in which the mean mass density is 200 times that of present-day mean mass density). This scaling relation is shown in Equation \ref{eq:halomass} and is the mean relation derived from stacked weak lensing profiles in \citet{murata_mass-richness_2019}, where $M_\textrm{pivot}=3\times10^{14}h^{-1}M_\odot$ and $z_{\textrm{pivot}}=0.6$.

\begin{align}
    0.83\Big(\frac{M_{200}}{M_{\textrm{pivot}}}\Big) = \ln(N) &- 3.36 + 0.2\ln\Big(\frac{1+z}{1+z_\textrm{pivot}}\Big) \label{eq:halomass} \\ 
    &- 3.51\Big[\ln\Big(\frac{1+z}{1+z_\textrm{pivot}}\Big)\Big]^2 \nonumber
\end{align}

In order to ensure that our CAMIRA and GAMA samples are as comparable as possible, we calculate $M_{200}$ for the GAMA groups using a scaling relation shown in Equation \ref{eq:gama_halomass} calibrated with weak lensing measurements from \citet{viola_dark_2015}. This relates the total $r$-band luminosity of the group given by the G$^3$Cv10 catalogue to the group's halo mass. 

\begin{equation}
    \frac{M_{200}}{10^{14}h^{-1}M_\odot}=0.95\Big(\frac{L_\textrm{grp}}{10^{11.5}h^{-2}L_\odot}\Big)^{1.16}
    \label{eq:gama_halomass}
\end{equation}

There is one outlier group where this calculation gives an unrealistic group halo mass of $10^{11.6}h^{-1}M_\odot$, so we remove this group from our halo mass analysis. The halo mass distribution across the GAMA and CAMIRA samples are shown in Figure \ref{fig:params}.

\begin{figure}
    \centering
    \includegraphics[width=\linewidth]{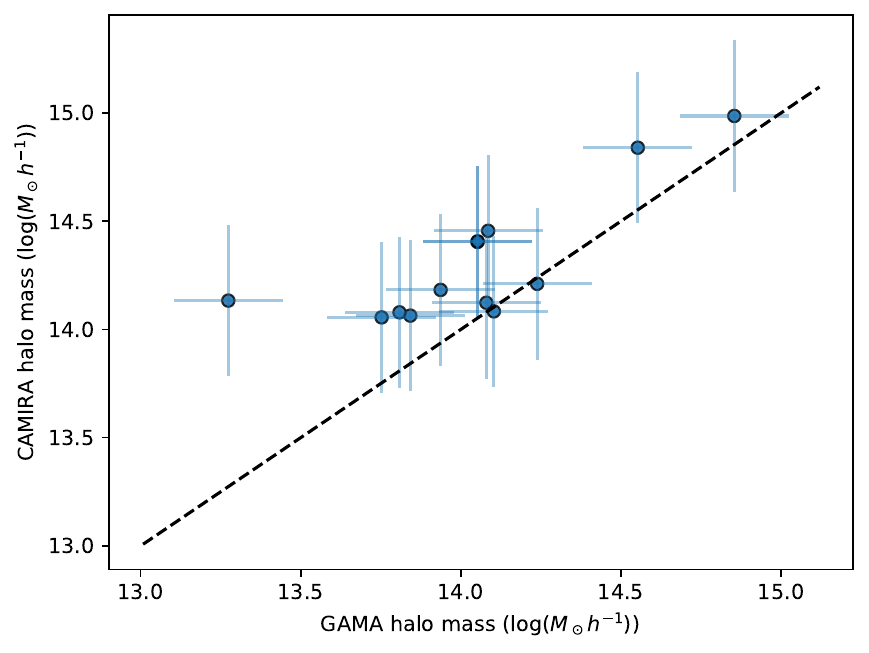}
    \caption{Halo masses for clusters identified by both CAMIRA and GAMA calculated with the CAMIRA \citep{murata_mass-richness_2019} and GAMA \citep{viola_dark_2015} scaling relations. The mean difference in halo masses is -0.25 dex.}
    \label{fig:gama_camira_halomass_comparison}
\end{figure}

We compare the agreement of the two scaling relations using the 12 clusters present in both the GAMA and CAMIRA catalogues as shown in Figure \ref{fig:gama_camira_halomass_comparison}. The offset in halo masses of the clusters calculated with the GAMA-calibrated scaling relation from the halo masses calculated with the CAMIRA-calibrated scaling relation has a mean of $-0.25\pm0.23$ dex, and a maximum absolute difference of 0.86 dex, meaning that there does appear to be a systematic offset between the two relations. Due to the fact that we only have very few clusters in both samples, we refrain from applying an offset calculated from this small sample to our wider sample in this work. However, we have tested the effect of scaling the GAMA group halo masses up by 0.25 dex to correct for the offset implied by these clusters, and find that our conclusions remain the same.

The magnitude gap, $\Delta m_{12}$, is the difference in $r$-band magnitude between the brightest and second brightest galaxy within $0.5R_{200}$ of the cluster (e.g. \citealp{dariush_mass_2010, golden-marx_hierarchical_2025}). Clusters that are actively merging are more likely to have a smaller magnitude gap as bright and massive galaxies infall and start merging with the BCG due to dynamical friction. This leads to significant growth in the BCG relative to the surrounding satellite galaxies. The magnitude gap thus traces the hierarchical formation of the BCG (e.g. \citealp{tremaine_test_1977, oliva-altamirano_galaxy_2014, golden-marx_hierarchical_2025}). A cluster with a BCG that is very dominant over the other galaxies indicates an older cluster that has already completed a number of significant merging events, with no current major mergers ongoing. $\Delta m_{12}$ has been found to correlate with earlier formation times in simulations (e.g. \citealp{donghia_formation_2005, dariush_mass_2010, yoo_spatial_2024, kimmig_intra-cluster_2025}). It has also been found in observations to correlate with more relaxed clusters where relaxation is measured with X-ray properties (e.g. \citealp{casas_optical_2024}) or mass concentration (e.g. \citealp{vitorelli_mass_2018}).

Since $\Delta m_{12}$ is an optical photometric indicator of dynamical state, it is suited to our sample since we only have photometric data available for our entire sample of groups/clusters. To calculate the magnitude gap, we need information about the group and cluster member galaxies. 

Both CAMIRA and GAMA provide membership catalogues. We use the GAMA membership catalogue as is, as these members are identified spectroscopically. However, since CAMIRA is a red-sequence cluster finder, its membership catalogue only identifies red-sequence members. In order to have more complete membership information, we perform a reclassification of members for the CAMIRA clusters using additional information. To do this, we make use of spectroscopic redshifts from the Dark Energy Spectroscopic Instrument (DESI; \citealp{desi_collaboration_overview_2022}) Early Data Release (EDR; \citealp{desi_collaboration_early_2024}). We make a cut in spectroscopic redshift and distance from the centre of the cluster in order to classify DESI galaxies as cluster members. We relate the halo mass of the cluster calculated using Equation \ref{eq:halomass} to its velocity dispersion $\sigma_{\textrm{cl}}$ and its $R_{200}$ (the spherical radius within which the mean mass density is 200 times that of present-day mean mass density) using Equations \ref{eq:sigmacl} and \ref{eq:r200} respectively (e.g. \citealp{finn_h-derived_2005}). 

\begin{align}
    \Big(\frac{\sigma_\textrm{cl}}{1000 \textrm{ km s}^{-1}}\Big)^3=\frac{\sqrt{\Omega_{\Lambda}+\Omega_M(1+z)^3}}{1.2\times10^{15} M_{200}}M_\odot \label{eq:sigmacl}\\
    R_{200} = 1.73 \frac{\sigma_\textrm{cl}}{1000\textrm{ km s}^{-1}}\frac{1}{\sqrt{\Omega_{\Lambda}+\Omega_M(1+z)^3}} \textrm{ Mpc} \label{eq:r200}
\end{align}

We then define the redshift limit as the redshift range that places the galaxy within $3\sigma_{\textrm{cl}}$, and the radius limit as $0.5R_{200}$, as this information will be used for calculating $\Delta m_{12}$ which is usually measured within $0.5R_{200}$ (e.g. \citealp{golden-marx_hierarchical_2025}).

Although this effectively identifies high confidence members, the DESI survey is not entirely complete, and can only be applied to CAMIRA clusters that have a spectroscopic redshift available for the BCG (76 out of 101 CAMIRA clusters), because we use this to calculate the limiting redshifts. We therefore combine this information with CAMIRA membership information, and classify a galaxy as a member of a CAMIRA cluster if it meets any of the following conditions:
\begin{enumerate}
    \item It is identified as the cluster BCG by the CAMIRA algorithm.
    \item The CAMIRA catalogue defines it as a member with membership probability $P_\textrm{mem}>0.8$, and it does not have a DESI spectroscopic redshift that excludes it from being part of the cluster according to the redshift limit defined above.
    \item It has a DESI spectroscopic redshift that meets the redshift limit and a position that meets the radius limit defined above.
\end{enumerate}

Of the 101 clusters in our CAMIRA sample, 75 are covered by the DESI survey, and 49 of these have new membership information from DESI in the form of new members and/or excluded CAMIRA members. In Appendix \ref{appendix:membership_examples}, we show examples of how the additional DESI information changes the membership assignment and magnitude gap calculations.

To calculate the magnitude gap, we obtain HSC-SSP $r$-band Kron photometry for all cluster members, and calculate the difference in magnitude between the first and second brightest galaxies in the cluster. Of the 49 CAMIRA clusters that have additional membership information from DESI, 26 result in a different magnitude gap when including the DESI spectroscopic information. 18 of these are because of a new brightest or second brightest member that was not in the CAMIRA catalogue, and the remaining 8 are due to a brightest or second brightest member that is excluded from being part of the cluster from spectroscopic information. The mean difference between the magnitude gap calculated with and without the additional DESI data is $\overline{\Delta m_{12}} = (-0.03\pm 0.44)$ mag. The distribution of magnitude gaps in our sample is shown in Figure \ref{fig:params}, ranging from 0.01 to 2.5 with a mean of 0.71. We still include magnitude gaps calculated for the CAMIRA clusters that were not covered by DESI, as the difference in magnitude gaps when including DESI information is on average much smaller than the mean magnitude gap value and therefore will overall remain unbiased. We have tested the effect of including only the 75 CAMIRA clusters that are covered by DESI and find that our conclusions are unchanged. There are 15 groups and clusters that do not have a valid $\Delta m_{12}$ value, 14 from the GAMA sample and 1 from the CAMIRA sample, and therefore we do not use them when exploring relationships with $\Delta m_{12}$. These are invalid because they do not have an identified member other than the BCG within $0.5R_{200}$. 

\begin{figure*}
    \centering
    \includegraphics[width=\textwidth]{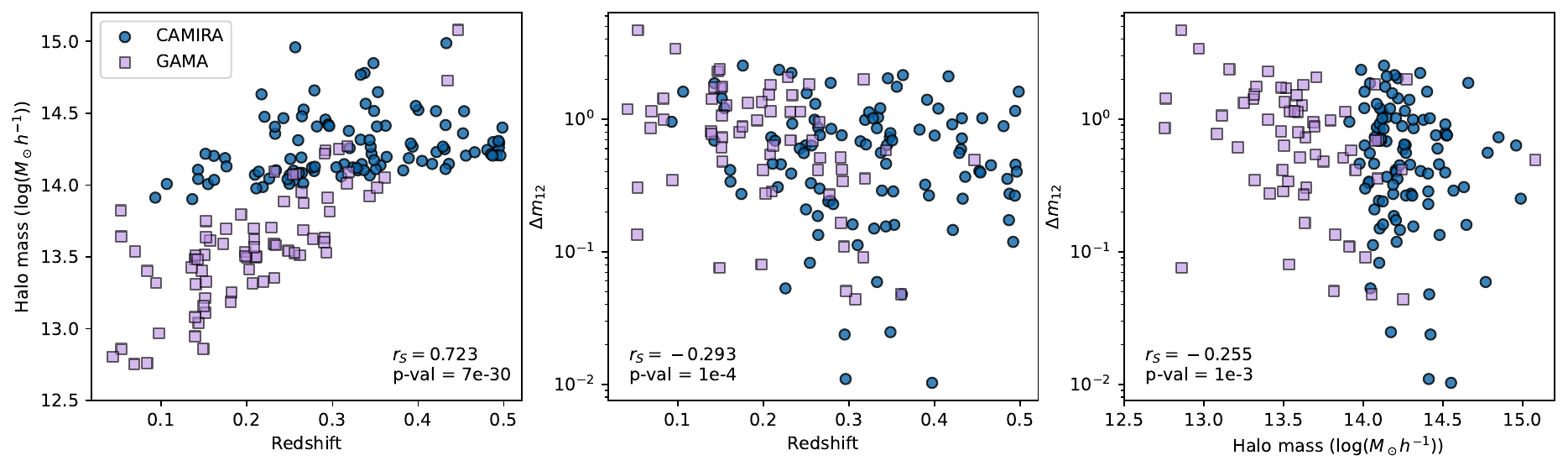}
    \caption{Correlation between redshift, halo mass, and magnitude gap parameters for the CAMIRA and GAMA samples. Spearman correlation coefficients and p-values for the combined sample are shown on each panel. Each of these parameters have significant correlations at the $>3\sigma$ level with each other.}
    \label{fig:params_corr}
\end{figure*}

Figure \ref{fig:params_corr} shows the relationships between our studied cluster parameters (redshift, halo mass, and magnitude gap). The Spearman correlation coefficients, $r_S$, and p-values are also indicated on the plots. All of these parameters have significant correlations at the $>3\sigma$ level with each other (i.e. p-val $<0.003$), although the strength of the magnitude gap-redshift correlation and magnitude gap-halo mass correlations are quite weak. In our analysis of the ICL fraction and its relationship to any of these parameters, we must first ensure that the correlations between the other parameters are removed.

\subsection{Measurement of the ICL fraction}
\citet{canepa_measuring_2025} presented a regression machine learning model (MICL; \citealp{canepa_lpcanmicl_2024}) to predict the ICL fraction from input HSC-SSP galaxy cluster images. The model consists of a convolutional neural network followed by dense linear layers that output parameters to a probability distribution, from which the mode is extracted as the most probable ICL fraction, and $1\sigma$ errors are estimated from the full distribution. The model was trained first on a large artificially generated dataset, consisting of 50,000 images centred on luminous red galaxies (LRGs) from the HSC-SSP survey, with a randomised exponential profile injected into the image standing in as the ``ICL". The model's learning was then transferred from this artificial dataset onto real cluster images through fine-tuning on our CAMIRA sample of 101 galaxy clusters.

The target fractions for the fine-tuning sample were generated through manual measurement of the images using the surface brightness cut method, adopting a threshold of $\mu_r=26$ mag arcsec$^{-2}$. The process for manually measuring the images is discussed in detail in \citet{canepa_measuring_2025}. Briefly, we prepare the images through visual inspection, manually creating masks for bright stars that were not automatically identified in HSC-SSP, and manually estimating and subtracting any image background, checking for both 2D gradients and constant sky background. Cluster members are identified as galaxies that have a photometric redshift within $3\sigma$ of the cluster redshift, while all other objects in the image are masked using a two-step masking process which first masks bright extended sources (e.g. foreground objects), and then masks dim compact sources (e.g. background objects). We then calculate the surface brightness limits of the image following \citet{roman_galactic_2020}. Finally, we correct the surface brightness cut for cosmological dimming and by applying a $k$-correction calculated as a function of redshift using the ``$k$-corrections calculator"\footnote{http://kcor.sai.msu.ru} \citep{chilingarian_analytical_2010} assuming the colour of the system within the ICL region to be $g-r=0.7$ \citep{martinez-lombilla_galaxy_2023}. The ICL is then identified as the unmasked flux in the image below the surface brightness cut and above the calculated surface brightness limit of the image, while the total cluster light is all unmasked flux in the image above the calculated surface brightness limit. The measurements are all conducted within a fixed radius of 300 kpc. By training the model on these manually measured fractions, the model learns to mimic the manual measurements on images which are given to the model without any background subtraction or masking. See \citet{canepa_measuring_2025} for more discussion on the architecture, training, and performance of this model, as well as the potential impact of our choice of a fixed physical radius as opposed to a radius scaled to the mass of the system, such as $R_{500}$.

In this work, we make use of the version of the model that was fine-tuned on all 101 of the available CAMIRA clusters. This model version is run on the CAMIRA clusters, as well as our sample of 75 GAMA groups to expand our sample of ICL fractions to lower halo masses. Running the same model version on all images ensures that the predictions for both the GAMA groups and CAMIRA clusters are made with exactly the same model parameters to ensure an unbiased comparison. We use HSC-SSP PDR2 imaging to be consistent with what the model was previously trained on. The GAMA groups constitute an unseen sample of images for the model. Therefore, although we are using identical imaging, due to the slight differences in this input sample, we need to check the model's performance on these previously unseen images.

To check the model's predictions, we manually measured 10 of the GAMA groups at varying predicted ICL fractions. We randomly selected the 10 groups by binning the clusters in equally spaced bins over the range of the predicted ICL fractions, and randomly selecting a group from each bin. We follow the manual measurement method presented in \citet{canepa_measuring_2025} in order to maintain consistency with the measurements that the model has been trained to replicate.  

\begin{figure}
    \centering
    \includegraphics[width=\linewidth]{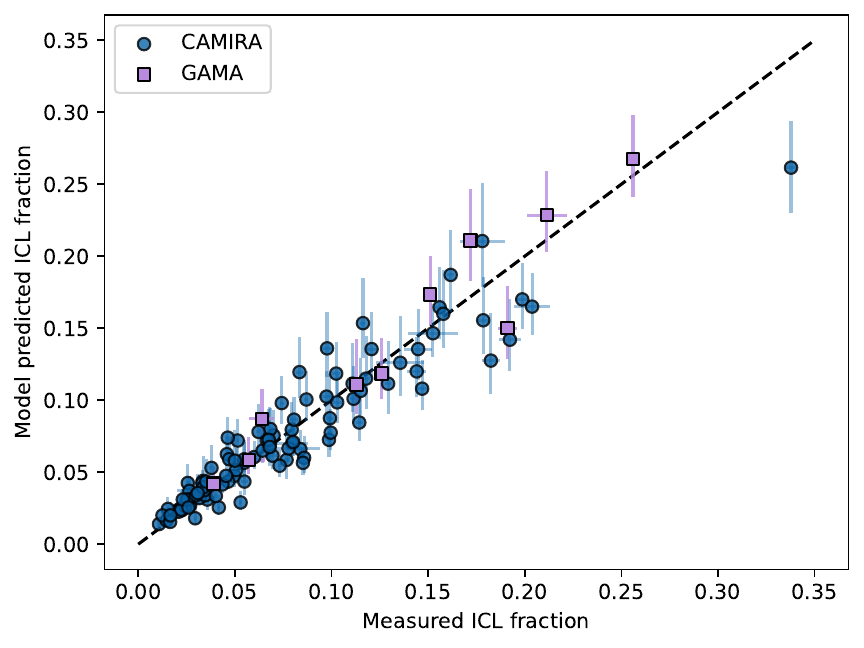}
    \caption{Model predicted ICL fractions for 10 GAMA groups and the CAMIRA sample of 101 clusters, compared to manually measured ICL fractions. The dashed line indicates a one-to-one-relationship. Vertical error bars are taken from the $1\sigma$ confidence intervals calculated from the model's output probability distribution. The model does not exhibit any particular bias in its predictions on either sample, and we find that it provides an accurate estimation of its uncertainties on its predicted fractions.}
    \label{fig:model_compare}
\end{figure}

Figure \ref{fig:model_compare} compares the model's predictions with our manual measurements of these clusters from \citet{canepa_measuring_2025} for the 101 CAMIRA clusters, and our new manual measurements for the 10 GAMA groups. We find that the model's prediction is consistent within $1\sigma$ uncertainties for 6 out of 10 of the GAMA groups, and within $2\sigma$ uncertainties for all 10 GAMA groups. The mean scatter of the predictions of the ICL fractions from the 10 GAMA groups is $-0.0067$, the standard deviation is 0.021, and the maximum absolute difference in these predictions is 0.041. For the CAMIRA clusters, 62.3\% of measurements are within the $1\sigma$ uncertainties output by the model, and $92\%$ of measurements are within $2\sigma$ uncertainties. The scatter has a mean of 0.0018, standard deviation of 0.018, and a maximum absolute difference of 0.072. This indicates that the model is not exhibiting any particular bias in its predictions on either sample, and is successfully predicting the manually measured ICL fractions. It also indicates that the model provides an accurate estimation of its uncertainties on the manually measured fraction. 

\begin{figure*}
    \centering
    \includegraphics[width=\textwidth]{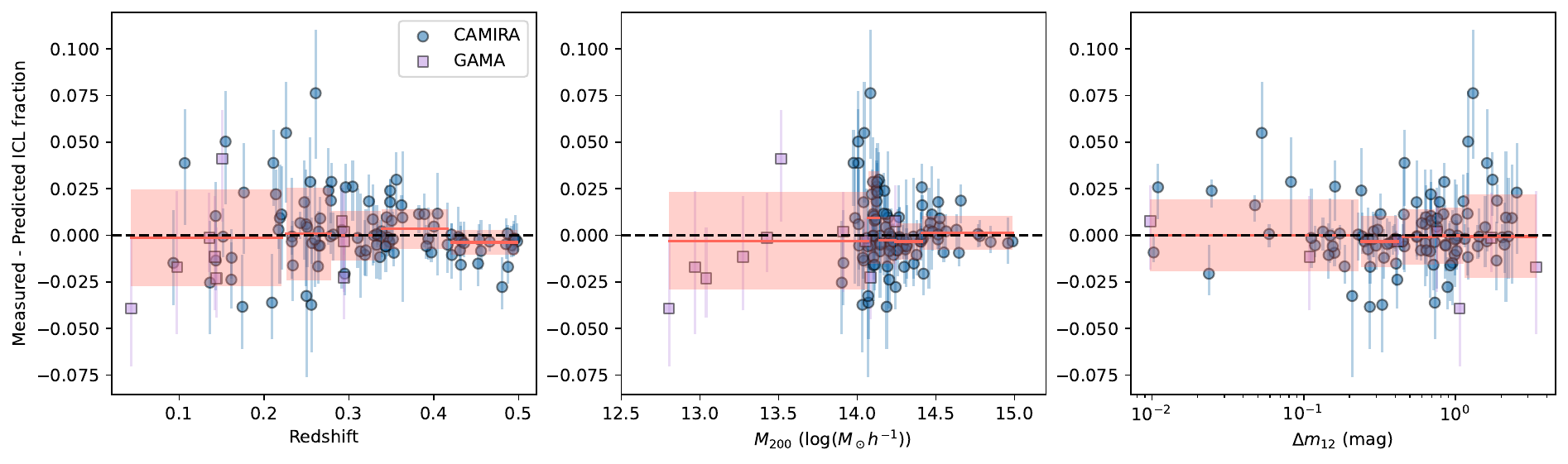}
    \caption{Model residuals as a function of redshift (left), halo mass (centre), and magnitude gap (right). Red boxes denote the five bins with equal numbers of clusters, with the vertical extent of each bin showing one standard deviation and the centre line showing the median. For all bins, the residuals are consistent with 0 within one standard deviation, indicating that the model's performance is not dependent on these parameters.}
    \label{fig:model_residuals}
\end{figure*}

Finally, we check that there is no trend with the accuracy of the model's predictions with any of our studied parameters (redshift, halo mass, and magnitude gap). We bin each of the cluster parameters into five bins, with equal numbers of clusters within each bin. We then calculate the median and standard deviation of the model's residuals within each bin, and find that in all bins, the residuals are consistent with 0 within one standard deviation, indicating that the model's performance is not dependent on any of these parameters. Figure \ref{fig:model_residuals} shows these results for each of the parameters. 

While there is some scatter in the model's predictions, meaning that on a single cluster basis the model may not be perfectly accurate, these tests show that the predictions are unbiased for both the GAMA and CAMIRA samples. This means that the ICL fraction trends that we observe for a large sample will be directly comparable to measurements made manually with the surface brightness cut method. 

\section{Results}
\label{sec:results}
\begin{figure*}
    \centering
    \includegraphics[width=0.9\linewidth]{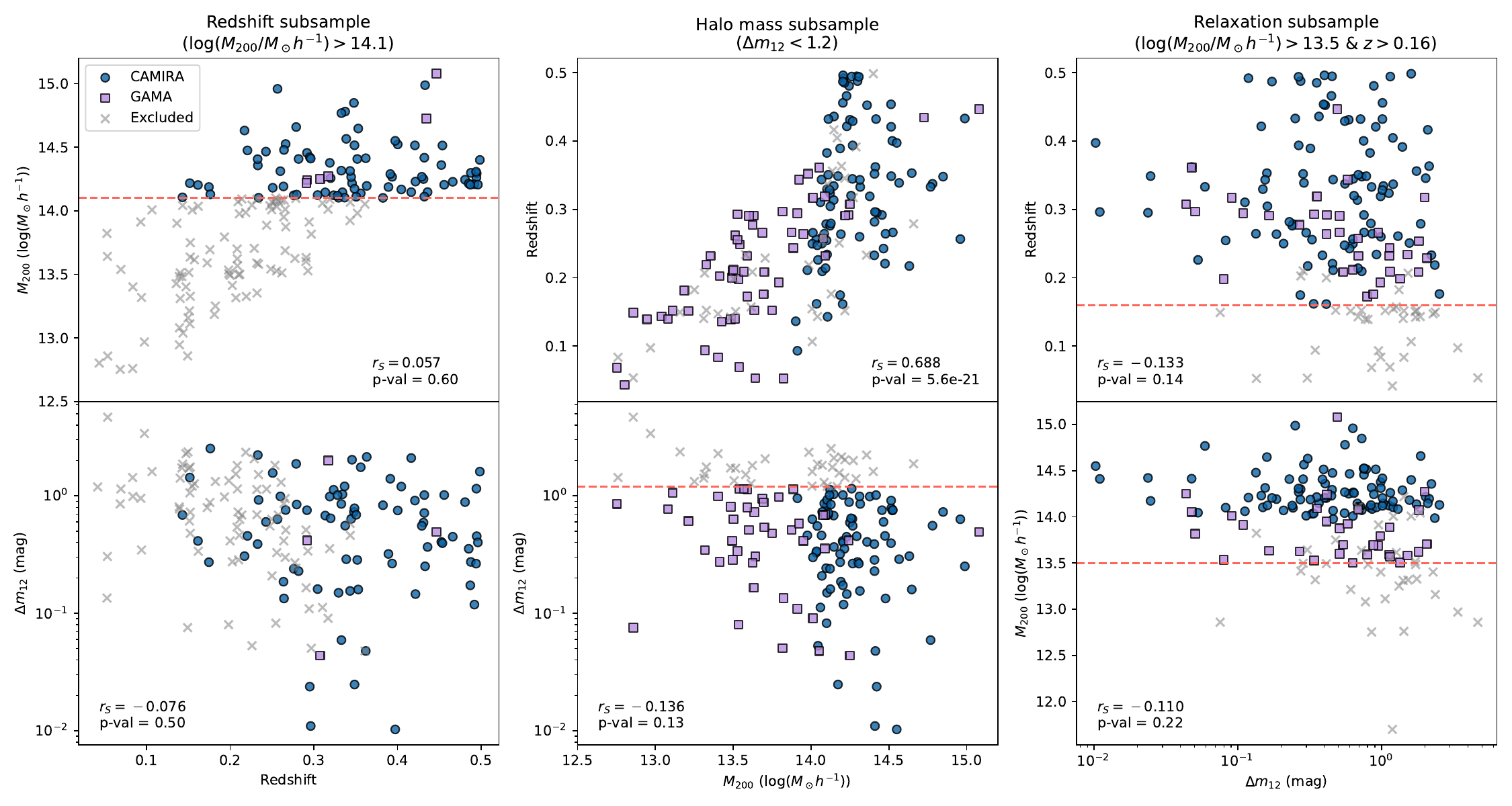}
    \caption{Sample cuts made for each subsample as in Table \ref{tab:subsamples}, shown by the red dashed lines, and how these cuts affect the correlations with the other two parameters. The left column shows the redshift subsample (used in Section \ref{sec:redshift}), the middle column shows the halo mass subsample (used in Section \ref{sec:halomass}), and the right column shows the relaxation subsample (used in Section \ref{sec:relaxation}). Spearman coefficients and p-values after these sample cuts are noted on each panel. In the case of the halo mass subsample (centre column), a simple cut is insufficient to remove the correlation with redshift (top centre panel), so we account for this differently as discussed in Section \ref{sec:halomass}.}
    \label{fig:subsamples}
\end{figure*}

In this section, we present the observed trends of ICL fraction with redshift, halo mass, and magnitude gap within our samples. Each parameter has significant correlations with the others as shown in Figure \ref{fig:params_corr}, so for our analysis of the ICL fraction trends with each parameter, we make cuts in our sample to ensure the significance of the Spearman correlations with the other parameters is under $2\sigma$ (p-value $>0.05$). These cuts are made to construct subsamples where the other parameters do not correlate with the parameter of interest, enabling us to isolate the effect of each parameter on the ICL fraction individually. In the case of the halo mass subsample (Figure \ref{fig:subsamples}, centre column), we are unable to make a simple cut that removes the correlation between halo mass and redshift. We therefore do not make a cut in this plane, and instead handle the correlation with redshift differently in that analysis, as discussed in Section \ref{sec:halomass}. Table \ref{tab:subsamples} and Figure \ref{fig:subsamples} show each subsample, and how these sample cuts affect the correlations between parameters shown in Figure \ref{fig:params_corr}. The subsamples and cuts are discussed further in Sections \ref{sec:redshift}, \ref{sec:halomass}, and \ref{sec:relaxation}. 

\begin{table*}
    \centering
    \begin{tabular}{c|c|c|c|c}
        \textbf{Subsample name} & \textbf{Definition} & \textbf{Number of GAMA groups} & \textbf{Number of CAMIRA clusters} & \textbf{Total sample size} \\
        \hline
        Redshift subsample & $\log(M_{200}/(\textrm{M}_\odot h^{-1}))>14.1$ & 6 & 79 & 85 \\
        \hline
        Halo mass subsample & $\Delta m_{12} < 1.2$ & 57 & 83 & 140\\
        \hline
        Relaxation subsample & \makecell{$z > 0.16$ and \\$\log(M_{200}/(\textrm{M}_\odot h^{-1})) > 13.5$} & 33 & 94 & 127
    \end{tabular}
    \caption{Subsamples created for the analysis of ICL fraction trends, as detailed in Sections \ref{sec:redshift}, \ref{sec:halomass}, and \ref{sec:relaxation}. The effect of these cuts on correlations with the other two parameters is shown in Figure \ref{fig:subsamples}.}
    \label{tab:subsamples}
\end{table*}

\subsection{Redshift}
\label{sec:redshift}
Figure \ref{fig:params_corr} shows that there is a significant correlation between redshift and halo mass, and redshift and magnitude gap within our sample. By restricting our sample to clusters with $\log(M_{200}/(\textrm{M}_\odot h^{-1}))>14.1$, we construct a subsample in which cluster redshift does not correlate significantly with halo mass or magnitude gap, allowing us to isolate the effect of the cluster redshift on the ICL fractions. We denote this as the redshift subsample as shown in Table \ref{tab:subsamples}. The p-values of the Spearman correlations between redshift and mass, and redshift and $\Delta m_{12}$ are 0.10 and 0.62 respectively, shown in Figure \ref{fig:subsamples}. This cut results in a subsample of 85 groups and clusters. 

\begin{figure}
    \centering
    \includegraphics[width=\linewidth]{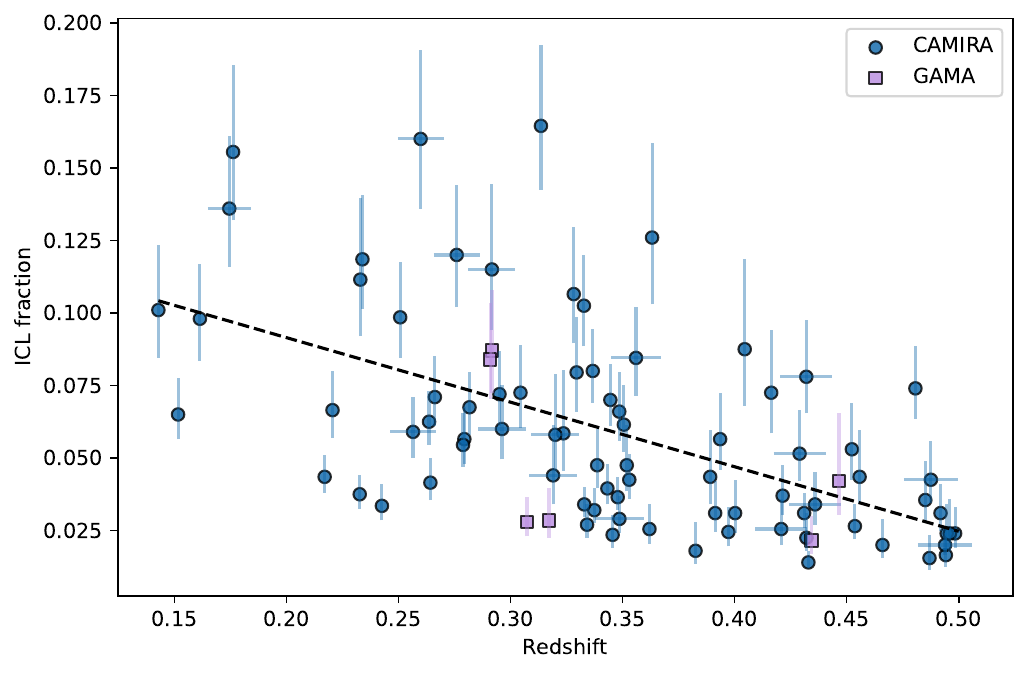}
    \caption{The ICL fraction's dependence on redshift in our redshift subsample. The Spearman correlation coefficient is $r_S=-0.604$, with a p-value$=9\times10^{-10}$. The dashed line shows the line of best fit to these points, which has equation $f_{\textrm{ICL}}=(-0.223\pm 0.035)z +(0.136\pm 0.013)$.}
    \label{fig:iclfrac_redshift}
\end{figure}

Figure \ref{fig:iclfrac_redshift} shows the relationship between ICL fraction and cluster redshift for the redshift subsample. We find that there is a decreasing trend such that the ICL fraction is lower with increasing redshift. A Spearman correlation coefficient of $r_S=-0.604$ shows this to be significant at a p-value $=9\times10^{-10}$ level.

\subsubsection{Observational effects}
Redshift is a parameter that could have a significant effect on the visibility of the ICL and therefore the trend that we find with ICL fraction. We need to account for these observational effects---cosmological surface brightness dimming and the passive aging of the stellar populations in the cluster member galaxies and the ICL---when considering what the true trend of ICL fraction with redshift may be. 

Cosmological surface brightness dimming means that both the ICL and the cluster galaxies appear dimmer at higher redshifts. They are both dimmed by the same $(1+z)^{-4}$ factor, which is corrected for when applying the surface brightness cut. However, this does not account for the surface brightness limits of the image, meaning that for a given depth, clusters that are higher in redshift have more of their ICL falling below the surface brightness limit of the image, lowering the ICL fraction. On the other hand, as stars age within both the BCG and the ICL, they become fainter in the $r$-band, so the same BCG and ICL stellar populations observed at a higher redshift will appear brighter. This will cause some ICL to be elevated above the surface brightness cut, and be counted as BCG light instead, pushing the ICL fraction further down. A very small amount of light will also be elevated above the surface brightness limits of the image due to the brightening of these stellar populations, slightly increasing the ICL fraction. This is further complicated by the fact that the BCG and ICL may have differing stellar populations, causing them to brighten differently. These effects together could be responsible for artificially lowering the ICL fraction at higher redshifts.

In order to investigate how these effects change the observed ICL fraction, we create a model cluster, and simulate what this system would look like if we observed it at different redshifts. We model the BCG and ICL with a double S\'ersic (\citeyear{sersic_atlas_1968}) profile, using the parameters measured for the cluster Abell 85 (hereafter A85) from \citet{montes_buildup_2021}. The A85 satellite galaxies are also modelled with S\'ersic functions, using parameters from \citet{owers_sami_2019}. 

To calculate the change in magnitude due to the de-aging of the stellar populations, we need an age and metallicity for each of these components. For the satellite galaxies, we use single stellar population parameters from the Sydney-AAO Multi-object Integral field spectrograph (SAMI) Galaxy Survey Data Release 3 \citep{croom_sami_2021} for their age and metallicity at the redshift of the cluster, $z=0.0549$ \citep{owers_sami_2017}. Given that we do not have accurate stellar population parameters for the ICL, we assume two cases: firstly, that the BCG and the ICL are the same age, and secondly, that the BCG and ICL are different ages. The latter case is likely to be more realistic, as several studies have found the presence of a radial gradient in BCG and ICL ages (e.g. \citealp{montes_intracluster_2014, montes_intracluster_2018, morishita_characterizing_2017, jimenez-teja_unveiling_2018}). For the same age case, we assume an age of 12 Gyr (at $z\approx 0$) for both the BCG and the ICL, and calculate the metallicity of each pixel in the BCG/ICL image using its $g-i$ colour from \citet{montes_buildup_2021}. For the differing age case, we use age and metallicity gradients measured in \citet{montes_intracluster_2018} from two typical massive clusters -- MACSJ1149.5+2223 (hereafter M1149) and Abell S1063 (hereafter AS1063), again assuming that the BCG has an age of 12 Gyr at $z\approx0$. M1149 has a difference in age between BCG and ICL of roughly 2 Gyr, whereas AS1063 has a difference of roughly 4 Gyr. 

To de-age the stellar populations, we use the \citet{bruzual_stellar_2003} models computed using the Padova 1994 isochrones. We expand the grid of models from seven metallicities between -1.79 $\leq$ [Fe/H] $\leq$ $+0.26$ and 50 ages from 0.03 to 17.8 Gyr to 200 metallicities and 200 ages by linearly interpolating the original SSPs, choosing a Chabrier (\citeyear{chabrier_galactic_2003}) IMF. The S\'ersic parameters of the BCG and ICL in \citet{montes_buildup_2021} are measured in the $g$ and $i$ bands, so to make the results more comparable to ours, which are measured in $r$ band images, for each pixel in the $i$ band model we calculate the expected flux in the $r$ band for the assumed age and metallicity. 

We are then ready to place the model at varying observed redshifts. Firstly, we scale the flux of the original image using the distance modulus to calculate the change in magnitude at the target redshift. We then rebin the image by calculating the conversion from kpc to arcsec for the target redshift and assuming HSC's pixel scale of 0.168 pix arcsec$^{-2}$. Together, this gives the model the correct angular size, and also creates the cosmological surface brightness dimming effect. Finally, using the ages of each component, we calculate the age of this stellar population at the target redshift and find the change in magnitude due to the de-aging of these stellar populations using our SSP models, and modify the flux of the model to apply this change in magnitude.

We also apply background Gaussian noise to the final image, using the equation from \citet{roman_galactic_2020} to calculate the standard deviation of the background based on the expected surface brightness limit of the image, $\mu_r\sim29.8$ mag arcsec$^{-2}$ using the surface brightness limits of the Ultradeep layer of HSC-SSP measured by \citet{martinez-lombilla_galaxy_2023}. 

We then measure the ICL fraction in the resultant image using the same surface brightness threshold method used for the manual measurements of our images, so that it is directly comparable to our sample. We do this by simply applying a threshold at 26 mag arcsec$^{-2}$, $k$-corrected and corrected for cosmological surface brightness dimming. This will remove the cosmological surface brightness dimming effect, except where the ICL has become undetectable due to falling below the surface brightness limit of the image. However, it does not correct for the brightening of the stars due to the de-aging of the stellar populations, which leads to some light that would be assigned to the ICL at low redshifts being instead assigned to the BCG at higher redshifts, lowering the ICL fraction. 

\begin{figure}
    \centering
    \includegraphics[width=\linewidth]{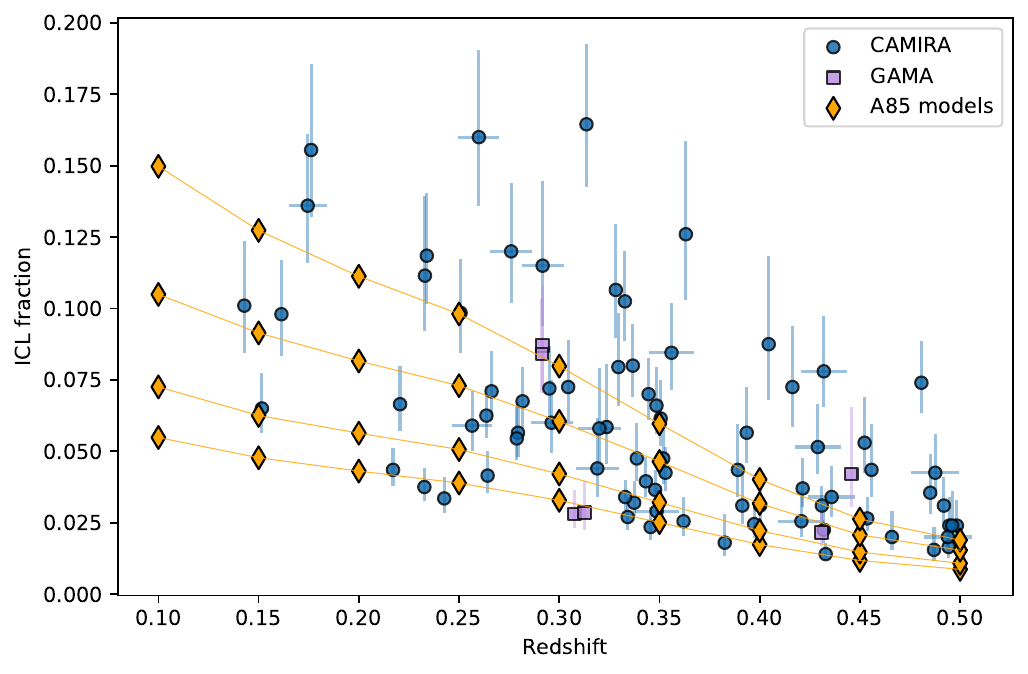}
    \caption{ICL fractions measured in A85 models with different BCG and ICL total magnitudes leading to different starting ICL fractions. All other model parameters are kept the same, and we assume the M1149 metallicity profile for these models.}
    \label{fig:dimming_frac_comparison}
\end{figure}

Figure \ref{fig:dimming_frac_comparison} shows the measured ICL fractions for our modelled A85 cluster with different starting ICL fractions, along with our observed data. The different starting ICL fractions at $z=0$ were created by changing the total magnitude of the ICL and BCG profiles, but leaving the cluster member profiles and other BCG/ICL parameters the same. We use the M1149 assumed metallicity profile for these models. There is clearly a significant impact from these observational effects on the measured ICL fractions, where they decrease as the redshift of the observations is increased. We also see that there is a tendency for all models to converge at a low ICL fraction as the redshift is increased. This indicates that the measured ICL fraction may have significantly less discriminatory power at these higher redshifts, and could also explain the apparent higher scatter we observe in our data at lower redshifts compared to a lower scatter at higher redshifts.

We also measured ICL fractions for the models made with the three different age profile assumptions (no age difference, M1149 profile, AS1063 profile), but find that these differing age assumptions have only a very minimal effect on the ICL fractions measured, with the maximum difference in ICL fractions occurring between the same age case and the AS1063 case of just 0.003.

The observational effects that we have applied to our simplistic model of A85 do appear to provide a plausible explanation for the shape and strength of the relationship between ICL fraction and redshift that we observe in our data. It is possible that most or all of the trend that we observe in our measured ICL fractions could be explained by observational effects, rather than a true decrease in ICL fraction with redshift. However, the simplicity of this model, combined with the large scatter in our observed values prevents any strong conclusions about whether these observational effects can account for all of the trend that we observe. Although it does indicate that more consideration and research is needed into the effects of redshift and how this may be influencing observational trends. 

\subsection{Halo mass}
\label{sec:halomass}

To account for the trend with magnitude gap and halo mass seen in Figure \ref{fig:params_corr}, we make a cut to exclude all clusters with $\Delta m_{12} > 1.2$. This cut is chosen to ensure that the resulting subsample will not have a significant Spearman correlation between halo mass and magnitude gap, meaning that any potential effect of magnitude gap will not influence our analysis of the effect of halo mass on ICL fraction. As shown in Figure \ref{fig:subsamples}, this subsample has a p-value = 0.13 for the Spearman correlation between $\Delta m_{12}$ and $M_{200}$, and consists of 166 groups and clusters. We call the resulting sample the halo mass subsample as shown in Table \ref{tab:subsamples}. Due to the strength of the correlation between halo mass and redshift, we are unable to make a cut in redshift large enough to remove the remaining trend between redshift and halo mass without removing a significant portion of our sample, so we instead account for this by performing a partial Spearman correlation, controlling for redshift. 

\begin{figure*}
    \centering
    \includegraphics[width=\linewidth]{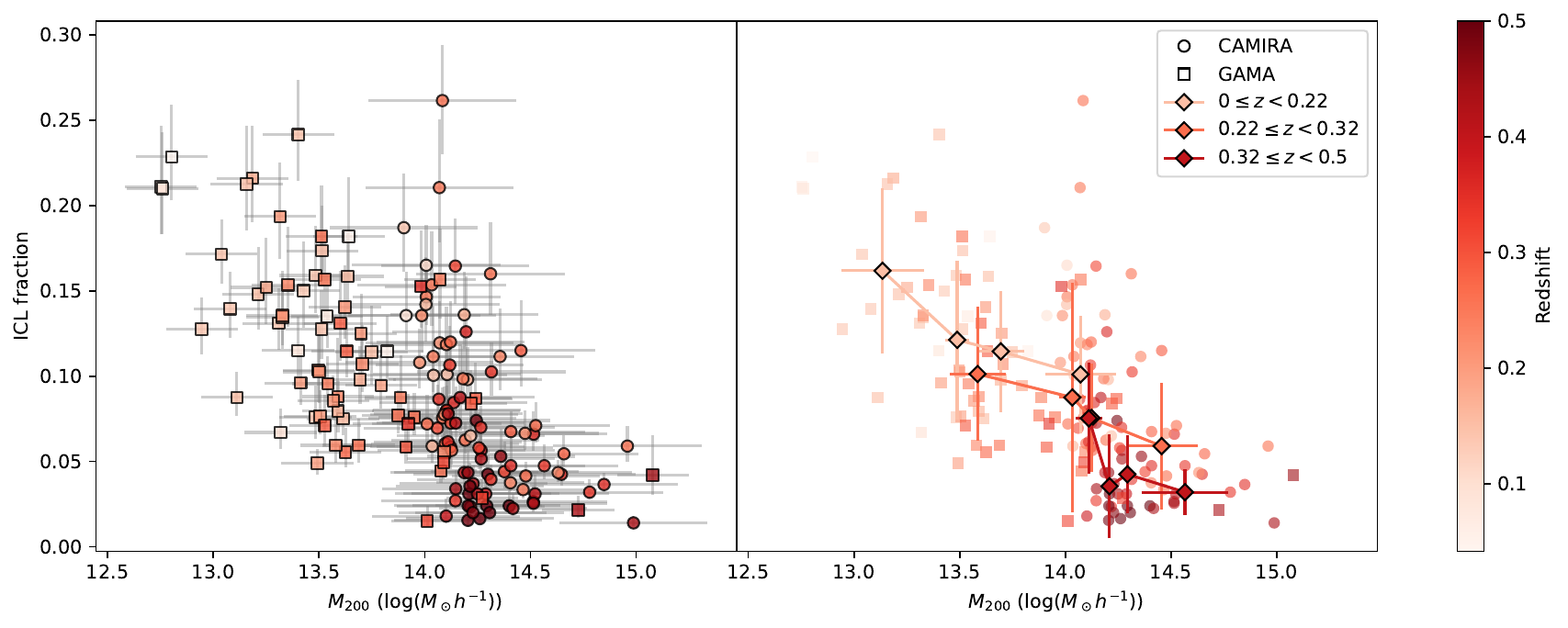}
    \caption{The ICL fraction's dependence on halo mass in our halo mass subsample. Points are coloured by redshift. There is a significant decreasing correlation ($r_S=-0.613$ with p-value = $10^{-13}$), but redshift also has a strong effect. The partial Spearman correlation coefficient, controlling for redshift, is -0.330, with p-value = $8\times 10^{-5}$, indicating that there is a weak, but significant, decreasing trend between ICL fraction and halo mass once the influence of redshift has been removed. The right panel bins the points into three redshift bins with equal numbers of samples in each bin, shown by the lines coloured by median redshift within the bin. Within each redshift bin, points are binned by halo mass with equal numbers of samples in each bin, and diamonds show the median halo mass and ICL fraction within that bin. This also shows qualitatively that although each redshift bin has consistently lower ICL fractions and higher halo masses than the previous one due to the ICL fraction-redshift and halo mass-redshift correlations, within these smaller redshift ranges we still observe a negative relationship between ICL fraction and halo mass in all bins.}
    \label{fig:iclfrac-halomass}
\end{figure*}

Figure \ref{fig:iclfrac-halomass} shows the trend of ICL fraction with halo mass in our sample, coloured by the redshift of the groups and clusters. There is a significant decreasing trend here, with a Spearman correlation of $r_S=-0.600$ and a p-value = $10^{-13}$. However there is also a strong effect due to the redshift evident in the gradient of the data points in the left panel of Figure \ref{fig:iclfrac-halomass}, so it is likely that the trend is at least partially due to the ICL fraction-redshift trend seen in Figure \ref{fig:iclfrac_redshift}. The right panel of Figure \ref{fig:iclfrac-halomass} shows qualitatively that when looking within smaller redshift bins where the ICL fraction-redshift correlation should have less impact, we still observe a systematic negative relationship between ICL fraction and halo mass within all redshift bins.

To quantitatively disentangle the trend with halo mass from the redshift of these clusters, we calculate the partial Spearman correlation between ICL fraction and halo mass in this halo mass subsample, controlling for redshift. We find that the partial Spearman correlation coefficient is $-0.330$, with a p-value of $8\times 10^{-5}$. This indicates that there is a remaining highly significant, but weak decreasing trend between ICL fraction and halo mass, in excess of what can be explained by the redshifts of these clusters.

One consideration to make is that the model was not trained on groups with halo masses lower than 13.9 M$_\odot h^{-1}$, which is the lowest halo mass in the CAMIRA sample. Although we previously confirmed that the residuals of the model are consistent with 0 over the full halo mass range of the GAMA+CAMIRA samples (see Figure \ref{fig:model_residuals}), the model's performance in this low halo mass region is only tested by a few points. This means that it is still possible that the model could be subtly biased within this region, given that the four lowest halo mass points have negative residuals. A systematic overprediction by the model could have an impact on the strength of the trend with halo mass that we have observed. To address this, we conservatively remove all groups and clusters with a halo mass $<13.9$ M$_\odot h^{-1}$, such that all remaining points are within the halo mass range that the model was trained on (leaving a sample of 97 clusters). Recalculating the partial Spearman correlation coefficient, controlling for redshift as before, we find a coefficient of $-0.329$, with a p-value of $0.001$. Therefore, even when not considering any clusters outside of the training mass range, we find that the correlation between ICL fraction and halo mass remains weakly negative, and still has a $>3\sigma$ significance. Our conclusions thus remain unchanged regardless of the model's performance at these lower halo masses.

\subsection{Cluster relaxation}
\label{sec:relaxation}

In this work, we use $\Delta m_{12}$ as an optical indicator of cluster relaxation. In order to remove the known dependence of $\Delta m_{12}$ on redshift and halo mass shown in Figure \ref{fig:params_corr}, we restrict this sample to clusters with $z > 0.16$ and $M_{200}>13.5$. These values are chosen to construct a subsample of clusters such that there is no significant Spearman correlation between magnitude gap and the other parameters, with p-values for the Spearman correlation of $\Delta m_{12}$ with redshift and $\Delta m_{12}$ with halo mass of 0.14 and 0.21 respectively as shown in Figure \ref{fig:subsamples}. This allows us to isolate the effect of magnitude gap on ICL fraction within this subsample. These cuts result in a subsample of 127 groups and clusters, and is called the relaxation subsample in Table \ref{tab:subsamples}.

\begin{figure}
    \centering
    \includegraphics[width=\linewidth]{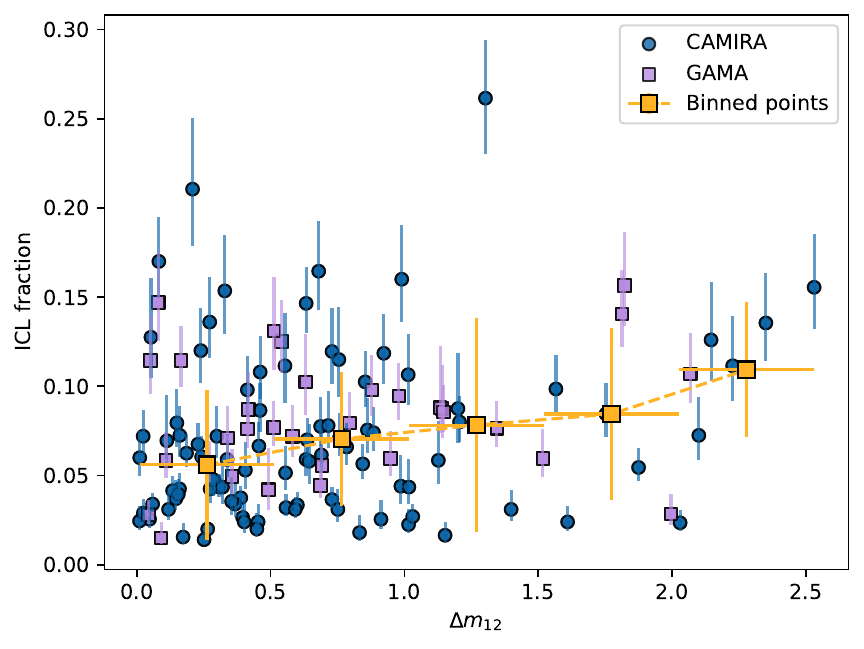}
    \caption{The ICL fraction's dependence on magnitude gap in our relaxation subsample. The yellow points show the median of ICL fractions in five bins equally spaced in $\Delta m_{12}$. The Spearman correlation coefficient is $r_S=0.226$, with p-value $=0.01$.}
    \label{fig:iclfrac_m12}
\end{figure}

Figure \ref{fig:iclfrac_m12} shows this sample of ICL fractions plotted against the $\Delta m_{12}$ of the cluster. The Spearman correlation coefficient is 0.226, with p-value 0.01. This indicates a very weak, but significant at the $2\sigma$ level, correlation between ICL fraction and magnitude gap. The binned points also show a weak systematic increasing trend. Clusters that have $\Delta m_{12} < 0.5$ have a median ICL fraction of $0.05\pm0.04$, whereas clusters with $\Delta m_{12} > 1.5$ have a median ICL fraction of $0.10\pm0.05$. Although there is a large amount of scatter present in this trend, our results suggest that more relaxed clusters have, on average, slightly higher ICL fractions than dynamically active clusters. 

\section{Discussion}
\label{sec:discussion}
\subsection{Redshift trend}

The relationship of the ICL fraction with redshift has been previously studied in both simulations and observations. Unfortunately, comparisons between other studies are not straightforward due to the differences in choice of measurement method, imaging band, and survey depth, all of which can profoundly affect the measurement of the ICL fraction.

\begin{figure}
    \centering
    \includegraphics[width=\linewidth]{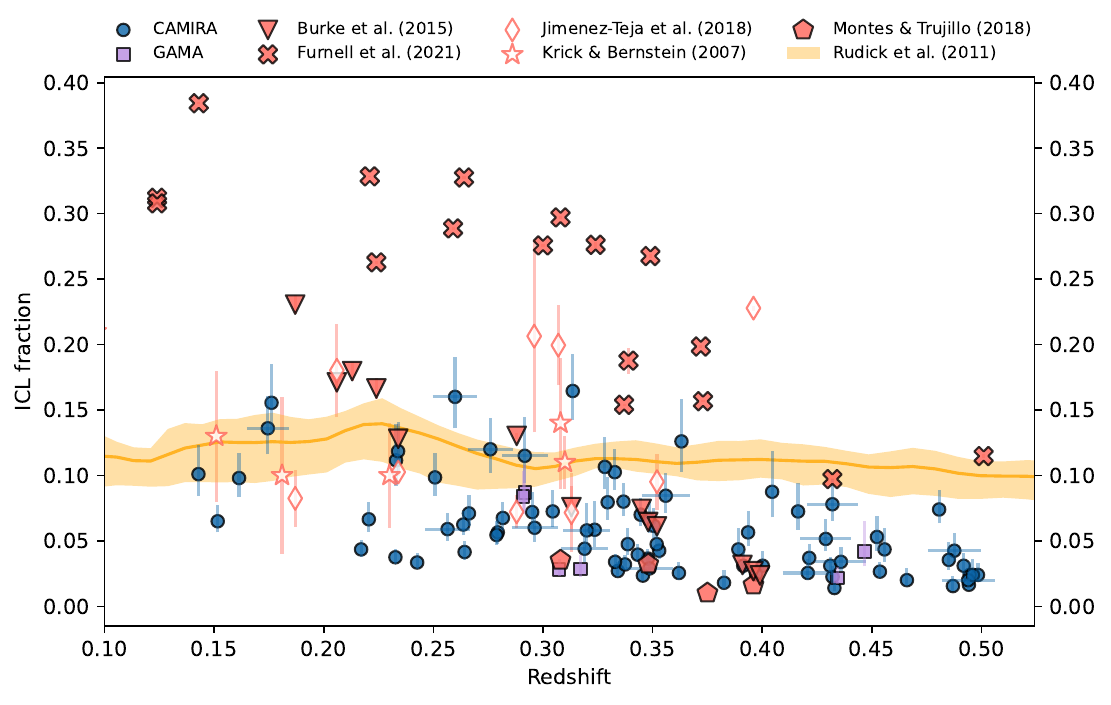}
    \caption{Our results compared to results from the literature. Points are from observational studies (\citealp{krick_diffuse_2007, burke_coevolution_2015, jimenez-teja_unveiling_2018, montes_intracluster_2018, furnell_growth_2021}). Closed points indicate measurements that are taken with the surface brightness threshold method (directly comparable to ours), whereas open points indicate other measurement methods. The orange line shows the result for one simulated cluster from \citet{rudick_quantity_2011}.}
    \label{fig:redshift_comparison}
\end{figure}

Figure \ref{fig:redshift_comparison} shows our results compared to a number of previous studies that have clusters within our redshift range. For \citet{krick_diffuse_2007}, we plot their ICL fractions measured in the r-band, and for \citet{jimenez-teja_unveiling_2018}, we plot their measurements in the F606W filter, as these are the most comparable with our r-band measurements. Most studies, both observational and simulated, find that the ICL fraction has a negative correlation with redshift, and argue that the ICL builds up in clusters over time, which qualitatively agrees with our observational data. However, the strength of this relationship and the exact values measured for the ICL varies widely between studies. 

The observational studies by \citet{burke_coevolution_2015} and \citet{furnell_growth_2021} find a much stronger evolution with redshift than we find here. This discrepancy can be explained by the differences between our methods. Although they also use the surface brightness threshold method, they use a threshold of $\mu_B=25$ mag arcsec$^{-2}$. Assuming an age of 2 Gyr and [Fe/H] of -0.4 for the ICL (e.g. \citealp{montes_intracluster_2018}) giving a colour of $B-r=0.74$ \citep{vazdekis_uv-extended_2016}, this corresponds to a surface brightness threshold of 24.26 mag arcsec$^{-2}$ in the $r$ band, which is significantly brighter than our threshold of $\mu_r=26$ mag arcsec$^{-2}$. A higher threshold will naturally lead to a higher ICL fraction as more light from the brighter parts of the cluster will be included as ICL. \citet{rudick_quantity_2011} also found that brighter surface brightness thresholds show stronger redshift evolution when measuring the same clusters, which would explain the discrepant strength of our trend compared with these other studies.

Other observational studies show only slight (\citealp{krick_diffuse_2007, montes_intracluster_2018}), or no (\citealp{jimenez-teja_unveiling_2018, joo_intracluster_2023}) evolution in the ICL fraction with redshift. The measurements of \citet{montes_intracluster_2018} are most comparable with ours in terms of measurement method, as they use images that are 600 kpc on each side, and a threshold of $\mu_V=26$ mag arcsec$^{-2}$. The ICL fractions that they find are consistent although slightly lower on average than ours (within the redshift range of their clusters, our fractions have an average value of $0.056\pm0.033$ compared to an average value among their clusters of 0.018), which could be explained by our measurements being made in different bands (their images are in V-band as opposed to our $r$-band images). The trend that they find is qualitatively compatible with ours. \citet{krick_diffuse_2007} and \citet{jimenez-teja_unveiling_2018} have differing measurement methods to ours, with \citet{krick_diffuse_2007} fitting a de Vaucouleurs profile to the ICL distribution to disentangle it from the BCG light, and \citet{jimenez-teja_unveiling_2018} using a multi-galaxy fitting method (CICLE; \citealp{jimenez-teja_disentangling_2016}) out to varying radii (ranging from 63.6 kpc to 626 kpc). These methods have significant differences to the surface brightness threshold method, in particular that they are able to capture ICL light in projection with the BCG and other satellite galaxies, making it difficult to compare directly to our ICL fractions. We also note that all clusters measured by \citet{jimenez-teja_unveiling_2018} and six out of ten clusters measured by \citet{krick_diffuse_2007} have higher halo masses than almost all of our clusters ($>10^{15}\textrm{M}_\odot h^{-1}$). However, their measurements, and in particular \citet{krick_diffuse_2007} appear to be broadly consistent with ours, although they each only have a few points.

We have found from our analysis of the effect of cosmological surface brightness dimming and passive aging that these effects can have a considerable impact on the measured ICL fraction, and implies that the real trend with redshift is far weaker than what is implied directly by observations, or could even be non-existent. More research is needed into the extent to which these effects could be impacting the ICL fractions in a diverse range of clusters. However, our analysis indicates that this could bring the observational measurements more in line with the significantly more modest evolution of the ICL fraction that is predicted from several simulations (e.g. \citealp{rudick_quantity_2011, canas_stellar_2020, contini_connection_2024}).

A modest but consistent buildup of the ICL over time is indicative of ICL production being an ongoing process, such as due to tidal stripping of galaxies within groups and clusters (e.g. \citealp{rudick_formation_2006}). We find that the actual trend of ICL buildup over time could be very weak, which indicates that in the halo mass range we have studied, the interplay between the disruption of galaxies via galaxy-galaxy interactions and the accretion of new groups with lower ICL fractions results, on average, in little evolution of the ICL fraction over time (e.g. \citealp{canas_stellar_2020}). However, the fraction for individual clusters varies quite a lot at any particular redshift, indicative of the stochastic nature of ICL formation, which is dependent on an individual cluster's accretion history.

\subsection{Halo mass trend}
\begin{figure}
    \centering
    \includegraphics[width=\linewidth]{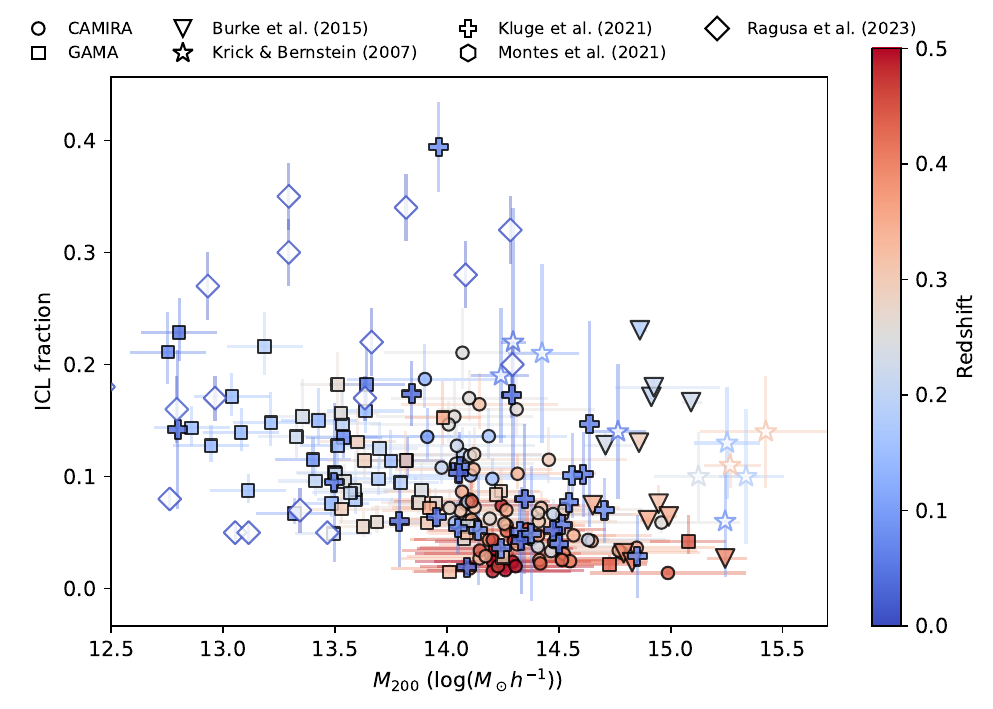}
    \caption{Our results compared to results from the literature (\citealp{krick_diffuse_2007, burke_coevolution_2015, kluge_photometric_2021, montes_buildup_2021, ragusa_does_2023}), coloured by redshift. Closed points indicate measurements that are taken with the surface brightness threshold method (more comparable to ours), whereas open points indicate other measurement methods.}
    \label{fig:halomass_comparison}
\end{figure}

Figure \ref{fig:halomass_comparison} shows several other studies from the literature within our halo mass and redshift range (\citealp{krick_diffuse_2007, burke_coevolution_2015, kluge_photometric_2021, montes_buildup_2021, ragusa_does_2023}) compared to our results, coloured by redshift. For clarity, we only plot ICL fractions from \citet{kluge_photometric_2021} that have an uncertainty of $<0.1$, excluding 46 of 170 points. 

Most observational measurements find that there is no evidence of a trend between ICL fraction and halo mass (e.g. \citealp{krick_diffuse_2007, burke_coevolution_2015, furnell_growth_2021, montes_faint_2022, ragusa_does_2023}). Among these studies, there is a very large scatter in measured ICL fractions at all halo masses. As noted previously, some inconsistencies in exact values measured are not unexpected considering the different systematics involved in different measurement methods and observational characteristics of the data. \citet{montes_buildup_2021} and \citet{kluge_photometric_2021}, whose methods are most comparable with ours, find fractions that are consistent with the values we have measured. At higher halo masses, the fractions measured by \citet{kluge_photometric_2021} scatter significantly higher than our measurements. However, this can be attributed to the fact that these clusters are all at lower redshifts, which due to the ICL fraction-redshift trend would lead to higher ICL fractions. Given that we have found that there is a weak excess trend of ICL fraction with decreasing halo mass once the trend with redshift has been removed ($r_S=-0.330$), it is possible that previous studies have not had the sample sizes necessary to isolate this trend.

\citet{gonzalez_census_2007, gonzalez_galaxy_2013} find that the BCG+ICL makes up a larger fraction of stellar mass at lower halo masses, and confirm that this result is not driven purely by the BCG by using various circular annuli to measure luminosity fractions at different radii. However, we cannot compare their results to ours as they do not directly measure ICL fractions in their clusters. Similarly, \citet{demaio_lost_2018} find that although the total stellar mass of the BCG+ICL goes up with increasing total mass, it increases more slowly than total cluster halo mass, suggesting that low-mass environments produce ICL more efficiently, however they also do not specifically measure ICL fractions.

Simulations, which do not suffer from the same sample size issues as many observational measurements of the ICL, report contradictory trends in ICL fraction with halo mass, even within the same study. Overall, the trend of ICL fraction with halo mass as measured in simulations appears to be highly dependent on the method used to measure those ICL fractions. In particular, \citet{canas_stellar_2020} used several methods (a fixed physical aperture, an aperture scaled by the spherical radius that encloses 50\% of the stellar mass of the central galaxy $R_{50}$, and a dynamical method) to measure the ICL fraction in the same clusters in the Horizon-AGN hydrodynamical simulation \citep{dubois_dancing_2014}. In the group and cluster halo mass range, they found no or very little relationship with halo mass using the dynamical method and when using a fixed 30 kpc aperture. However, they found an increasing relationship using a fixed 100 kpc aperture, and using the aperture scaled by $R_{50}$, they found a slightly decreasing relationship with halo mass. Similarly, \citet{montenegro-taborda_stellar_2025} found either no relationship or a very slightly increasing relationship with halo masses in the IllustrisTNG hydrodynamical simulation \citep{pillepich_first_2018} when using fixed aperture or scaled aperture methods (as a factor of $R_{200}$). However, when they defined the aperture as a factor of the half light radius, they found a slightly decreasing relationship with increasing halo mass. 

Given that simulations appear to show that there is a very strong dependence of ICL fraction-halo mass trend on measurement method, which each introduce different assumptions and can have different levels of contamination from the BCG, it is more reasonable to compare our results to simulations that have measured the ICL fraction by making mock images and mimicking observational measurement methods. \citet{brough_preparing_2024} found overall no trend in ICL fraction with halo mass when using observational measurements on mock images (using the surface brightness threshold method, composite model methods, 2D fitting methods, and wavelet decomposition methods). However, their clusters occupy only a very small range in halo mass ($14.0 < \log(M_{200}/\textrm{M}_\odot) < 14.5$), which may not be enough to reveal a trend. \citet{cui_characterizing_2014} used a surface brightness threshold method, and finds a slightly decreasing relationship with halo mass in a set of zoom-in simulations that include AGN feedback. \citet{tang_investigation_2018} also find a decreasing relationship with halo mass using their surface brightness threshold method. 

The weakly decreasing relationship between ICL fraction and halo mass that we have found appears to be consistent with expectations from comparable simulated studies. However, given the apparent issues with a disappearance or reversal of this trend from simulations that compared multiple different measurement methods, it is possible that this trend is specific to the surface brightness threshold method that we have chosen. 

If the ICL fraction does decrease with increasing halo mass, this could imply that the dominant formation channel for ICL production is one that is more efficient in lower-mass haloes. For example, if tidal stripping is the dominant formation channel, then groups could be more efficient at stripping these stars into the ICL due to the longer interaction times within galaxy groups. This is the scenario suggested by, for example, \citet{demaio_lost_2018}, who do find in their analysis of ICL colour gradients that tidal stripping is responsible for the bulk of the ICL production in the groups that they studied. \citet{khalid_characterizing_2024} find in their analysis of simulated mock images that galaxy tidal features, which are visible indications of ongoing or past galaxy mergers, are more likely to occur in group-mass haloes ($< 10^{13} \textrm{M}_\odot$, with a peak at $\sim 10^{12.7} \textrm{M}_\odot$) than more massive haloes, which could imply an enhancement of interactions at these lower halo masses, leading to a greater rate of ICL production, although it could also imply that tidal features are longer-lived in low mass haloes than in clusters, resulting in more detections. \citet{bahe_disruption_2019} also find that satellite galaxies are less likely to be totally disrupted in more massive halos, and that up to 90\% of mass disruption in clusters occurs through pre-processing in lower mass sub-groups, so a scenario where total disruption of satellites is also an important contributor to the ICL is also compatible with our results. On the other hand, this would indicate that cluster-only processes such as harassment (e.g. \citealp{moore_galaxy_1996}) may not be as important for ICL formation.

However, we note that the negative correlation between ICL fraction and halo mass that we find is weak in strength ($r_S= -0.33$), meaning that the potential difference in ICL production efficiency from low to high mass haloes does not appear to be very large, and there could be other effects at play. Although interaction times are longer in group mass haloes, making them better at stripping stars, clusters have hosted more interactions over time, have more available fuel to produce ICL, and accrete ``pre-processed" ICL from infalling groups (e.g. \citealp{mihos_interactions_2004, ragusa_does_2023}). The relationship between host halo mass and ICL fraction, which depends on the luminosity of both the ICL and the cluster, is therefore not precisely tied to the environmental efficiencies of the interactions that produce it. Our results indicate that, per unit cluster luminosity, group environments produce more ICL than cluster environments, although there is clearly a large scatter.

The other possibility as mentioned is that the trend we find is due specifically to our measurement methods, and using another method may not reveal this same trend. It is possible that different measurement methods are probing different portions of the ICL, or are contaminated by the BCG to different levels. These could therefore find discrepant trends in halo mass if other formation channels, which are affected differently by environment, contribute ICL stars to locations in the cluster that are less effectively probed by our measurement method (e.g. \citealp{harris_quantifying_2017, butler_intracluster_2025}). For example, \citet{montenegro-taborda_growth_2023}, in their study of simulated IllustrisTNG clusters, found that accreted stars from major mergers are found closer to the centre, while stars that were stripped from surviving satellites dominate at larger radii. This has also been seen observationally by \citet{montes_new_2022}, who find through analysis of colour profiles and morphological properties that the ICL in their studied cluster originates from a major merger in the inner 100 kpc, as opposed to tidal stripping of Milky Way-like satellites at $>150$ kpc from the centre. A limitation of the surface brightness threshold method is that some ICL is lost in projection with the BCG and other satellite galaxies. This could mean that it is less sensitive to the ICL that is located more centrally within the cluster, but more sensitive to ICL located further out. In order to test this, we need larger observational samples complete in both redshift and halo mass, measured using other methods to compare to.

\subsection{Cluster relaxation trend}
In observations, the analysis of the dependence of ICL fraction on cluster relaxation is complicated by the fact that proxies are necessary in order to measure the cluster relaxation. This makes direct comparison between studies more challenging, as well as the difficulties stemming from the differences in ICL fraction measurement method. 

\citet{golden-marx_hierarchical_2025} also use the magnitude gap (although they use $\Delta m_{14}$) as an estimate for cluster relaxation, and find that the ICL fraction increases with increasing magnitude gap. They find a significantly stronger trend than we do in our sample. This could be attributed to differing measurement methods (they use a fixed aperture of 30 kpc to separate the BCG and ICL, and measure out to a 150 kpc radius), and the fact that they use stacked measurements. \citet{ragusa_does_2023} also finds a weakly increasing trend between ICL fraction and cluster relaxation, measured by the fraction of early type galaxies within the cluster. 

Simulations are generally in agreement that more relaxed clusters show higher ICL fractions (e.g. \citealp{contini_connection_2024, contreras-santos_characterising_2024, montenegro-taborda_stellar_2025, kimmig_intra-cluster_2025}). In particular, \citet{canas_stellar_2020} measure the relaxation of their clusters using the mass gap between the most massive and second most massive galaxy, which is comparable to the observational magnitude gap measurement. At the higher end of total stellar masses in their sample ($M_{*,\textrm{tot}}>10^{11.6}$), they find that more relaxed clusters have slightly higher ICL fractions on average, in agreement with our results. They also find significant scatter in the relation, and given a relatively few number of high stellar mass systems, the trend is quite weak. This appears to agree with the nature of the trend that we find in our results, where the difference between ICL fractions of relaxed and unrelaxed clusters as measured by their magnitude gaps is small.

Our results suggest that, on average, relaxed clusters are more likely to host slightly larger fractions of ICL. This is because relaxed clusters have had more time to disrupt cluster member galaxies and add to the ICL, whereas dynamically active clusters that result from recent mergers would still consist of gravitationally bound infalling objects that have not had the time to be tidally stripped and contribute stars to the ICL. 

The trend appears to be marginal in this sample, in contrast to some observational studies that find a stronger trend. There are a few potential reasons for this. One is that the magnitude gap is an imperfect measure of the dynamical state of clusters, meaning that there could be a significant amount of extra noise due to cluster members that have been incorrectly classified, especially considering that many clusters in our sample only have photometric member information. With more data, ideally with spectroscopic memberships, it is possible that a trend will become clearer.

\cite{casas_optical_2024} found in observations that the projected distance between the brightest and second brightest galaxy $d_{12}$ can be combined with $\Delta m_{14}$ (the magnitude gap between the brightest and fourth brightest cluster galaxy) to identify and remove a large fraction of unrelaxed clusters. Although they use $\Delta m_{14}$, they found it to be only a marginal improvement over $\Delta m_{12}$. We calculated $d_{12}$ for our clusters, but find no correlation between $d_{12}$ and the ICL fraction.

Another possibility is that the surface brightness threshold method is somewhat biased against this trend. Relaxed clusters are more concentrated, a property which also correlates with the ICL fraction in simulations (e.g. \citealp{contini_connection_2024, contreras-santos_characterising_2024}). If the ICL is highly concentrated towards the centre (i.e. around the BCG), this could mean that with the surface brightness threshold method we are losing proportionally more light in projection with the BCG or that more of the ICL is misclassified as the BCG. It is possible that the observational studies that find stronger trends are using measurement methods that are more suited to identifying ICL in relaxed clusters.

\section{Summary and conclusions}
\label{sec:summary}

In this work, we use the machine learning model presented in \citet{canepa_measuring_2025} to create a large sample of 177 ICL fractions using the CAMIRA cluster catalogue (101 clusters) and the GAMA group catalogue (76 groups). For this sample we analysed the ICL fraction's dependence on cluster redshift, halo mass, and relaxation. 

We find a decreasing trend of ICL fraction with redshift with a Spearman correlation of $-0.604$, which appears to indicate that the ICL is building up consistently over time. However, we also find through modelling of a cluster observed at different redshifts that a significant amount of this trend could be plausibly explained by observational effects as opposed to a real increase of the ICL fraction in these clusters. This indicates that the real trend of ICL fraction over time could be much weaker than what is implied by observations, and could bring the trend more in line with results from some simulations that indicate no or very weak evolution in the ICL over time. More research is needed using more complex models of a variety of different types of clusters in order to evaluate the extent to which these observational effects are influencing the observed trends.

We find evidence of a weak decreasing trend between ICL fraction and halo mass, with a partial Spearman correlation coefficient of $-0.330$ in excess of what can be explained by the higher redshifts of higher halo mass clusters. Although this could be an indication that ICL production is more efficient in lower mass clusters, it could also be explained as being due to our choice of measurement method.

We find a marginal positive correlation between ICL fraction and magnitude gap with a Spearman correlation of 0.226. The lowest magnitude gap clusters ($<0.5$) have a median ICL fraction of $0.05\pm0.04$, while clusters with the highest magnitude gaps ($>1.5$) have a median ICL fraction of $0.10\pm0.05$. There is a large amount of scatter in the relation, however it indicates that on average relaxed clusters are more likely to host proportionally larger amounts of ICL than dynamically active clusters, in agreement with other observational and simulated results. 

Taken together, our results indicate that the ICL could form mainly due to galaxy-galaxy interactions such as tidal stripping, which strip stars more efficiently in lower halo mass galaxy groups. This increases the ICL fraction slowly and consistently in clusters over time in the absence of recent mergers and accretion of other groups, leading to slightly higher ICL fractions in more relaxed clusters. Cluster-only processes are likely to not contribute much to the formation of the ICL, otherwise we would expect the ICL fraction in higher halo mass clusters to be elevated. However, after accounting for observational effects and controlling for cluster relaxation and halo mass, we find plausibly no buildup in the ICL fraction over redshift, although this is a simple model and there could be other effects at play. Therefore, while in individual clusters the ICL increases slightly over time in the absence of new mergers as the cluster relaxes, when this effect is removed there may not be much increase in the ICL fraction with redshift since $z\sim0.5$.  

In observations, we need more understanding of the effects that observational systematics are having on our measurements. We also need large observational samples measured with other observational methods, as the effect of differing measurement methods on the trends we have observed here are still a significant uncertainty. Simulations have found that measurement method can significantly affect observed trends, particularly with halo mass. As each measurement method has its own set of assumptions and limitations, making comparisons between them on a large-scale sample such as this could help us to confirm our physical interpretations, rule out specific measurement systematics, and understand how these different measurement methods probe the ICL. 

\section*{Acknowledgements}


SB acknowledges funding support from the Australian Research Council through a Discovery Project DP190101943. MM acknowledges support from grant RYC2022-036949-I financed by the MICIU/AEI/10.13039/501100011033 and by ESF+, grant CNS2024-154592 financed by MICIU/AEI/10.13039/501100011033 and program Unidad de Excelencia Mar\'{i}a de Maeztu CEX2020-001058-M. NAH acknowledges support from the Leverhulme Trust, and the UK Science and Technology Facilities Council (STFC) under grant ST/X000982/1.

This research includes computations using the computational cluster Katana supported by Research Technology Services at UNSW Sydney.

The Hyper Suprime-Cam (HSC) collaboration includes the astronomical communities of Japan and Taiwan, and Princeton University. The HSC instrumentation and software were developed by the National Astronomical Observatory of Japan (NAOJ), the Kavli Institute for the Physics and Mathematics of the Universe (Kavli IPMU), the University of Tokyo, the High Energy Accelerator Research Organization (KEK), the Academia Sinica Institute for Astronomy and Astrophysics in Taiwan (ASIAA), and Princeton University. Funding was contributed by the FIRST program from the Japanese Cabinet Office, the Ministry of Education, Culture, Sports, Science and Technology (MEXT), the Japan Society for the Promotion of Science (JSPS), Japan Science and Technology Agency (JST), the Toray Science Foundation, NAOJ, Kavli IPMU, KEK, ASIAA, and Princeton University. 

This paper makes use of software developed for Vera C. Rubin Observatory. We thank the Rubin Observatory for making their code available as free software at http://pipelines.lsst.io/.

This paper is based on data collected at the Subaru Telescope and retrieved from the HSC data archive system, which is operated by the Subaru Telescope and Astronomy Data Center (ADC) at NAOJ. Data analysis was in part carried out with the cooperation of Center for Computational Astrophysics (CfCA), NAOJ. We are honored and grateful for the opportunity of observing the Universe from Maunakea, which has the cultural, historical and natural significance in Hawaii. 

\section*{Data Availability}
All data used in this work is publicly available at \url{https://hsc-release.mtk.nao.ac.jp/doc/} (HSC-SSP data), \url{https://www.gama-survey.org/} (GAMA data), and \url{https://data.desi.lbl.gov/doc/} (DESI data). The machine learning model and instructions for its use are available on GitHub and can be downloaded from \url{https://github.com/lpcan/MICL/tree/v1.0} \citep{canepa_lpcanmicl_2024}.
 



\bibliographystyle{mnras}
\bibliography{references} 

@article{aihara_second_2019,
	title = {Second data release of the {Hyper} {Suprime}-{Cam} {Subaru} {Strategic} {Program}},
	volume = {71},
	issn = {0004-6264},
	url = {https://doi.org/10.1093/pasj/psz103},
	doi = {10.1093/pasj/psz103},
	number = {6},
	urldate = {2023-10-19},
	journal = {Publications of the Astronomical Society of Japan},
	author = {Aihara, Hiroaki and AlSayyad, Yusra and Ando, Makoto and Armstrong, Robert and Bosch, James and Egami, Eiichi and Furusawa, Hisanori and Furusawa, Junko and Goulding, Andy and Harikane, Yuichi and Hikage, Chiaki and Ho, Paul T P and Hsieh, Bau-Ching and Huang, Song and Ikeda, Hiroyuki and Imanishi, Masatoshi and Ito, Kei and Iwata, Ikuru and Jaelani, Anton T and Kakuma, Ryota and Kawana, Kojiro and Kikuta, Satoshi and Kobayashi, Umi and Koike, Michitaro and Komiyama, Yutaka and Li, Xiangchong and Liang, Yongming and Lin, Yen-Ting and Luo, Wentao and Lupton, Robert and Lust, Nate B and MacArthur, Lauren A and Matsuoka, Yoshiki and Mineo, Sogo and Miyatake, Hironao and Miyazaki, Satoshi and More, Surhud and Murata, Ryoma and Namiki, Shigeru V and Nishizawa, Atsushi J and Oguri, Masamune and Okabe, Nobuhiro and Okamoto, Sakurako and Okura, Yuki and Ono, Yoshiaki and Onodera, Masato and Onoue, Masafusa and Osato, Ken and Ouchi, Masami and Shibuya, Takatoshi and Strauss, Michael A and Sugiyama, Naoshi and Suto, Yasushi and Takada, Masahiro and Takagi, Yuhei and Takata, Tadafumi and Takita, Satoshi and Tanaka, Masayuki and Terai, Tsuyoshi and Toba, Yoshiki and Uchiyama, Hisakazu and Utsumi, Yousuke and Wang, Shiang-Yu and Wang, Wenting and Yamada, Yoshihiko},
	month = dec,
	year = {2019},
	pages = {114},
}

@article{aihara_third_2022,
	title = {Third data release of the {Hyper} {Suprime}-{Cam} {Subaru} {Strategic} {Program}},
	volume = {74},
	copyright = {https://academic.oup.com/journals/pages/open\_access/funder\_policies/chorus/standard\_publication\_model},
	issn = {0004-6264, 2053-051X},
	url = {https://academic.oup.com/pasj/article/74/2/247/6528503},
	doi = {10.1093/pasj/psab122},
	language = {en},
	number = {2},
	urldate = {2024-08-09},
	journal = {Publications of the Astronomical Society of Japan},
	author = {Aihara, Hiroaki and AlSayyad, Yusra and Ando, Makoto and Armstrong, Robert and Bosch, James and Egami, Eiichi and Furusawa, Hisanori and Furusawa, Junko and Harasawa, Sumiko and Harikane, Yuichi and Hsieh, Bau-Ching and Ikeda, Hiroyuki and Ito, Kei and Iwata, Ikuru and Kodama, Tadayuki and Koike, Michitaro and Kokubo, Mitsuru and Komiyama, Yutaka and Li, Xiangchong and Liang, Yongming and Lin, Yen-Ting and Lupton, Robert H and Lust, Nate B and MacArthur, Lauren A and Mawatari, Ken and Mineo, Sogo and Miyatake, Hironao and Miyazaki, Satoshi and More, Surhud and Morishima, Takahiro and Murayama, Hitoshi and Nakajima, Kimihiko and Nakata, Fumiaki and Nishizawa, Atsushi J and Oguri, Masamune and Okabe, Nobuhiro and Okura, Yuki and Ono, Yoshiaki and Osato, Ken and Ouchi, Masami and Pan, Yen-Chen and Plazas Malagón, Andrés A and Price, Paul A and Reed, Sophie L and Rykoff, Eli S and Shibuya, Takatoshi and Simunovic, Mirko and Strauss, Michael A and Sugimori, Kanako and Suto, Yasushi and Suzuki, Nao and Takada, Masahiro and Takagi, Yuhei and Takata, Tadafumi and Takita, Satoshi and Tanaka, Masayuki and Tang, Shenli and Taranu, Dan S and Terai, Tsuyoshi and Toba, Yoshiki and Turner, Edwin L and Uchiyama, Hisakazu and Vijarnwannaluk, Bovornpratch and Waters, Christopher Z and Yamada, Yoshihiko and Yamamoto, Naoaki and Yamashita, Takuji},
	month = apr,
	year = {2022},
	pages = {247--272},
}

@article{driver_galaxy_2011,
	title = {Galaxy and {Mass} {Assembly} ({GAMA}): survey diagnostics and core data release: {GAMA}},
	volume = {413},
	issn = {00358711},
	shorttitle = {Galaxy and {Mass} {Assembly} ({GAMA})},
	url = {https://academic.oup.com/mnras/article-lookup/doi/10.1111/j.1365-2966.2010.18188.x},
	doi = {10.1111/j.1365-2966.2010.18188.x},
	language = {en},
	number = {2},
	urldate = {2025-01-16},
	journal = {Monthly Notices of the Royal Astronomical Society},
	author = {Driver, S. P. and Hill, D. T. and Kelvin, L. S. and Robotham, A. S. G. and Liske, J. and Norberg, P. and Baldry, I. K. and Bamford, S. P. and Hopkins, A. M. and Loveday, J. and Peacock, J. A. and Andrae, E. and Bland-Hawthorn, J. and Brough, S. and Brown, M. J. I. and Cameron, E. and Ching, J. H. Y. and Colless, M. and Conselice, C. J. and Croom, S. M. and Cross, N. J. G. and De Propris, R. and Dye, S. and Drinkwater, M. J. and Ellis, S. and Graham, Alister W. and Grootes, M. W. and Gunawardhana, M. and Jones, D. H. and Van Kampen, E. and Maraston, C. and Nichol, R. C. and Parkinson, H. R. and Phillipps, S. and Pimbblet, K. and Popescu, C. C. and Prescott, M. and Roseboom, I. G. and Sadler, E. M. and Sansom, A. E. and Sharp, R. G. and Smith, D. J. B. and Taylor, E. and Thomas, D. and Tuffs, R. J. and Wijesinghe, D. and Dunne, L. and Frenk, C. S. and Jarvis, M. J. and Madore, B. F. and Meyer, M. J. and Seibert, M. and Staveley-Smith, L. and Sutherland, W. J. and Warren, S. J.},
	month = may,
	year = {2011},
	pages = {971--995},
}

@article{gladders_new_2000,
	title = {A {New} {Method} {For} {Galaxy} {Cluster} {Detection}. {I}. {The} {Algorithm}},
	volume = {120},
	issn = {0004-6256},
	url = {https://ui.adsabs.harvard.edu/abs/2000AJ....120.2148G},
	doi = {10.1086/301557},
	urldate = {2024-02-12},
	journal = {The Astronomical Journal},
	author = {Gladders, Michael D. and Yee, H. K. C.},
	month = oct,
	year = {2000},
	note = {ADS Bibcode: 2000AJ....120.2148G},
	pages = {2148--2162},
}

@article{huang_weak_2020,
	title = {Weak lensing reveals a tight connection between dark matter halo mass and the distribution of stellar mass in massive galaxies},
	volume = {492},
	copyright = {https://academic.oup.com/journals/pages/open\_access/funder\_policies/chorus/standard\_publication\_model},
	issn = {0035-8711, 1365-2966},
	url = {https://academic.oup.com/mnras/article/492/3/3685/5658706},
	doi = {10.1093/mnras/stz3314},
	language = {en},
	number = {3},
	urldate = {2024-08-09},
	journal = {Monthly Notices of the Royal Astronomical Society},
	author = {Huang, Song and Leauthaud, Alexie and Hearin, Andrew and Behroozi, Peter and Bradshaw, Christopher and Ardila, Felipe and Speagle, Joshua and Tenneti, Ananth and Bundy, Kevin and Greene, Jenny and Sifón, Cristóbal and Bahcall, Neta},
	month = mar,
	year = {2020},
	pages = {3685--3707},
}

@article{li_reaching_2022,
	title = {Reaching for the {Edge} {I}: probing the outskirts of massive galaxies with {HSC}, {DECaLS}, {SDSS}, and {Dragonfly}},
	volume = {515},
	copyright = {https://academic.oup.com/journals/pages/open\_access/funder\_policies/chorus/standard\_publication\_model},
	issn = {0035-8711, 1365-2966},
	shorttitle = {Reaching for the {Edge} {I}},
	url = {https://academic.oup.com/mnras/article/515/4/5335/6652119},
	doi = {10.1093/mnras/stac2121},
	language = {en},
	number = {4},
	urldate = {2024-08-09},
	journal = {Monthly Notices of the Royal Astronomical Society},
	author = {Li, Jiaxuan and Huang, Song and Leauthaud, Alexie and Moustakas, John and Danieli, Shany and Greene, Jenny E and Abraham, Roberto and Ardila, Felipe and Kado-Fong, Erin and Lokhorst, Deborah and Lupton, Robert and Price, Paul},
	month = aug,
	year = {2022},
	pages = {5335--5357},
}

@misc{mihos_deep_2019,
	title = {Deep {Imaging} of {Diffuse} {Light} {Around} {Galaxies} and {Clusters}: {Progress} and {Challenges}},
	shorttitle = {Deep {Imaging} of {Diffuse} {Light} {Around} {Galaxies} and {Clusters}},
	url = {https://ui.adsabs.harvard.edu/abs/2019arXiv190909456M},
	doi = {10.48550/arXiv.1909.09456},
	urldate = {2024-02-25},
	author = {Mihos, J. Christopher},
	month = sep,
	year = {2019},
	note = {Publication Title: arXiv e-prints
ADS Bibcode: 2019arXiv190909456M},
}

@article{miyazaki_hyper_2018,
	title = {Hyper {Suprime}-{Cam}: {System} design and verification of image quality},
	volume = {70},
	issn = {0004-6264},
	shorttitle = {Hyper {Suprime}-{Cam}},
	url = {https://doi.org/10.1093/pasj/psx063},
	doi = {10.1093/pasj/psx063},
	number = {SP1},
	urldate = {2023-10-19},
	journal = {Publications of the Astronomical Society of Japan},
	author = {Miyazaki, Satoshi and Komiyama, Yutaka and Kawanomoto, Satoshi and Doi, Yoshiyuki and Furusawa, Hisanori and Hamana, Takashi and Hayashi, Yusuke and Ikeda, Hiroyuki and Kamata, Yukiko and Karoji, Hiroshi and Koike, Michitaro and Kurakami, Tomio and Miyama, Shoken and Morokuma, Tomoki and Nakata, Fumiaki and Namikawa, Kazuhito and Nakaya, Hidehiko and Nariai, Kyoji and Obuchi, Yoshiyuki and Oishi, Yukie and Okada, Norio and Okura, Yuki and Tait, Philip and Takata, Tadafumi and Tanaka, Yoko and Tanaka, Masayuki and Terai, Tsuyoshi and Tomono, Daigo and Uraguchi, Fumihiro and Usuda, Tomonori and Utsumi, Yousuke and Yamada, Yoshihiko and Yamanoi, Hitomi and Aihara, Hiroaki and Fujimori, Hiroki and Mineo, Sogo and Miyatake, Hironao and Oguri, Masamune and Uchida, Tomohisa and Tanaka, Manobu M and Yasuda, Naoki and Takada, Masahiro and Murayama, Hitoshi and Nishizawa, Atsushi J and Sugiyama, Naoshi and Chiba, Masashi and Futamase, Toshifumi and Wang, Shiang-Yu and Chen, Hsin-Yo and Ho, Paul T P and Liaw, Eric J Y and Chiu, Chi-Fang and Ho, Cheng-Lin and Lai, Tsang-Chih and Lee, Yao-Cheng and Jeng, Dun-Zen and Iwamura, Satoru and Armstrong, Robert and Bickerton, Steve and Bosch, James and Gunn, James E and Lupton, Robert H and Loomis, Craig and Price, Paul and Smith, Steward and Strauss, Michael A and Turner, Edwin L and Suzuki, Hisanori and Miyazaki, Yasuhito and Muramatsu, Masaharu and Yamamoto, Koei and Endo, Makoto and Ezaki, Yutaka and Ito, Noboru and Kawaguchi, Noboru and Sofuku, Satoshi and Taniike, Tomoaki and Akutsu, Kotaro and Dojo, Naoto and Kasumi, Kazuyuki and Matsuda, Toru and Imoto, Kohei and Miwa, Yoshinori and Suzuki, Masayuki and Takeshi, Kunio and Yokota, Hideo},
	month = jan,
	year = {2018},
	pages = {S1},
}

@article{murata_mass-richness_2019,
	title = {The mass-richness relation of optically-selected clusters from weak gravitational lensing and abundance with {Subaru} {HSC} first-year data},
	volume = {71},
	issn = {0004-6264, 2053-051X},
	url = {http://arxiv.org/abs/1904.07524},
	doi = {10.1093/pasj/psz092},
	number = {5},
	urldate = {2025-01-06},
	journal = {Publications of the Astronomical Society of Japan},
	author = {Murata, Ryoma and Oguri, Masamune and Nishimichi, Takahiro and Takada, Masahiro and Mandelbaum, Rachel and More, Surhud and Shirasaki, Masato and Nishizawa, Atsushi J. and Osato, Ken},
	month = oct,
	year = {2019},
	note = {arXiv:1904.07524 [astro-ph]},
	pages = {107},
}

@article{oguri_cluster_2014,
	title = {A cluster finding algorithm based on the multiband identification of red sequence galaxies},
	volume = {444},
	issn = {0035-8711},
	url = {https://doi.org/10.1093/mnras/stu1446},
	doi = {10.1093/mnras/stu1446},
	number = {1},
	urldate = {2023-10-19},
	journal = {Monthly Notices of the Royal Astronomical Society},
	author = {Oguri, Masamune},
	month = oct,
	year = {2014},
	pages = {147--161},
}

@article{oguri_optically-selected_2018,
	title = {An optically-selected cluster catalog at redshift 0.1 {\textless} z {\textless} 1.1 from the {Hyper} {Suprime}-{Cam} {Subaru} {Strategic} {Program} {S16A} data},
	volume = {70},
	issn = {0004-6264},
	url = {https://doi.org/10.1093/pasj/psx042},
	doi = {10.1093/pasj/psx042},
	number = {SP1},
	urldate = {2023-10-19},
	journal = {Publications of the Astronomical Society of Japan},
	author = {Oguri, Masamune and Lin, Yen-Ting and Lin, Sheng-Chieh and Nishizawa, Atsushi J and More, Anupreeta and More, Surhud and Hsieh, Bau-Ching and Medezinski, Elinor and Miyatake, Hironao and Jian, Hung-Yu and Lin, Lihwai and Takada, Masahiro and Okabe, Nobuhiro and Speagle, Joshua S and Coupon, Jean and Leauthaud, Alexie and Lupton, Robert H and Miyazaki, Satoshi and Price, Paul A and Tanaka, Masayuki and Chiu, I-Non and Komiyama, Yutaka and Okura, Yuki and Tanaka, Manobu M and Usuda, Tomonori},
	month = jan,
	year = {2018},
	pages = {S20},
}

@article{robotham_galaxy_2011,
	title = {Galaxy and {Mass} {Assembly} ({GAMA}): the {GAMA} galaxy group catalogue ({G3Cv1}): {GAMA}: the {GAMA} galaxy group catalogue ({G3Cv1})},
	volume = {416},
	issn = {00358711},
	shorttitle = {Galaxy and {Mass} {Assembly} ({GAMA})},
	url = {https://academic.oup.com/mnras/article-lookup/doi/10.1111/j.1365-2966.2011.19217.x},
	doi = {10.1111/j.1365-2966.2011.19217.x},
	language = {en},
	number = {4},
	urldate = {2025-01-16},
	journal = {Monthly Notices of the Royal Astronomical Society},
	author = {Robotham, A. S. G. and Norberg, P. and Driver, S. P. and Baldry, I. K. and Bamford, S. P. and Hopkins, A. M. and Liske, J. and Loveday, J. and Merson, A. and Peacock, J. A. and Brough, S. and Cameron, E. and Conselice, C. J. and Croom, S. M. and Frenk, C. S. and Gunawardhana, M. and Hill, D. T. and Jones, D. H. and Kelvin, L. S. and Kuijken, K. and Nichol, R. C. and Parkinson, H. R. and Pimbblet, K. A. and Phillipps, S. and Popescu, C. C. and Prescott, M. and Sharp, R. G. and Sutherland, W. J. and Taylor, E. N. and Thomas, D. and Tuffs, R. J. and Van Kampen, E. and Wijesinghe, D.},
	month = oct,
	year = {2011},
	pages = {2640--2668},
}

@article{viola_dark_2015,
	title = {Dark matter halo properties of {GAMA} galaxy groups from 100 square degrees of {KiDS} weak lensing data},
	volume = {452},
	issn = {0035-8711, 1365-2966},
	url = {https://academic.oup.com/mnras/article-lookup/doi/10.1093/mnras/stv1447},
	doi = {10.1093/mnras/stv1447},
	language = {en},
	number = {4},
	urldate = {2025-01-16},
	journal = {Monthly Notices of the Royal Astronomical Society},
	author = {Viola, M. and Cacciato, M. and Brouwer, M. and Kuijken, K. and Hoekstra, H. and Norberg, P. and Robotham, A. S. G. and Van Uitert, E. and Alpaslan, M. and Baldry, I. K. and Choi, A. and De Jong, J. T. A. and Driver, S. P. and Erben, T. and Grado, A. and Graham, Alister W. and Heymans, C. and Hildebrandt, H. and Hopkins, A. M. and Irisarri, N. and Joachimi, B. and Loveday, J. and Miller, L. and Nakajima, R. and Schneider, P. and Sifón, C. and Verdoes Kleijn, G.},
	month = oct,
	year = {2015},
	pages = {3529--3550},
}

@article{contini_origin_2021,
	title = {On the {Origin} and {Evolution} of the {Intra}-{Cluster} {Light}: {A} {Brief} {Review} of the {Most} {Recent} {Developments}},
	volume = {9},
	shorttitle = {On the {Origin} and {Evolution} of the {Intra}-{Cluster} {Light}},
	url = {https://ui.adsabs.harvard.edu/abs/2021Galax...9...60C},
	doi = {10.3390/galaxies9030060},
	urldate = {2024-02-25},
	journal = {Galaxies},
	author = {Contini, Emanuele},
	month = aug,
	year = {2021},
	note = {ADS Bibcode: 2021Galax...9...60C},
	pages = {60},
}

@article{montes_faint_2022,
	title = {The faint light in groups and clusters of galaxies},
	volume = {6},
	copyright = {2022 Springer Nature Limited},
	issn = {2397-3366},
	url = {https://www.nature.com/articles/s41550-022-01616-z},
	doi = {10.1038/s41550-022-01616-z},
	language = {en},
	number = {3},
	urldate = {2023-10-19},
	journal = {Nature Astronomy},
	author = {Montes, Mireia},
	month = mar,
	year = {2022},
	note = {Number: 3
Publisher: Nature Publishing Group},
	pages = {308--316},
}

@article{rudick_formation_2006,
	title = {The {Formation} and {Evolution} of {Intracluster} {Light}},
	volume = {648},
	issn = {0004-637X},
	url = {https://ui.adsabs.harvard.edu/abs/2006ApJ...648..936R},
	doi = {10.1086/506176},
	urldate = {2024-02-25},
	journal = {The Astrophysical Journal},
	author = {Rudick, Craig S. and Mihos, J. Christopher and McBride, Cameron},
	month = sep,
	year = {2006},
	note = {ADS Bibcode: 2006ApJ...648..936R},
	pages = {936--946},
}

@article{contini_formation_2014,
	title = {On the formation and physical properties of the intracluster light in hierarchical galaxy formation models},
	volume = {437},
	issn = {0035-8711},
	url = {https://doi.org/10.1093/mnras/stt2174},
	doi = {10.1093/mnras/stt2174},
	number = {4},
	urldate = {2023-10-19},
	journal = {Monthly Notices of the Royal Astronomical Society},
	author = {Contini, E. and De Lucia, G. and Villalobos, {\'A}. and Borgani, S.},
	month = feb,
	year = {2014},
	pages = {3787--3802},
}

@article{lidman_importance_2013,
	title = {The importance of major mergers in the build up of stellar mass in brightest cluster galaxies at z = 1},
	volume = {433},
	issn = {0035-8711},
	url = {https://doi.org/10.1093/mnras/stt777},
	doi = {10.1093/mnras/stt777},
	number = {1},
	urldate = {2025-05-14},
	journal = {Monthly Notices of the Royal Astronomical Society},
	author = {Lidman, C. and Iacobuta, G. and Bauer, A. E. and Barrientos, L. F. and Cerulo, P. and Couch, W. J. and Delaye, L. and Demarco, R. and Ellingson, E. and Faloon, A. J. and Gilbank, D. and Huertas-Company, M. and Mei, S. and Meyers, J. and Muzzin, A. and Noble, A. and Nantais, J. and Rettura, A. and Rosati, P. and Sánchez-Janssen, R. and Strazzullo, V. and Webb, T. M. A. and Wilson, G. and Yan, R. and Yee, H. K. C.},
	month = jul,
	year = {2013},
	pages = {825--837},
}

@article{puchwein_intracluster_2010,
	title = {Intracluster stars in simulations with active galactic nucleus feedback},
	volume = {406},
	issn = {0035-8711},
	url = {https://doi.org/10.1111/j.1365-2966.2010.16786.x},
	doi = {10.1111/j.1365-2966.2010.16786.x},
	number = {2},
	urldate = {2023-10-19},
	journal = {Monthly Notices of the Royal Astronomical Society},
	author = {Puchwein, Ewald and Springel, Volker and Sijacki, Debora and Dolag, Klaus},
	month = aug,
	year = {2010},
	pages = {936--951},
}

@article{groenewald_close_2017,
	title = {The close pair fraction of {BCGs} since z = 0.5: major mergers dominate recent {BCG} stellar mass growth},
	volume = {467},
	issn = {0035-8711},
	shorttitle = {The close pair fraction of {BCGs} since z = 0.5},
	url = {https://doi.org/10.1093/mnras/stx340},
	doi = {10.1093/mnras/stx340},
	number = {4},
	urldate = {2025-05-14},
	journal = {Monthly Notices of the Royal Astronomical Society},
	author = {Groenewald, Danièl N. and Skelton, Rosalind E. and Gilbank, David G. and Loubser, S. Ilani},
	month = jun,
	year = {2017},
	pages = {4101--4117},
}

@article{brough_preparing_2024,
	title = {Preparing for low surface brightness science with the {Vera} {C}. {Rubin} {Observatory}: a comparison of observable and simulated intracluster light fractions},
	volume = {528},
	issn = {0035-8711},
	shorttitle = {Preparing for low surface brightness science with the {Vera} {C}. {Rubin} {Observatory}},
	url = {https://ui.adsabs.harvard.edu/abs/2024MNRAS.528..771B},
	doi = {10.1093/mnras/stad3810},
	urldate = {2024-02-12},
	journal = {Monthly Notices of the Royal Astronomical Society},
	author = {Brough, Sarah and Ahad, Syeda Lammim and Bahé, Yannick M. and Ellien, Amaël and Gonzalez, Anthony H. and Jiménez-Teja, Yolanda and Kimmig, Lucas C. and Martin, Garreth and Martínez-Lombilla, Cristina and Montes, Mireia and Pillepich, Annalisa and Ragusa, Rossella and Remus, Rhea-Silvia and Collins, Chris A. and Knapen, Johan H. and Mihos, J. Christopher},
	month = feb,
	year = {2024},
	note = {ADS Bibcode: 2024MNRAS.528..771B},
	pages = {771--795},
}

@article{burke_coevolution_2015,
	title = {Coevolution of brightest cluster galaxies and intracluster light using {CLASH}},
	volume = {449},
	issn = {0035-8711},
	url = {https://doi.org/10.1093/mnras/stv450},
	doi = {10.1093/mnras/stv450},
	number = {3},
	urldate = {2023-10-19},
	journal = {Monthly Notices of the Royal Astronomical Society},
	author = {Burke, Claire and Hilton, Matt and Collins, Chris},
	month = may,
	year = {2015},
	pages = {2353--2367},
}

@article{feldmeier_deep_2004,
	title = {Deep {CCD} {Surface} {Photometry} of {Galaxy} {Clusters}. {II}. {Searching} for {Intracluster} {Starlight} in {Non}-{cD} clusters},
	volume = {609},
	issn = {0004-637X},
	url = {https://ui.adsabs.harvard.edu/abs/2004ApJ...609..617F},
	doi = {10.1086/421313},
	urldate = {2024-09-10},
	journal = {The Astrophysical Journal},
	author = {Feldmeier, John J. and Mihos, J. Christopher and Morrison, Heather L. and Harding, Paul and Kaib, Nathan and Dubinski, John},
	month = jul,
	year = {2004},
	note = {Publisher: IOP
ADS Bibcode: 2004ApJ...609..617F},
	pages = {617--637},
}

@article{golden-marx_hierarchical_2025,
	title = {The hierarchical growth of bright central galaxies and intracluster light as traced by the magnitude gap},
	volume = {538},
	copyright = {https://creativecommons.org/licenses/by/4.0/},
	issn = {0035-8711, 1365-2966},
	url = {https://academic.oup.com/mnras/article/538/2/622/8011550},
	doi = {10.1093/mnras/staf277},
	language = {en},
	number = {2},
	urldate = {2025-04-22},
	journal = {Monthly Notices of the Royal Astronomical Society},
	author = {Golden-Marx, Jesse B and Zhang, Y and Ogando, R L C and Yanny, B and da Silva Pereira, M E and Hilton, M and Aguena, M and Allam, S and Andrade-Oliveira, F and Bacon, D and Brooks, D and Carnero Rosell, A and Carretero, J and Cheng, T -Y and da Costa, L N and De Vicente, J and Desai, S and Doel, P and Everett, S and Ferrero, I and Frieman, J and García-Bellido, J and Gatti, M and Giannini, G and Gruen, D and Gruendl, R A and Gutierrez, G and Hinton, S R and Hollowood, D L and Honscheid, K and James, D J and Kuehn, K and Lee, S and Mena-Fernández, J and Menanteau, F and Miquel, R and Mohr, J and Palmese, A and Pieres, A and Plazas Malagón, A A and Samuroff, S and Sanchez, E and Schubnell, M and Sevilla-Noarbe, I and Smith, M and Suchyta, E and Tarle, G and Vikram, V and Walker, A R and Weaverdyck, N and Wiseman, P},
	month = mar,
	year = {2025},
	pages = {622--638},
}

@article{morishita_characterizing_2017,
	title = {Characterizing {Intracluster} {Light} in the {Hubble} {Frontier} {Fields}},
	volume = {846},
	issn = {0004-637X},
	url = {https://dx.doi.org/10.3847/1538-4357/aa8403},
	doi = {10.3847/1538-4357/aa8403},
	language = {en},
	number = {2},
	urldate = {2023-10-19},
	journal = {The Astrophysical Journal},
	author = {Morishita, Takahiro and Abramson, Louis E. and Treu, Tommaso and Schmidt, Kasper B. and Vulcani, Benedetta and Wang, Xin},
	month = sep,
	year = {2017},
	note = {Publisher: The American Astronomical Society},
	pages = {139},
}

@article{jimenez-teja_disentangling_2016,
	title = {{DISENTANGLING} {THE} {ICL} {WITH} {THE} {CHEFs}: {ABELL} 2744 {AS} {A} {CASE} {STUDY}},
	volume = {820},
	issn = {0004-637X},
	shorttitle = {{DISENTANGLING} {THE} {ICL} {WITH} {THE} {CHEFs}},
	url = {https://dx.doi.org/10.3847/0004-637X/820/1/49},
	doi = {10.3847/0004-637X/820/1/49},
	language = {en},
	number = {1},
	urldate = {2023-10-19},
	journal = {The Astrophysical Journal},
	author = {Jiménez-Teja, Y. and Dupke, R.},
	month = mar,
	year = {2016},
	note = {Publisher: The American Astronomical Society},
	pages = {49},
}

@article{ellien_dawis_2021,
	title = {{DAWIS}: a detection algorithm with wavelets for intracluster light studies},
	volume = {649},
	copyright = {© A. Ellien et al. 2021},
	issn = {0004-6361, 1432-0746},
	shorttitle = {{DAWIS}},
	url = {https://www.aanda.org/articles/aa/abs/2021/05/aa38419-20/aa38419-20.html},
	doi = {10.1051/0004-6361/202038419},
	language = {en},
	urldate = {2023-10-19},
	journal = {Astronomy \& Astrophysics},
	author = {Ellien, A. and Slezak, E. and Martinet, N. and Durret, F. and Adami, C. and Gavazzi, R. and Rabaça, C. R. and Rocha, C. Da and Pereira, D. N. Epitácio},
	month = may,
	year = {2021},
	note = {Publisher: EDP Sciences},
	pages = {A38},
}

@article{contreras-santos_characterising_2024,
	title = {Characterising the intra-cluster light in {The} {Three} {Hundred} simulations},
	volume = {683},
	issn = {0004-6361},
	url = {https://ui.adsabs.harvard.edu/abs/2024A&A...683A..59C},
	doi = {10.1051/0004-6361/202348474},
	urldate = {2024-06-23},
	journal = {Astronomy and Astrophysics},
	author = {Contreras-Santos, A. and Knebe, A. and Cui, W. and Alonso Asensio, I. and Dalla Vecchia, C. and Cañas, R. and Haggar, R. and Mostoghiu Paun, R. A. and Pearce, F. R. and Rasia, E.},
	month = mar,
	year = {2024},
	note = {ADS Bibcode: 2024A\&A...683A..59C},
	pages = {A59},
}

@misc{montenegro-taborda_stellar_2025,
	title = {The stellar mass composition of galaxy clusters and dependencies on dark matter halo properties},
	url = {http://arxiv.org/abs/2502.07927},
	doi = {10.48550/arXiv.2502.07927},
	urldate = {2025-02-25},
	publisher = {arXiv},
	author = {Montenegro-Taborda, Daniel and Avila-Reese, Vladimir and Rodriguez-Gomez, Vicente and Manuwal, Aditya and Cervantes-Sodi, Bernardo},
	month = feb,
	year = {2025},
	note = {arXiv:2502.07927 [astro-ph]},
}

@article{remus_outer_2017,
	title = {The {Outer} {Halos} of {Very} {Massive} {Galaxies}: {BCGs} and their {DSC} in the {Magneticum} {Simulations}},
	volume = {5},
	copyright = {http://creativecommons.org/licenses/by/3.0/},
	issn = {2075-4434},
	shorttitle = {The {Outer} {Halos} of {Very} {Massive} {Galaxies}},
	url = {https://www.mdpi.com/2075-4434/5/3/49},
	doi = {10.3390/galaxies5030049},
	language = {en},
	number = {3},
	urldate = {2025-05-14},
	journal = {Galaxies},
	author = {Remus, Rhea-Silvia and Dolag, Klaus and Hoffmann, Tadziu L.},
	month = sep,
	year = {2017},
	note = {Number: 3
Publisher: Multidisciplinary Digital Publishing Institute},
	pages = {49},
}

@article{montes_intracluster_2018,
	title = {Intracluster light at the {Frontier} – {II}. {The} {Frontier} {Fields} {Clusters}},
	volume = {474},
	issn = {0035-8711},
	url = {https://doi.org/10.1093/mnras/stx2847},
	doi = {10.1093/mnras/stx2847},
	number = {1},
	urldate = {2023-10-19},
	journal = {Monthly Notices of the Royal Astronomical Society},
	author = {Montes, Mireia and Trujillo, Ignacio},
	month = feb,
	year = {2018},
	pages = {917--932},
}

@article{rudick_quantity_2011,
	title = {{THE} {QUANTITY} {OF} {INTRACLUSTER} {LIGHT}: {COMPARING} {THEORETICAL} {AND} {OBSERVATIONAL} {MEASUREMENT} {TECHNIQUES} {USING} {SIMULATED} {CLUSTERS}},
	volume = {732},
	issn = {0004-637X},
	shorttitle = {{THE} {QUANTITY} {OF} {INTRACLUSTER} {LIGHT}},
	url = {https://dx.doi.org/10.1088/0004-637X/732/1/48},
	doi = {10.1088/0004-637X/732/1/48},
	language = {en},
	number = {1},
	urldate = {2023-10-19},
	journal = {The Astrophysical Journal},
	author = {Rudick, Craig S. and Mihos, J. Christopher and McBride, Cameron K.},
	month = apr,
	year = {2011},
	note = {Publisher: The American Astronomical Society},
	pages = {48},
}

@article{tang_investigation_2018,
	title = {An {Investigation} of {Intracluster} {Light} {Evolution} {Using} {Cosmological} {Hydrodynamical} {Simulations}},
	volume = {859},
	issn = {0004-637X},
	url = {https://dx.doi.org/10.3847/1538-4357/aabd78},
	doi = {10.3847/1538-4357/aabd78},
	language = {en},
	number = {2},
	urldate = {2023-10-19},
	journal = {The Astrophysical Journal},
	author = {Tang, Lin and Lin, Weipeng and Cui, Weiguang and Kang, Xi and Wang, Yang and Contini, E. and Yu, Yu},
	month = may,
	year = {2018},
	note = {Publisher: The American Astronomical Society},
	pages = {85},
}

@article{canas_stellar_2020,
	title = {From stellar haloes to intracluster light: the physics of the {Intra}-{Halo} {Stellar} {Component} in cosmological hydrodynamical simulations},
	volume = {494},
	issn = {0035-8711},
	shorttitle = {From stellar haloes to intracluster light},
	url = {https://doi.org/10.1093/mnras/staa1027},
	doi = {10.1093/mnras/staa1027},
	number = {3},
	urldate = {2023-10-19},
	journal = {Monthly Notices of the Royal Astronomical Society},
	author = {Cañas, Rodrigo and Lagos, Claudia del P and Elahi, Pascal J and Power, Chris and Welker, Charlotte and Dubois, Yohan and Pichon, Christophe},
	month = may,
	year = {2020},
	pages = {4314--4333},
}

@article{contini_connection_2024,
	title = {The {Connection} between the {Intracluster} {Light} and its {Host} {Halo}: {Formation} {Time} and {Contribution} from {Different} {Channels}},
	volume = {167},
	issn = {0004-6256, 1538-3881},
	shorttitle = {The {Connection} between the {Intracluster} {Light} and its {Host} {Halo}},
	url = {https://iopscience.iop.org/article/10.3847/1538-3881/ad0894},
	doi = {10.3847/1538-3881/ad0894},
	language = {en},
	number = {1},
	urldate = {2024-03-13},
	journal = {The Astronomical Journal},
	author = {Contini, Emanuele and Rhee, Jinsu and Han, San and Jeon, Seyoung and Yi, Sukyoung K.},
	month = jan,
	year = {2024},
	pages = {7},
}

@article{furnell_growth_2021,
	title = {The growth of intracluster light in {XCS}-{HSC} galaxy clusters from 0.1 {\textless} z {\textless} 0.5},
	volume = {502},
	issn = {0035-8711},
	url = {https://doi.org/10.1093/mnras/stab065},
	doi = {10.1093/mnras/stab065},
	number = {2},
	urldate = {2023-10-19},
	journal = {Monthly Notices of the Royal Astronomical Society},
	author = {Furnell, Kate E and Collins, Chris A and Kelvin, Lee S and Baldry, Ivan K and James, Phil A and Manolopoulou, Maria and Mann, Robert G and Giles, Paul A and Bermeo, Alberto and Hilton, Matthew and Wilkinson, Reese and Romer, A Kathy and Vergara, Carlos and Bhargava, Sunayana and Stott, John P and Mayers, Julian and Viana, Pedro},
	month = apr,
	year = {2021},
	pages = {2419--2437},
}

@article{krick_diffuse_2007,
	title = {Diffuse {Optical} {Light} in {Galaxy} {Clusters}. {II}. {Correlations} with {Cluster} {Properties}},
	volume = {134},
	issn = {1538-3881},
	url = {https://iopscience.iop.org/article/10.1086/518787/meta},
	doi = {10.1086/518787},
	language = {en},
	number = {2},
	urldate = {2023-10-19},
	journal = {The Astronomical Journal},
	author = {Krick, J. E. and Bernstein, R. A.},
	month = jun,
	year = {2007},
	note = {Publisher: IOP Publishing},
	pages = {466},
}

@article{jimenez-teja_unveiling_2018,
	title = {Unveiling the {Dynamical} {State} of {Massive} {Clusters} through the {ICL} {Fraction}},
	volume = {857},
	issn = {0004-637X, 1538-4357},
	url = {https://iopscience.iop.org/article/10.3847/1538-4357/aab70f},
	doi = {10.3847/1538-4357/aab70f},
	language = {en},
	number = {2},
	urldate = {2024-06-23},
	journal = {The Astrophysical Journal},
	author = {Jiménez-Teja, Yolanda and Dupke, Renato and Benítez, Narciso and Koekemoer, Anton M. and Zitrin, Adi and Umetsu, Keiichi and Ziegler, Bodo L. and Frye, Brenda L. and Ford, Holland and Bouwens, Rychard J. and Bradley, Larry D. and Broadhurst, Thomas and Coe, Dan and Donahue, Megan and Graves, Genevieve J. and Grillo, Claudio and Infante, Leopoldo and Jouvel, Stephanie and Kelson, Daniel D. and Lahav, Ofer and Lazkoz, Ruth and Lemze, Dorom and Maoz, Dan and Medezinski, Elinor and Melchior, Peter and Meneghetti, Massimo and Mercurio, Amata and Merten, Julian and Molino, Alberto and Moustakas, Leonidas A. and Nonino, Mario and Ogaz, Sara and Riess, Adam G. and Rosati, Piero and Sayers, Jack and Seitz, Stella and Zheng, Wei},
	month = apr,
	year = {2018},
	pages = {79},
}

@article{purcell_shredded_2007,
	title = {Shredded {Galaxies} as the {Source} of {Diffuse} {Intrahalo} {Light} on {Varying} {Scales}},
	volume = {666},
	issn = {0004-637X},
	url = {https://iopscience.iop.org/article/10.1086/519787/meta},
	doi = {10.1086/519787},
	language = {en},
	number = {1},
	urldate = {2023-10-19},
	journal = {The Astrophysical Journal},
	author = {Purcell, Chris W. and Bullock, James S. and Zentner, Andrew R.},
	month = sep,
	year = {2007},
	note = {Publisher: IOP Publishing},
	pages = {20},
}

@article{cui_characterizing_2014,
	title = {Characterizing diffused stellar light in simulated galaxy clusters},
	volume = {437},
	issn = {0035-8711},
	url = {https://doi.org/10.1093/mnras/stt1940},
	doi = {10.1093/mnras/stt1940},
	number = {1},
	urldate = {2023-10-19},
	journal = {Monthly Notices of the Royal Astronomical Society},
	author = {Cui, Weiguang and Murante, G. and Monaco, P. and Borgani, S. and Granato, G. L. and Killedar, M. and De Lucia, G. and Presotto, V. and Dolag, K.},
	month = jan,
	year = {2014},
	pages = {816--830},
}

@article{proctor_identifying_2024,
	title = {Identifying the discs, bulges, and intra-halo light of simulated galaxies through structural decomposition},
	volume = {527},
	issn = {0035-8711},
	url = {https://ui.adsabs.harvard.edu/abs/2024MNRAS.527.2624P},
	doi = {10.1093/mnras/stad3341},
	urldate = {2024-02-12},
	journal = {Monthly Notices of the Royal Astronomical Society},
	author = {Proctor, Katy L. and Lagos, Claudia del P. and Ludlow, Aaron D. and Robotham, Aaron S. G.},
	month = jan,
	year = {2024},
	note = {ADS Bibcode: 2024MNRAS.527.2624P},
	pages = {2624--2638},
}

@article{ragusa_does_2023,
	title = {Does the virial mass drive the intra-cluster light?: {Relationship} between the {ICL} and \textit{{M}} $_{\textrm{vir}}$ from {VEGAS}},
	volume = {670},
	issn = {0004-6361, 1432-0746},
	shorttitle = {Does the virial mass drive the intra-cluster light?},
	url = {https://www.aanda.org/10.1051/0004-6361/202245530},
	doi = {10.1051/0004-6361/202245530},
	language = {en},
	urldate = {2024-03-21},
	journal = {Astronomy \& Astrophysics},
	author = {Ragusa, R. and Iodice, E. and Spavone, M. and Montes, M. and Forbes, D. A. and Brough, S. and Mirabile, M. and Cantiello, M. and Paolillo, M. and Schipani, P.},
	month = feb,
	year = {2023},
	pages = {L20},
}

@article{kluge_photometric_2021,
	title = {Photometric {Dissection} of {Intracluster} {Light} and {Its} {Correlations} with {Host} {Cluster} {Properties}},
	volume = {252},
	issn = {0067-0049},
	url = {https://dx.doi.org/10.3847/1538-4365/abcda6},
	doi = {10.3847/1538-4365/abcda6},
	language = {en},
	number = {2},
	urldate = {2023-10-19},
	journal = {The Astrophysical Journal Supplement Series},
	author = {Kluge, M. and Bender, R. and Riffeser, A. and Goessl, C. and Hopp, U. and Schmidt, M. and Ries, C.},
	month = feb,
	year = {2021},
	note = {Publisher: The American Astronomical Society},
	pages = {27},
}

@article{ivezic_lsst_2019,
	title = {{LSST}: {From} {Science} {Drivers} to {Reference} {Design} and {Anticipated} {Data} {Products}},
	volume = {873},
	issn = {0004-637X},
	shorttitle = {{LSST}},
	url = {https://dx.doi.org/10.3847/1538-4357/ab042c},
	doi = {10.3847/1538-4357/ab042c},
	language = {en},
	number = {2},
	urldate = {2023-10-19},
	journal = {The Astrophysical Journal},
	author = {Ivezić, {\v{Z}}eljko and Kahn, Steven M. and Tyson, J. Anthony and Abel, Bob and Acosta, Emily and Allsman, Robyn and Alonso, David and AlSayyad, Yusra and Anderson, Scott F. and Andrew, John and Angel, James Roger P. and Angeli, George Z. and Ansari, Reza and Antilogus, Pierre and Araujo, Constanza and Armstrong, Robert and Arndt, Kirk T. and Astier, Pierre and Aubourg, Éric and Auza, Nicole and Axelrod, Tim S. and Bard, Deborah J. and Barr, Jeff D. and Barrau, Aurelian and Bartlett, James G. and Bauer, Amanda E. and Bauman, Brian J. and Baumont, Sylvain and Bechtol, Ellen and Bechtol, Keith and Becker, Andrew C. and Becla, Jacek and Beldica, Cristina and Bellavia, Steve and Bianco, Federica B. and Biswas, Rahul and Blanc, Guillaume and Blazek, Jonathan and Blandford, Roger D. and Bloom, Josh S. and Bogart, Joanne and Bond, Tim W. and Booth, Michael T. and Borgland, Anders W. and Borne, Kirk and Bosch, James F. and Boutigny, Dominique and Brackett, Craig A. and Bradshaw, Andrew and Brandt, William Nielsen and Brown, Michael E. and Bullock, James S. and Burchat, Patricia and Burke, David L. and Cagnoli, Gianpietro and Calabrese, Daniel and Callahan, Shawn and Callen, Alice L. and Carlin, Jeffrey L. and Carlson, Erin L. and Chandrasekharan, Srinivasan and Charles-Emerson, Glenaver and Chesley, Steve and Cheu, Elliott C. and Chiang, Hsin-Fang and Chiang, James and Chirino, Carol and Chow, Derek and Ciardi, David R. and Claver, Charles F. and Cohen-Tanugi, Johann and Cockrum, Joseph J. and Coles, Rebecca and Connolly, Andrew J. and Cook, Kem H. and Cooray, Asantha and Covey, Kevin R. and Cribbs, Chris and Cui, Wei and Cutri, Roc and Daly, Philip N. and Daniel, Scott F. and Daruich, Felipe and Daubard, Guillaume and Daues, Greg and Dawson, William and Delgado, Francisco and Dellapenna, Alfred and Peyster, Robert de and Val-Borro, Miguel de and Digel, Seth W. and Doherty, Peter and Dubois, Richard and Dubois-Felsmann, Gregory P. and Durech, Josef and Economou, Frossie and Eifler, Tim and Eracleous, Michael and Emmons, Benjamin L. and Neto, Angelo Fausti and Ferguson, Henry and Figueroa, Enrique and Fisher-Levine, Merlin and Focke, Warren and Foss, Michael D. and Frank, James and Freemon, Michael D. and Gangler, Emmanuel and Gawiser, Eric and Geary, John C. and Gee, Perry and Geha, Marla and Gessner, Charles J. B. and Gibson, Robert R. and Gilmore, D. Kirk and Glanzman, Thomas and Glick, William and Goldina, Tatiana and Goldstein, Daniel A. and Goodenow, Iain and Graham, Melissa L. and Gressler, William J. and Gris, Philippe and Guy, Leanne P. and Guyonnet, Augustin and Haller, Gunther and Harris, Ron and Hascall, Patrick A. and Haupt, Justine and Hernandez, Fabio and Herrmann, Sven and Hileman, Edward and Hoblitt, Joshua and Hodgson, John A. and Hogan, Craig and Howard, James D. and Huang, Dajun and Huffer, Michael E. and Ingraham, Patrick and Innes, Walter R. and Jacoby, Suzanne H. and Jain, Bhuvnesh and Jammes, Fabrice and Jee, M. James and Jenness, Tim and Jernigan, Garrett and Jevremović, Darko and Johns, Kenneth and Johnson, Anthony S. and Johnson, Margaret W. G. and Jones, R. Lynne and Juramy-Gilles, Claire and Jurić, Mario and Kalirai, Jason S. and Kallivayalil, Nitya J. and Kalmbach, Bryce and Kantor, Jeffrey P. and Karst, Pierre and Kasliwal, Mansi M. and Kelly, Heather and Kessler, Richard and Kinnison, Veronica and Kirkby, David and Knox, Lloyd and Kotov, Ivan V. and Krabbendam, Victor L. and Krughoff, K. Simon and Kubánek, Petr and Kuczewski, John and Kulkarni, Shri and Ku, John and Kurita, Nadine R. and Lage, Craig S. and Lambert, Ron and Lange, Travis and Langton, J. Brian and Guillou, Laurent Le and Levine, Deborah and Liang, Ming and Lim, Kian-Tat and Lintott, Chris J. and Long, Kevin E. and Lopez, Margaux and Lotz, Paul J. and Lupton, Robert H. and Lust, Nate B. and MacArthur, Lauren A. and Mahabal, Ashish and Mandelbaum, Rachel and Markiewicz, Thomas W. and Marsh, Darren S. and Marshall, Philip J. and Marshall, Stuart and May, Morgan and McKercher, Robert and McQueen, Michelle and Meyers, Joshua and Migliore, Myriam and Miller, Michelle and Mills, David J. and Miraval, Connor and Moeyens, Joachim and Moolekamp, Fred E. and Monet, David G. and Moniez, Marc and Monkewitz, Serge and Montgomery, Christopher and Morrison, Christopher B. and Mueller, Fritz and Muller, Gary P. and Arancibia, Freddy Muñoz and Neill, Douglas R. and Newbry, Scott P. and Nief, Jean-Yves and Nomerotski, Andrei and Nordby, Martin and O’Connor, Paul and Oliver, John and Olivier, Scot S. and Olsen, Knut and O’Mullane, William and Ortiz, Sandra and Osier, Shawn and Owen, Russell E. and Pain, Reynald and Palecek, Paul E. and Parejko, John K. and Parsons, James B. and Pease, Nathan M. and Peterson, J. Matt and Peterson, John R. and Petravick, Donald L. and Petrick, M. E. Libby and Petry, Cathy E. and Pierfederici, Francesco and Pietrowicz, Stephen and Pike, Rob and Pinto, Philip A. and Plante, Raymond and Plate, Stephen and Plutchak, Joel P. and Price, Paul A. and Prouza, Michael and Radeka, Veljko and Rajagopal, Jayadev and Rasmussen, Andrew P. and Regnault, Nicolas and Reil, Kevin A. and Reiss, David J. and Reuter, Michael A. and Ridgway, Stephen T. and Riot, Vincent J. and Ritz, Steve and Robinson, Sean and Roby, William and Roodman, Aaron and Rosing, Wayne and Roucelle, Cecille and Rumore, Matthew R. and Russo, Stefano and Saha, Abhijit and Sassolas, Benoit and Schalk, Terry L. and Schellart, Pim and Schindler, Rafe H. and Schmidt, Samuel and Schneider, Donald P. and Schneider, Michael D. and Schoening, William and Schumacher, German and Schwamb, Megan E. and Sebag, Jacques and Selvy, Brian and Sembroski, Glenn H. and Seppala, Lynn G. and Serio, Andrew and Serrano, Eduardo and Shaw, Richard A. and Shipsey, Ian and Sick, Jonathan and Silvestri, Nicole and Slater, Colin T. and Smith, J. Allyn and Smith, R. Chris and Sobhani, Shahram and Soldahl, Christine and Storrie-Lombardi, Lisa and Stover, Edward and Strauss, Michael A. and Street, Rachel A. and Stubbs, Christopher W. and Sullivan, Ian S. and Sweeney, Donald and Swinbank, John D. and Szalay, Alexander and Takacs, Peter and Tether, Stephen A. and Thaler, Jon J. and Thayer, John Gregg and Thomas, Sandrine and Thornton, Adam J. and Thukral, Vaikunth and Tice, Jeffrey and Trilling, David E. and Turri, Max and Berg, Richard Van and Berk, Daniel Vanden and Vetter, Kurt and Virieux, Francoise and Vucina, Tomislav and Wahl, William and Walkowicz, Lucianne and Walsh, Brian and Walter, Christopher W. and Wang, Daniel L. and Wang, Shin-Yawn and Warner, Michael and Wiecha, Oliver and Willman, Beth and Winters, Scott E. and Wittman, David and Wolff, Sidney C. and Wood-Vasey, W. Michael and Wu, Xiuqin and Xin, Bo and Yoachim, Peter and Zhan, Hu},
	month = mar,
	year = {2019},
	note = {Publisher: The American Astronomical Society},
	pages = {111},
}

@article{mellier_euclid_2025,
	title = {Euclid - {I}. {Overview} of the {Euclid} mission},
	volume = {697},
	copyright = {© The Authors 2025},
	issn = {0004-6361, 1432-0746},
	url = {https://www.aanda.org/articles/aa/abs/2025/05/aa50810-24/aa50810-24.html},
	doi = {10.1051/0004-6361/202450810},
	language = {en},
	urldate = {2025-05-14},
	journal = {Astronomy \& Astrophysics},
	author = {Mellier, Y. and Abdurro’uf, Abdurro’uf and Barroso, J. A. Acevedo and Achúcarro, A. and Adamek, J. and Adam, R. and Addison, G. E. and Aghanim, N. and Aguena, M. and Ajani, V. and Akrami, Y. and Al-Bahlawan, A. and Alavi, A. and Albuquerque, I. S. and Alestas, G. and Alguero, G. and Allaoui, A. and Allen, S. W. and Allevato, V. and Alonso-Tetilla, A. V. and Altieri, B. and Alvarez-Candal, A. and Alvi, S. and Amara, A. and Amendola, L. and Amiaux, J. and Andika, I. T. and Andreon, S. and Andrews, A. and Angora, G. and Angulo, R. E. and Annibali, F. and Anselmi, A. and Anselmi, S. and Arcari, S. and Archidiacono, M. and Aricò, G. and Arnaud, M. and Arnouts, S. and Asgari, M. and Asorey, J. and Atayde, L. and Atek, H. and Atrio-Barandela, F. and Aubert, M. and Aubourg, E. and Auphan, T. and Auricchio, N. and Aussel, B. and Aussel, H. and Avelino, P. P. and Avgoustidis, A. and Avila, S. and Awan, S. and Azzollini, R. and Baccigalupi, C. and Bachelet, E. and Bacon, D. and Baes, M. and Bagley, M. B. and Bahr-Kalus, B. and Balaguera-Antolinez, A. and Balbinot, E. and Balcells, M. and Baldi, M. and Baldry, I. and Balestra, A. and Ballardini, M. and Ballester, O. and Balogh, M. and Bañados, E. and Barbier, R. and Bardelli, S. and Baron, M. and Barreiro, T. and Barrena, R. and Barriere, J.-C. and Barros, B. J. and Barthelemy, A. and Bartolo, N. and Basset, A. and Battaglia, P. and Battisti, A. J. and Baugh, C. M. and Baumont, L. and Bazzanini, L. and Beaulieu, J.-P. and Beckmann, V. and Belikov, A. N. and Bel, J. and Bellagamba, F. and Bella, M. and Bellini, E. and Benabed, K. and Bender, R. and Benevento, G. and Bennett, C. L. and Benson, K. and Bergamini, P. and Bermejo-Climent, J. R. and Bernardeau, F. and Bertacca, D. and Berthe, M. and Berthier, J. and Bethermin, M. and Beutler, F. and Bevillon, C. and Bhargava, S. and Bhatawdekar, R. and Bianchi, D. and Bisigello, L. and Biviano, A. and Blake, R. P. and Blanchard, A. and Blazek, J. and Blot, L. and Bosco, A. and Bodendorf, C. and Boenke, T. and Böhringer, H. and Boldrini, P. and Bolzonella, M. and Bonchi, A. and Bonici, M. and Bonino, D. and Bonino, L. and Bonvin, C. and Bon, W. and Booth, J. T. and Borgani, S. and Borlaff, A. S. and Borsato, E. and Bosco, A. and Bose, B. and Botticella, M. T. and Boucaud, A. and Bouche, F. and Boucher, J. S. and Boutigny, D. and Bouvard, T. and Bouwens, R. and Bouy, H. and Bowler, R. a. A. and Bozza, V. and Bozzo, E. and Branchini, E. and Brando, G. and Brau-Nogue, S. and Brekke, P. and Bremer, M. N. and Brescia, M. and Breton, M.-A. and Brinchmann, J. and Brinckmann, T. and Brockley-Blatt, C. and Brodwin, M. and Brouard, L. and Brown, M. L. and Bruton, S. and Bucko, J. and Buddelmeijer, H. and Buenadicha, G. and Buitrago, F. and Burger, P. and Burigana, C. and Busillo, V. and Busonero, D. and Cabanac, R. and Cabayol-Garcia, L. and Cagliari, M. S. and Caillat, A. and Caillat, L. and Calabrese, M. and Calabro, A. and Calderone, G. and Calura, F. and Quevedo, B. Camacho and Camera, S. and Campos, L. and Cañas-Herrera, G. and Candini, G. P. and Cantiello, M. and Capobianco, V. and Cappellaro, E. and Cappelluti, N. and Cappi, A. and Caputi, K. I. and Cara, C. and Carbone, C. and Cardone, V. F. and Carella, E. and Carlberg, R. G. and Carle, M. and Carminati, L. and Caro, F. and Carrasco, J. M. and Carretero, J. and Carrilho, P. and Duque, J. Carron and Carry, B. and Carvalho, A. and Carvalho, C. S. and Casas, R. and Casas, S. and Casenove, P. and Casey, C. M. and Cassata, P. and Castander, F. J. and Castelao, D. and Castellano, M. and Castiblanco, L. and Castignani, G. and Castro, T. and Cavet, C. and Cavuoti, S. and Chabaud, P.-Y. and Chambers, K. C. and Charles, Y. and Charlot, S. and Chartab, N. and Chary, R. and Chaumeil, F. and Cho, H. and Chon, G. and Ciancetta, E. and Ciliegi, P. and Cimatti, A. and Cimino, M. and Cioni, M.-R. L. and Claydon, R. and Cleland, C. and Clément, B. and Clements, D. L. and Clerc, N. and Clesse, S. and Codis, S. and Cogato, F. and Colbert, J. and Cole, R. E. and Coles, P. and Collett, T. E. and Collins, R. S. and Colodro-Conde, C. and Colombo, C. and Combes, F. and Conforti, V. and Congedo, G. and Conseil, S. and Conselice, C. J. and Contarini, S. and Contini, T. and Conversi, L. and Cooray, A. R. and Copin, Y. and Corasaniti, P.-S. and Corcho-Caballero, P. and Corcione, L. and Cordes, O. and Corpace, O. and Correnti, M. and Costanzi, M. and Costille, A. and Courbin, F. and Mifsud, L. Courcoult and Courtois, H. M. and Cousinou, M.-C. and Covone, G. and Cowell, T. and Cragg, C. and Cresci, G. and Cristiani, S. and Crocce, M. and Cropper, M. and Crouzet, P. E. and Csizi, B. and Cuby, J.-G. and Cucchetti, E. and Cucciati, O. and Cuillandre, J.-C. and Cunha, P. a. C. and Cuozzo, V. and Daddi, E. and D’Addona, M. and Dafonte, C. and Dagoneau, N. and Dalessandro, E. and Dalton, G. B. and D’Amico, G. and Dannerbauer, H. and Danto, P. and Das, I. and Silva, A. Da and Silva, R. da and Doumerg, W. d’Assignies and Daste, G. and Davies, J. E. and Davini, S. and Dayal, P. and Boer, T. de and Decarli, R. and Caro, B. De and Degaudenzi, H. and Degni, G. and Jong, J. T. A. de and Bella, L. F. de la and Torre, S. de la and Delhaise, F. and Delley, D. and Delucchi, G. and Lucia, G. De and Denniston, J. and Paolis, F. De and Petris, M. De and Derosa, A. and Desai, S. and Desjacques, V. and Despali, G. and Desprez, G. and Vicente-Albendea, J. De and Deville, Y. and Dias, J. D. F. and Díaz-Sánchez, A. and Diaz, J. J. and Domizio, S. Di and Diego, J. M. and Ferdinando, D. Di and Giorgio, A. M. Di and Dimauro, P. and Dinis, J. and Dolag, K. and Dolding, C. and Dole, H. and Sánchez, H. Domínguez and Doré, O. and Dournac, F. and Douspis, M. and Dreihahn, H. and Droge, B. and Dryer, B. and Dubath, F. and Duc, P.-A. and Ducret, F. and Duffy, C. and Dufresne, F. and Duncan, C. a. J. and Dupac, X. and Duret, V. and Durrer, R. and Durret, F. and Dusini, S. and Ealet, A. and Eggemeier, A. and Eisenhardt, P. R. M. and Elbaz, D. and Elkhashab, M. Y. and Ellien, A. and Endicott, J. and Enia, A. and Erben, T. and Vigo, J. A. Escartin and Escoffier, S. and Sanz, I. Escudero and Essert, J. and Ettori, S. and Ezziati, M. and Fabbian, G. and Fabricius, M. and Fang, Y. and Farina, A. and Farina, M. and Farinelli, R. and Farrens, S. and Faustini, F. and Feltre, A. and Ferguson, A. M. N. and Ferrando, P. and Ferrari, A. G. and Ferré-Mateu, A. and Ferreira, P. G. and Ferreras, I. and Ferrero, I. and Ferriol, S. and Ferruit, P. and Filleul, D. and Finelli, F. and Finkelstein, S. L. and Finoguenov, A. and Fiorini, B. and Flentge, F. and Focardi, P. and Fonseca, J. and Fontana, A. and Fontanot, F. and Fornari, F. and Fosalba, P. and Fossati, M. and Fotopoulou, S. and Fouchez, D. and Fourmanoit, N. and Frailis, M. and Fraix-Burnet, D. and Franceschi, E. and Franco, A. and Franzetti, P. and Freihoefer, J. and Frenk, C. S. and Frittoli, G. and Frugier, P.-A. and Frusciante, N. and Fumagalli, A. and Fumagalli, M. and Fumana, M. and Fu, Y. and Gabarra, L. and Galeotta, S. and Galluccio, L. and Ganga, K. and Gao, H. and García-Bellido, J. and Garcia, K. and Gardner, J. P. and Garilli, B. and Gaspar-Venancio, L.-M. and Gasparetto, T. and Gautard, V. and Gavazzi, R. and Gaztanaga, E. and Genolet, L. and Santos, R. Genova and Gentile, F. and George, K. and Gerbino, M. and Ghaffari, Z. and Giacomini, F. and Gianotti, F. and Gibb, G. P. S. and Gillard, W. and Gillis, B. and Ginolfi, M. and Giocoli, C. and Girardi, M. and Giri, S. K. and Goh, L. W. K. and Gómez-Alvarez, P. and Gonzalez-Perez, V. and Gonzalez, A. H. and Gonzalez, E. J. and Gonzalez, J. C. and Beauchamps, S. Gouyou and Gozaliasl, G. and Gracia-Carpio, J. and Grandis, S. and Granett, B. R. and Granvik, M. and Grazian, A. and Gregorio, A. and Grenet, C. and Grillo, C. and Grupp, F. and Gruppioni, C. and Gruppuso, A. and Guerbuez, C. and Guerrini, S. and Guidi, M. and Guillard, P. and Gutierrez, C. M. and Guttridge, P. and Guzzo, L. and Gwyn, S. and Haapala, J. and Haase, J. and Haddow, C. R. and Hailey, M. and Hall, A. and Hall, D. and Hamaus, N. and Haridasu, B. S. and Harnois-Déraps, J. and Harper, C. and Hartley, W. G. and Hasinger, G. and Hassani, F. and Hatch, N. A. and Haugan, S. V. H. and Häußler, B. and Heavens, A. and Heisenberg, L. and Helmi, A. and Helou, G. and Hemmati, S. and Henares, K. and Herent, O. and Hernández-Monteagudo, C. and Heuberger, T. and Hewett, P. C. and Heydenreich, S. and Hildebrandt, H. and Hirschmann, M. and Hjorth, J. and Hoar, J. and Hoekstra, H. and Holland, A. D. and Holliman, M. S. and Holmes, W. and Hook, I. and Horeau, B. and Hormuth, F. and Hornstrup, A. and Hosseini, S. and Hu, D. and Hudelot, P. and Hudson, M. J. and Huertas-Company, M. and Huff, E. M. and Hughes, A. C. N. and Humphrey, A. and Hunt, L. K. and Huynh, D. D. and Ibata, R. and Ichikawa, K. and Iglesias-Groth, S. and Ilbert, O. and Ilić, S. and Ingoglia, L. and Iodice, E. and Israel, H. and Israelsson, U. E. and Izzo, L. and Jablonka, P. and Jackson, N. and Jacobson, J. and Jafariyazani, M. and Jahnke, K. and Jain, B. and Jansen, H. and Jarvis, M. J. and Jasche, J. and Jauzac, M. and Jeffrey, N. and Jhabvala, M. and Jimenez-Teja, Y. and Muñoz, A. Jimenez and Joachimi, B. and Johansson, P. H. and Joudaki, S. and Jullo, E. and Kajava, J. J. E. and Kang, Y. and Kannawadi, A. and Kansal, V. and Karagiannis, D. and Kärcher, M. and Kashlinsky, A. and Kazandjian, M. V. and Keck, F. and Keihänen, E. and Kerins, E. and Kermiche, S. and Khalil, A. and Kiessling, A. and Kiiveri, K. and Kilbinger, M. and Kim, J. and King, R. and Kirkpatrick, C. C. and Kitching, T. and Kluge, M. and Knabenhans, M. and Knapen, J. H. and Knebe, A. and Kneib, J.-P. and Kohley, R. and Koopmans, L. V. E. and Koskinen, H. and Koulouridis, E. and Kou, R. and Kovács, A. and Kovačić, I. and Kowalczyk, A. and Koyama, K. and Kraljic, K. and Krause, O. and Kruk, S. and Kubik, B. and Kuchner, U. and Kuijken, K. and Kümmel, M. and Kunz, M. and Kurki-Suonio, H. and Lacasa, F. and Lacey, C. G. and Franca, F. La and Lagarde, N. and Lahav, O. and Laigle, C. and Marca, A. La and Marle, O. La and Lamine, B. and Lam, M. C. and Lançon, A. and Landt, H. and Langer, M. and Lapi, A. and Larcheveque, C. and Larsen, S. S. and Lattanzi, M. and Laudisio, F. and Laugier, D. and Laureijs, R. and Laurent, V. and Lavaux, G. and Lawrenson, A. and Lazanu, A. and Lazeyras, T. and Boulc’h, Q. Le and Brun, A. M. C. Le and Brun, V. Le and Leclercq, F. and Lee, S. and Graet, J. Le and Legrand, L. and Leirvik, K. N. and Jeune, M. Le and Lembo, M. and Mignant, D. Le and Lepinzan, M. D. and Lepori, F. and Reun, A. Le and Leroy, G. and Lesci, G. F. and Lesgourgues, J. and Leuzzi, L. and Levi, M. E. and Liaudat, T. I. and Libet, G. and Liebing, P. and Ligori, S. and Lilje, P. B. and Lin, C.-C. and Linde, D. and Linder, E. and Lindholm, V. and Linke, L. and Li, S.-S. and Liu, S. J. and Lloro, I. and Lobo, F. S. N. and Lodieu, N. and Lombardi, M. and Lombriser, L. and Lonare, P. and Longo, G. and López-Caniego, M. and Lopez, X. Lopez and Alvarez, J. Lorenzo and Loureiro, A. and Loveday, J. and Lusso, E. and Macias-Perez, J. and Maciaszek, T. and Maggio, G. and Magliocchetti, M. and Magnard, F. and Magnier, E. A. and Magro, A. and Mahler, G. and Mainetti, G. and Maino, D. and Maiorano, E. and Malavasi, N. and Mamon, G. A. and Mancini, C. and Mandelbaum, R. and Manera, M. and Manjón-García, A. and Mannucci, F. and Mansutti, O. and Outeiro, M. Manteiga and Maoli, R. and Maraston, C. and Marcin, S. and Marcos-Arenal, P. and Margalef-Bentabol, B. and Marggraf, O. and Marinucci, D. and Marinucci, M. and Markovic, K. and Marleau, F. R. and Marpaud, J. and Martignac, J. and Martín-Fleitas, J. and Martin-Moruno, P. and Martin, E. L. and Martinelli, M. and Martinet, N. and Martin, H. and Martins, C. J. a. P. and Marulli, F. and Massari, D. and Massey, R. and Masters, D. C. and Matarrese, S. and Matsuoka, Y. and Matthew, S. and Maughan, B. J. and Mauri, N. and Maurin, L. and Maurogordato, S. and McCarthy, K. and McConnachie, A. W. and McCracken, H. J. and McDonald, I. and McEwen, J. D. and McPartland, C. J. R. and Medinaceli, E. and Mehta, V. and Mei, S. and Melchior, M. and Melin, J.-B. and Ménard, B. and Mendes, J. and Mendez-Abreu, J. and Meneghetti, M. and Mercurio, A. and Merlin, E. and Metcalf, R. B. and Meylan, G. and Migliaccio, M. and Mignoli, M. and Miller, L. and Miluzio, M. and Milvang-Jensen, B. and Mimoso, J. P. and Miquel, R. and Miyatake, H. and Mobasher, B. and Mohr, J. J. and Monaco, P. and Monguió, M. and Montoro, A. and Mora, A. and Dizgah, A. Moradinezhad and Moresco, M. and Moretti, C. and Morgante, G. and Morisset, N. and Moriya, T. J. and Morris, P. W. and Mortlock, D. J. and Moscardini, L. and Mota, D. F. and Mottet, S. and Moustakas, L. A. and Moutard, T. and Müller, T. and Munari, E. and Murphree, G. and Murray, C. and Murray, N. and Musi, P. and Nadathur, S. and Nagam, B. C. and Nagao, T. and Naidoo, K. and Nakajima, R. and Nally, C. and Natoli, P. and Navarro-Alsina, A. and Girones, D. Navarro and Neissner, C. and Nersesian, A. and Nesseris, S. and Nguyen-Kim, H. N. and Nicastro, L. and Nichol, R. C. and Nielbock, M. and Niemi, S.-M. and Nieto, S. and Nilsson, K. and Noller, J. and Norberg, P. and Nouri-Zonoz, A. and Ntelis, P. and Nucita, A. A. and Nugent, P. and Nunes, N. J. and Nutma, T. and Ocampo, I. and Odier, J. and Oesch, P. A. and Oguri, M. and Oliveira, D. Magalhaes and Onoue, M. and Oosterbroek, T. and Oppizzi, F. and Ordenovic, C. and Osato, K. and Pacaud, F. and Pace, F. and Padilla, C. and Paech, K. and Pagano, L. and Page, M. J. and Palazzi, E. and Paltani, S. and Pamuk, S. and Pandolfi, S. and Paoletti, D. and Paolillo, M. and Papaderos, P. and Pardede, K. and Parimbelli, G. and Parmar, A. and Partmann, C. and Pasian, F. and Passalacqua, F. and Paterson, K. and Patrizii, L. and Pattison, C. and Paulino-Afonso, A. and Paviot, R. and Peacock, J. A. and Pearce, F. R. and Pedersen, K. and Peel, A. and Peletier, R. F. and Ibanez, M. Pellejero and Pello, R. and Penny, M. T. and Percival, W. J. and Perez-Garrido, A. and Perotto, L. and Pettorino, V. and Pezzotta, A. and Pezzuto, S. and Philippon, A. and Pierre, M. and Piersanti, O. and Pietroni, M. and Piga, L. and Pilo, L. and Pires, S. and Pisani, A. and Pizzella, A. and Pizzuti, L. and Plana, C. and Polenta, G. and Pollack, J. E. and Poncet, M. and Pöntinen, M. and Pool, P. and Popa, L. A. and Popa, V. and Popp, J. and Porciani, C. and Porth, L. and Potter, D. and Poulain, M. and Pourtsidou, A. and Pozzetti, L. and Prandoni, I. and Pratt, G. W. and Prezelus, S. and Prieto, E. and Pugno, A. and Quai, S. and Quilley, L. and Racca, G. D. and Raccanelli, A. and Rácz, G. and Radinović, S. and Radovich, M. and Ragagnin, A. and Ragnit, U. and Raison, F. and Ramos-Chernenko, N. and Ranc, C. and Rasera, Y. and Raylet, N. and Rebolo, R. and Refregier, A. and Reimberg, P. and Reiprich, T. H. and Renk, F. and Renzi, A. and Retre, J. and Revaz, Y. and Reylé, C. and Reynolds, L. and Rhodes, J. and Ricci, F. and Ricci, M. and Riccio, G. and Ricken, S. O. and Rissanen, S. and Risso, I. and Rix, H.-W. and Robin, A. C. and Rocca-Volmerange, B. and Rocci, P.-F. and Rodenhuis, M. and Rodighiero, G. and Monroy, M. Rodriguez and Rollins, R. P. and Romanello, M. and Roman, J. and Romelli, E. and Romero-Gomez, M. and Roncarelli, M. and Rosati, P. and Rosset, C. and Rossetti, E. and Roster, W. and Rottgering, H. J. A. and Rozas-Fernández, A. and Ruane, K. and Rubino-Martin, J. A. and Rudolph, A. and Ruppin, F. and Rusholme, B. and Sacquegna, S. and Sáez-Casares, I. and Saga, S. and Saglia, R. and Sahlén, M. and Saifollahi, T. and Sakr, Z. and Salvalaggio, J. and Salvaterra, R. and Salvati, L. and Salvato, M. and Salvignol, J.-C. and Sánchez, A. G. and Sanchez, E. and Sanders, D. B. and Sapone, D. and Saponara, M. and Sarpa, E. and Sarron, F. and Sartori, S. and Sartoris, B. and Sassolas, B. and Sauniere, L. and Sauvage, M. and Sawicki, M. and Scaramella, R. and Scarlata, C. and Scharré, L. and Schaye, J. and Schewtschenko, J. A. and Schindler, J.-T. and Schinnerer, E. and Schirmer, M. and Schmidt, F. and Schmidt, M. and Schneider, A. and Schneider, M. and Schneider, P. and Schöneberg, N. and Schrabback, T. and Schultheis, M. and Schulz, S. and Schuster, N. and Schwartz, J. and Sciotti, D. and Scodeggio, M. and Scognamiglio, D. and Scott, D. and Scottez, V. and Secroun, A. and Sefusatti, E. and Seidel, G. and Seiffert, M. and Sellentin, E. and Selwood, M. and Semboloni, E. and Sereno, M. and Serjeant, S. and Serrano, S. and Setnikar, G. and Shankar, F. and Sharples, R. M. and Short, A. and Shulevski, A. and Shuntov, M. and Sias, M. and Sikkema, G. and Silvestri, A. and Simon, P. and Sirignano, C. and Sirri, G. and Skottfelt, J. and Slezak, E. and Sluse, D. and Smith, G. P. and Smith, L. C. and Smith, R. E. and Smit, S. J. A. and Soldano, F. and Solheim, B. G. B. and Sorce, J. G. and Sorrenti, F. and Soubrie, E. and Spinoglio, L. and Mancini, A. Spurio and Stadel, J. and Stagnaro, L. and Stanco, L. and Stanford, S. A. and Starck, J.-L. and Stassi, P. and Steinwagner, J. and Stern, D. and Stone, C. and Strada, P. and Strafella, F. and Stramaccioni, D. and Surace, C. and Sureau, F. and Suyu, S. H. and Swindells, I. and Szafraniec, M. and Szapudi, I. and Taamoli, S. and Talia, M. and Tallada-Crespí, P. and Tanidis, K. and Tao, C. and Tarrío, P. and Tavagnacco, D. and Taylor, A. N. and Taylor, J. E. and Taylor, P. L. and Teixeira, E. M. and Tenti, M. and Idiago, P. Teodoro and Teplitz, H. I. and Tereno, I. and Tessore, N. and Testa, V. and Testera, G. and Tewes, M. and Teyssier, R. and Theret, N. and Thizy, C. and Thomas, P. D. and Toba, Y. and Toft, S. and Toledo-Moreo, R. and Tolstoy, E. and Tommasi, E. and Torbaniuk, O. and Torradeflot, F. and Tortora, C. and Tosi, S. and Tosti, S. and Trifoglio, M. and Troja, A. and Trombetti, T. and Tronconi, A. and Tsedrik, M. and Tsyganov, A. and Tucci, M. and Tutusaus, I. and Uhlemann, C. and Ulivi, L. and Urbano, M. and Vacher, L. and Vaillon, L. and Valageas, P. and Valdes, I. and Valentijn, E. A. and Valenziano, L. and Valieri, C. and Valiviita, J. and Broeck, M. Van den and Vassallo, T. and Vavrek, R. and Vega-Ferrero, J. and Venemans, B. and Venhola, A. and Ventura, S. and Kleijn, G. Verdoes and Vergani, D. and Verma, A. and Vernizzi, F. and Veropalumbo, A. and Verza, G. and Vescovi, C. and Vibert, D. and Viel, M. and Vielzeuf, P. and Viglione, C. and Viitanen, A. and Villaescusa-Navarro, F. and Vinciguerra, S. and Visticot, F. and Voggel, K. and Wietersheim-Kramsta, M. von and Vriend, W. J. and Wachter, S. and Walmsley, M. and Walth, G. and Walton, D. M. and Walton, N. A. and Wander, M. and Wang, L. and Wang, Y. and Weaver, J. R. and Weller, J. and Wetzstein, M. and Whalen, D. J. and Whittam, I. H. and Widmer, A. and Wiesmann, M. and Wilde, J. and Williams, O. R. and Winther, H.-A. and Wittje, A. and Wong, J. H. W. and Wright, A. H. and Yankelevich, V. and Yeung, H. W. and Yoon, M. and Youles, S. and Yung, L. Y. A. and Zacchei, A. and Zalesky, L. and Zamorani, G. and Vitorelli, A. Zamorano and Marc, M. Zanoni and Zennaro, M. and Zerbi, F. M. and Zinchenko, I. A. and Zoubian, J. and Zucca, E. and Zumalacarregui, M.},
	month = may,
	year = {2025},
	note = {Publisher: EDP Sciences},
	pages = {A1},
}

@misc{montes_intracluster_2019,
	title = {The intracluster light and its role in galaxy evolution in clusters},
	url = {http://arxiv.org/abs/1912.01616},
	language = {en},
	urldate = {2024-06-28},
	publisher = {arXiv},
	author = {Montes, Mireia},
	month = dec,
	year = {2019},
	note = {arXiv:1912.01616 [astro-ph]},
}

@misc{brough_vera_2020,
	title = {The {Vera} {Rubin} {Observatory} {Legacy} {Survey} of {Space} and {Time} and the {Low} {Surface} {Brightness} {Universe}},
	url = {http://arxiv.org/abs/2001.11067},
	language = {en},
	urldate = {2024-06-28},
	publisher = {arXiv},
	author = {Brough, Sarah and Collins, Chris and Demarco, Ricardo and Ferguson, Henry C. and Galaz, Gaspar and Holwerda, Benne and Martinez-Lombilla, Cristina and Mihos, Chris and Montes, Mireia},
	month = jan,
	year = {2020},
	note = {arXiv:2001.11067 [astro-ph]},
}

@article{desmons_detecting_2024,
	title = {Detecting galaxy tidal features using self-supervised representation learning},
	volume = {531},
	copyright = {https://creativecommons.org/licenses/by/4.0/},
	issn = {0035-8711, 1365-2966},
	url = {https://academic.oup.com/mnras/article/531/4/4070/7688470},
	doi = {10.1093/mnras/stae1402},
	language = {en},
	number = {4},
	urldate = {2024-06-28},
	journal = {Monthly Notices of the Royal Astronomical Society},
	author = {Desmons, Alice and Brough, Sarah and Lanusse, Francois},
	month = jun,
	year = {2024},
	pages = {4070--4084},
}

@article{hayat_self-supervised_2021,
	title = {Self-supervised {Representation} {Learning} for {Astronomical} {Images}},
	volume = {911},
	issn = {2041-8205},
	url = {https://dx.doi.org/10.3847/2041-8213/abf2c7},
	doi = {10.3847/2041-8213/abf2c7},
	language = {en},
	number = {2},
	urldate = {2023-10-19},
	journal = {The Astrophysical Journal Letters},
	author = {Hayat, Md Abul and Stein, George and Harrington, Peter and Lukić, Zarija and Mustafa, Mustafa},
	month = apr,
	year = {2021},
	note = {Publisher: The American Astronomical Society},
	pages = {L33},
}

@article{pearson_identifying_2019,
	title = {Identifying galaxy mergers in observations and simulations with deep learning},
	volume = {626},
	issn = {0004-6361, 1432-0746},
	url = {https://www.aanda.org/10.1051/0004-6361/201935355},
	doi = {10.1051/0004-6361/201935355},
	language = {en},
	urldate = {2024-02-22},
	journal = {Astronomy \& Astrophysics},
	author = {Pearson, W. J. and Wang, L. and Trayford, J. W. and Petrillo, C. E. and Van Der Tak, F. F. S.},
	month = jun,
	year = {2019},
	pages = {A49},
}

@article{canepa_measuring_2025,
	title = {Measuring the {Intracluster} {Light} {Fraction} with {Machine} {Learning}},
	volume = {980},
	issn = {0004-637X},
	url = {https://dx.doi.org/10.3847/1538-4357/adabc7},
	doi = {10.3847/1538-4357/adabc7},
	language = {en},
	number = {2},
	urldate = {2025-05-15},
	journal = {The Astrophysical Journal},
	author = {Canepa, Louisa and Brough, Sarah and Lanusse, Francois and Montes, Mireia and Hatch, Nina},
	month = feb,
	year = {2025},
	note = {Publisher: The American Astronomical Society},
	pages = {245},
}

@article{desi_collaboration_overview_2022,
	title = {Overview of the {Instrumentation} for the {Dark} {Energy} {Spectroscopic} {Instrument}},
	volume = {164},
	issn = {1538-3881},
	url = {https://dx.doi.org/10.3847/1538-3881/ac882b},
	doi = {10.3847/1538-3881/ac882b},
	language = {en},
	number = {5},
	urldate = {2025-05-15},
	journal = {The Astronomical Journal},
	author = {{DESI Collaboration} and Abareshi, B. and Aguilar, J. and Ahlen, S. and Alam, Shadab and Alexander, David M. and Alfarsy, R. and Allen, L. and Allende Prieto, C. and Alves, O. and Ameel, J. and Armengaud, E. and Asorey, J. and Aviles, Alejandro and Bailey, S. and Balaguera-Antolínez, A. and Ballester, O. and Baltay, C. and Bault, A. and Beltran, S. F. and Benavides, B. and BenZvi, S. and Berti, A. and Besuner, R. and Beutler, Florian and Bianchi, D. and Blake, C. and Blanc, P. and Blum, R. and Bolton, A. and Bose, S. and Bramall, D. and Brieden, S. and Brodzeller, A. and Brooks, D. and Brownewell, C. and Buckley-Geer, E. and Cahn, R. N. and Cai, Z. and Canning, R. and Capasso, R. and Carnero Rosell, A. and Carton, P. and Casas, R. and Castander, F. J. and Cervantes-Cota, J. L. and Chabanier, S. and Chaussidon, E. and Chuang, C. and Circosta, C. and Cole, S. and Cooper, A. P. and da Costa, L. and Cousinou, M.-C. and Cuceu, A. and Davis, T. M. and Dawson, K. and de la Cruz-Noriega, R. and de la Macorra, A. and de Mattia, A. and Della Costa, J. and Demmer, P. and Derwent, M. and Dey, A. and Dey, B. and Dhungana, G. and Ding, Z. and Dobson, C. and Doel, P. and Donald-McCann, J. and Donaldson, J. and Douglass, K. and Duan, Y. and Dunlop, P. and Edelstein, J. and Eftekharzadeh, S. and Eisenstein, D. J. and Enriquez-Vargas, M. and Escoffier, S. and Evatt, M. and Fagrelius, P. and Fan, X. and Fanning, K. and Fawcett, V. A. and Ferraro, S. and Ereza, J. and Flaugher, B. and Font-Ribera, A. and Forero-Romero, J. E. and Frenk, C. S. and Fromenteau, S. and Gänsicke, B. T. and Garcia-Quintero, C. and Garrison, L. and Gaztañaga, E. and Gerardi, F. and Gil-Marín, H. and Gontcho A Gontcho, S. and Gonzalez-Morales, Alma X. and Gonzalez-de-Rivera, G. and Gonzalez-Perez, V. and Gordon, C. and Graur, O. and Green, D. and Grove, C. and Gruen, D. and Gutierrez, G. and Guy, J. and Hahn, C. and Harris, S. and Herrera, D. and Herrera-Alcantar, Hiram K. and Honscheid, K. and Howlett, C. and Huterer, D. and Iršič, V. and Ishak, M. and Jelinsky, P. and Jiang, L. and Jimenez, J. and Jing, Y. P. and Joyce, R. and Jullo, E. and Juneau, S. and Karaçaylı, N. G. and Karamanis, M. and Karcher, A. and Karim, T. and Kehoe, R. and Kent, S. and Kirkby, D. and Kisner, T. and Kitaura, F. and Koposov, S. E. and Kovács, A. and Kremin, A. and Krolewski, Alex and L’Huillier, B. and Lahav, O. and Lambert, A. and Lamman, C. and Lan, Ting-Wen and Landriau, M. and Lane, S. and Lang, D. and Lange, J. U. and Lasker, J. and Le Guillou, L. and Leauthaud, A. and Le Van Suu, A. and Levi, Michael E. and Li, T. S. and Magneville, C. and Manera, M. and Manser, Christopher J. and Marshall, B. and Martini, Paul and McCollam, W. and McDonald, P. and Meisner, Aaron M. and Mena-Fernández, J. and Meneses-Rizo, J. and Mezcua, M. and Miller, T. and Miquel, R. and Montero-Camacho, P. and Moon, J. and Moustakas, J. and Mueller, E. and Muñoz-Gutiérrez, Andrea and Myers, Adam D. and Nadathur, S. and Najita, J. and Napolitano, L. and Neilsen, E. and Newman, Jeffrey A. and Nie, J. D. and Ning, Y. and Niz, G. and Norberg, P. and Noriega, Hernán E. and O’Brien, T. and Obuljen, A. and Palanque-Delabrouille, N. and Palmese, A. and Zhiwei, P. and Pappalardo, D. and PENG, X. and Percival, W. J. and Perruchot, S. and Pogge, R. and Poppett, C. and Porredon, A. and Prada, F. and Prochaska, J. and Pucha, R. and Pérez-Fernández, A. and Pérez-Ràfols, I. and Rabinowitz, D. and Raichoor, A. and Ramirez-Solano, S. and Ramírez-Pérez, César and Ravoux, C. and Reil, K. and Rezaie, M. and Rocher, A. and Rockosi, C. and Roe, N. A. and Roodman, A. and Ross, A. J. and Rossi, G. and Ruggeri, R. and Ruhlmann-Kleider, V. and Sabiu, C. G. and Gaines, S. and Said, K. and Saintonge, A. and Salas Catonga, Javier and Samushia, L. and Sanchez, E. and Saulder, C. and Schaan, E. and Schlafly, E. and Schlegel, D. and Schmoll, J. and Scholte, D. and Schubnell, M. and Secroun, A. and Seo, H. and Serrano, S. and Sharples, Ray M. and Sholl, Michael J. and Silber, Joseph Harry and Silva, D. R. and Sirk, M. and Siudek, M. and Smith, A. and Sprayberry, D. and Staten, R. and Stupak, B. and Tan, T. and Tarlé, Gregory and Tie, Suk Sien and Tojeiro, R. and Ureña-López, L. A. and Valdes, F. and Valenzuela, O. and Valluri, M. and Vargas-Magaña, M. and Verde, L. and Walther, M. and Wang, B. and Wang, M. S. and Weaver, B. A. and Weaverdyck, C. and Wechsler, R. and Wilson, Michael J. and Yang, J. and Yu, Y. and Yuan, S. and Yèche, Christophe and Zhang, H. and Zhang, K. and Zhao, Cheng and Zhou, Rongpu and Zhou, Zhimin and Zou, H. and Zou, J. and Zou, S. and Zu, Y.},
	month = oct,
	year = {2022},
	note = {Publisher: The American Astronomical Society},
	pages = {207},
}

@article{desi_collaboration_early_2024,
	title = {The {Early} {Data} {Release} of the {Dark} {Energy} {Spectroscopic} {Instrument}},
	volume = {168},
	issn = {1538-3881},
	url = {https://dx.doi.org/10.3847/1538-3881/ad3217},
	doi = {10.3847/1538-3881/ad3217},
	language = {en},
	number = {2},
	urldate = {2025-05-15},
	journal = {The Astronomical Journal},
	author = {{DESI Collaboration} and Adame, A. G. and Aguilar, J. and Ahlen, S. and Alam, S. and Aldering, G. and Alexander, D. M. and Alfarsy, R. and Allende Prieto, C. and Alvarez, M. and Alves, O. and Anand, A. and Andrade-Oliveira, F. and Armengaud, E. and Asorey, J. and Avila, S. and Aviles, A. and Bailey, S. and Balaguera-Antolínez, A. and Ballester, O. and Baltay, C. and Bault, A. and Bautista, J. and Behera, J. and Beltran, S. F. and BenZvi, S. and Beraldo e Silva, L. and Bermejo-Climent, J. R. and Berti, A. and Besuner, R. and Beutler, F. and Bianchi, D. and Blake, C. and Blum, R. and Bolton, A. S. and Brieden, S. and Brodzeller, A. and Brooks, D. and Brown, Z. and Buckley-Geer, E. and Burtin, E. and Cabayol-Garcia, L. and Cai, Z. and Canning, R. and Cardiel-Sas, L. and Carnero Rosell, A. and Castander, F. J. and Cervantes-Cota, J. L. and Chabanier, S. and Chaussidon, E. and Chaves-Montero, J. and Chen, S. and Chen, X. and Chuang, C. and Claybaugh, T. and Cole, S. and Cooper, A. P. and Cuceu, A. and Davis, T. M. and Dawson, K. and de Belsunce, R. and de la Cruz, R. and de la Macorra, A. and Della Costa, J. and de Mattia, A. and Demina, R. and Demirbozan, U. and DeRose, J. and Dey, A. and Dey, B. and Dhungana, G. and Ding, J. and Ding, Z. and Doel, P. and Doshi, R. and Douglass, K. and Edge, A. and Eftekharzadeh, S. and Eisenstein, D. J. and Elliott, A. and Ereza, J. and Escoffier, S. and Fagrelius, P. and Fan, X. and Fanning, K. and Fawcett, V. A. and Ferraro, S. and Flaugher, B. and Font-Ribera, A. and Forero-Romero, J. E. and Forero-Sánchez, D. and Frenk, C. S. and Gänsicke, B. T. and García, L. Á. and García-Bellido, J. and Garcia-Quintero, C. and Garrison, L. H. and Gil-Marín, H. and Golden-Marx, J. and Gontcho A Gontcho, S. and Gonzalez-Morales, A. X. and Gonzalez-Perez, V. and Gordon, C. and Graur, O. and Green, D. and Gruen, D. and Guy, J. and Hadzhiyska, B. and Hahn, C. and Han, J. J. and Hanif, M. M. S and Herrera-Alcantar, H. K. and Honscheid, K. and Hou, J. and Howlett, C. and Huterer, D. and Iršič, V. and Ishak, M. and Jacques, A. and Jana, A. and Jiang, L. and Jimenez, J. and Jing, Y. P. and Joudaki, S. and Joyce, R. and Jullo, E. and Juneau, S. and Karaçaylı, N. G. and Karim, T. and Kehoe, R. and Kent, S. and Khederlarian, A. and Kim, S. and Kirkby, D. and Kisner, T. and Kitaura, F. and Kizhuprakkat, N. and Kneib, J. and Koposov, S. E. and Kovács, A. and Kremin, A. and Krolewski, A. and L’Huillier, B. and Lahav, O. and Lambert, A. and Lamman, C. and Lan, T.-W. and Landriau, M. and Lang, D. and Lange, J. U. and Lasker, J. and Leauthaud, A. and Le Guillou, L. and Levi, M. E. and Li, T. S. and Linder, E. and Lyons, A. and Magneville, C. and Manera, M. and Manser, C. J. and Margala, D. and Martini, P. and McDonald, P. and Medina, G. E. and Medina-Varela, L. and Meisner, A. and Mena-Fernández, J. and Meneses-Rizo, J. and Mezcua, M. and Miquel, R. and Montero-Camacho, P. and Moon, J. and Moore, S. and Moustakas, J. and Mueller, E. and Mundet, J. and Muñoz-Gutiérrez, A. and Myers, A. D. and Nadathur, S. and Napolitano, L. and Neveux, R. and Newman, J. A. and Nie, J. and Nikutta, R. and Niz, G. and Norberg, P. and Noriega, H. E. and Paillas, E. and Palanque-Delabrouille, N. and Palmese, A. and Pan, Z. and Parkinson, D. and Penmetsa, S. and Percival, W. J. and Pérez-Fernández, A. and Pérez-Ràfols, I. and Pieri, M. and Poppett, C. and Porredon, A. and Pothier, S. and Prada, F. and Pucha, R. and Raichoor, A. and Ramírez-Pérez, C. and Ramirez-Solano, S. and Rashkovetskyi, M. and Ravoux, C. and Rocher, A. and Rockosi, C. and Ross, A. J. and Rossi, G. and Ruggeri, R. and Ruhlmann-Kleider, V. and Sabiu, C. G. and Said, K. and Saintonge, A. and Samushia, L. and Sanchez, E. and Saulder, C. and Schaan, E. and Schlafly, E. F. and Schlegel, D. and Scholte, D. and Schubnell, M. and Seo, H. and Shafieloo, A. and Sharples, R. and Sheu, W. and Silber, J. and Sinigaglia, F. and Siudek, M. and Slepian, Z. and Smith, A. and Soumagnac, M. T. and Sprayberry, D. and Stephey, L. and Suárez-Pérez, J. and Sun, Z. and Tan, T. and Tarlé, G. and Tojeiro, R. and Ureña-López, L. A. and Vaisakh, R. and Valcin, D. and Valdes, F. and Valluri, M. and Vargas-Magaña, M. and Variu, A. and Verde, L. and Walther, M. and Wang, B. and Wang, M. S. and Weaver, B. A. and Weaverdyck, N. and Wechsler, R. H. and White, M. and Xie, Y. and Yang, J. and Yèche, C. and Yu, J. and Yuan, S. and Zhang, H. and Zhang, Z. and Zhao, C. and Zheng, Z. and Zhou, R. and Zhou, Z. and Zou, H. and Zou, S. and Zu, Y.},
	month = jul,
	year = {2024},
	note = {Publisher: The American Astronomical Society},
	pages = {58},
}

@article{finn_h-derived_2005,
	title = {Hα-derived {Star} {Formation} {Rates} for {Three} z≃0.75 {EDisCS} {Galaxy} {Clusters}* **},
	volume = {630},
	issn = {0004-637X},
	url = {https://iopscience.iop.org/article/10.1086/431642/meta},
	doi = {10.1086/431642},
	language = {en},
	number = {1},
	urldate = {2025-05-16},
	journal = {The Astrophysical Journal},
	author = {Finn, Rose A. and Zaritsky, Dennis and Donald W. McCarthy, Jr and Poggianti, Bianca and Rudnick, Gregory and Halliday, Claire and Milvang-Jensen, Bo and Pelló, Roser and Simard, Luc},
	month = sep,
	year = {2005},
	note = {Publisher: IOP Publishing},
	pages = {206},
}

@article{casas_optical_2024,
	title = {Optical {Photometric} {Indicators} of {Galaxy} {Cluster} {Relaxation}},
	volume = {967},
	issn = {0004-637X},
	url = {https://dx.doi.org/10.3847/1538-4357/ad41de},
	doi = {10.3847/1538-4357/ad41de},
	language = {en},
	number = {1},
	urldate = {2025-05-19},
	journal = {The Astrophysical Journal},
	author = {Casas, Madeline C. and Putnam, Ky and Mantz, Adam B. and Allen, Steven W. and Somboonpanyakul, Taweewat},
	month = may,
	year = {2024},
	note = {Publisher: The American Astronomical Society},
	pages = {14},
}

@article{donghia_formation_2005,
	title = {The {Formation} of {Fossil} {Galaxy} {Groups} in the {Hierarchical} {Universe}},
	volume = {630},
	issn = {0004-637X},
	url = {https://iopscience.iop.org/article/10.1086/491651/meta},
	doi = {10.1086/491651},
	language = {en},
	number = {2},
	urldate = {2025-05-19},
	journal = {The Astrophysical Journal},
	author = {D’Onghia, E. and Sommer-Larsen, J. and Romeo, A. D. and Burkert, A. and Pedersen, K. and Portinari, L. and Rasmussen, J.},
	month = aug,
	year = {2005},
	note = {Publisher: IOP Publishing},
	pages = {L109},
}

@article{dariush_mass_2010,
	title = {The mass assembly of galaxy groups and the evolution of the magnitude gap},
	volume = {405},
	issn = {0035-8711},
	url = {https://doi.org/10.1111/j.1365-2966.2010.16569.x},
	doi = {10.1111/j.1365-2966.2010.16569.x},
	number = {3},
	urldate = {2025-05-19},
	journal = {Monthly Notices of the Royal Astronomical Society},
	author = {Dariush, Ali A. and Raychaudhury, Somak and Ponman, Trevor J. and Khosroshahi, Habib G. and Benson, Andrew J. and Bower, Richard G. and Pearce, Frazer},
	month = jul,
	year = {2010},
	pages = {1873--1887},
}

@article{vitorelli_mass_2018,
	title = {On mass concentrations and magnitude gaps of galaxy systems in the {CS82} survey},
	volume = {474},
	issn = {0035-8711},
	url = {https://doi.org/10.1093/mnras/stx2791},
	doi = {10.1093/mnras/stx2791},
	number = {1},
	urldate = {2025-05-19},
	journal = {Monthly Notices of the Royal Astronomical Society},
	author = {Vitorelli, André Z. and Cypriano, Eduardo S. and Makler, Martín and Pereira, Maria E. S. and Erben, Thomas and Moraes, Bruno},
	month = feb,
	year = {2018},
	pages = {866--875},
}

@article{yoo_spatial_2024,
	title = {Spatial {Distribution} of {Intracluster} {Light} versus {Dark} {Matter} in {Horizon} {Run} 5},
	volume = {965},
	issn = {0004-637X},
	url = {https://dx.doi.org/10.3847/1538-4357/ad2df8},
	doi = {10.3847/1538-4357/ad2df8},
	language = {en},
	number = {2},
	urldate = {2025-05-19},
	journal = {The Astrophysical Journal},
	author = {Yoo, Jaewon and Park, Changbom and Sabiu, Cristiano G. and Singh, Ankit and Ko, Jongwan and Lee, Jaehyun and Pichon, Christophe and Jee, M. James and Gibson, Brad K. and Snaith, Owain and Kim, Juhan and Shin, Jihye and Kim, Yonghwi and Kim, Hyowon},
	month = apr,
	year = {2024},
	note = {Publisher: The American Astronomical Society},
	pages = {145},
}

@misc{canepa_lpcanmicl_2024,
	title = {lpcan/{MICL}: {Paper} version},
	copyright = {Creative Commons Attribution 4.0 International},
	shorttitle = {lpcan/{MICL}},
	url = {https://zenodo.org/doi/10.5281/zenodo.14376238},
	urldate = {2024-12-18},
	publisher = {Zenodo},
	author = {Canepa, Louisa and Brough, Sarah and Lanusse, Francois and Montes, Mireia and Hatch, Nina},
	month = dec,
	year = {2024},
	doi = {10.5281/ZENODO.14376238},
}

@article{roman_galactic_2020,
	title = {Galactic cirri in deep optical imaging},
	volume = {644},
	copyright = {© ESO 2020},
	issn = {0004-6361, 1432-0746},
	url = {https://www.aanda.org/articles/aa/abs/2020/12/aa36111-19/aa36111-19.html},
	doi = {10.1051/0004-6361/201936111},
	language = {en},
	urldate = {2023-10-19},
	journal = {Astronomy \& Astrophysics},
	author = {Román, Javier and Trujillo, Ignacio and Montes, Mireia},
	month = dec,
	year = {2020},
	note = {Publisher: EDP Sciences},
	pages = {A42},
}

@article{chilingarian_analytical_2010,
	title = {Analytical approximations of {K}-corrections in optical and near-infrared bands: {Analytical} approximations of {K}-corrections},
	volume = {405},
	issn = {00358711, 13652966},
	shorttitle = {Analytical approximations of {K}-corrections in optical and near-infrared bands},
	url = {https://academic.oup.com/mnras/article-lookup/doi/10.1111/j.1365-2966.2010.16506.x},
	doi = {10.1111/j.1365-2966.2010.16506.x},
	language = {en},
	number = {3},
	urldate = {2024-06-23},
	journal = {Monthly Notices of the Royal Astronomical Society},
	author = {Chilingarian, Igor V. and Melchior, Anne-Laure and Zolotukhin, Ivan Yu.},
	month = may,
	year = {2010},
	pages = {1409--1420},
}

@article{martinez-lombilla_galaxy_2023,
	title = {Galaxy {And} {Mass} {Assembly} ({GAMA}): extended intragroup light in a group at z = 0.2 from deep {Hyper} {Suprime}-{Cam} images},
	volume = {518},
	issn = {0035-8711},
	shorttitle = {Galaxy {And} {Mass} {Assembly} ({GAMA})},
	url = {https://doi.org/10.1093/mnras/stac3119},
	doi = {10.1093/mnras/stac3119},
	number = {1},
	urldate = {2023-10-19},
	journal = {Monthly Notices of the Royal Astronomical Society},
	author = {Martínez-Lombilla, Cristina and Brough, Sarah and Montes, Mireia and Baena-Gallé, Roberto and Akhlaghi, Mohammad and Infante-Sainz, Raúl and Driver, Simon P and Holwerda, Benne W and Pimbblet, Kevin A and Robotham, Aaron S G},
	month = jan,
	year = {2023},
	pages = {1195--1213},
}

@article{owers_sami_2019,
	title = {The {SAMI} {Galaxy} {Survey}: {Quenching} of {Star} {Formation} in {Clusters} {I}. {Transition} {Galaxies}},
	volume = {873},
	issn = {0004-637X},
	shorttitle = {The {SAMI} {Galaxy} {Survey}},
	url = {https://dx.doi.org/10.3847/1538-4357/ab0201},
	doi = {10.3847/1538-4357/ab0201},
	language = {en},
	number = {1},
	urldate = {2025-05-20},
	journal = {The Astrophysical Journal},
	author = {Owers, Matt S. and Hudson, Michael J. and Oman, Kyle A. and Bland-Hawthorn, Joss and Brough, S. and Bryant, Julia J. and Cortese, Luca and Couch, Warrick J. and Croom, Scott M. and Sande, Jesse van de and Federrath, Christoph and Groves, Brent and Hopkins, A. M. and Lawrence, J. S. and Lorente, Nuria P. F. and McDermid, Richard M. and Medling, Anne M. and Richards, Samuel N. and Scott, Nicholas and Taranu, Dan S. and Welker, Charlotte and Yi, Sukyoung K.},
	month = mar,
	year = {2019},
	note = {Publisher: The American Astronomical Society},
	pages = {52},
}

@article{croom_sami_2021,
	title = {The {SAMI} {Galaxy} {Survey}: the third and final data release},
	volume = {505},
	issn = {0035-8711},
	shorttitle = {The {SAMI} {Galaxy} {Survey}},
	url = {https://doi.org/10.1093/mnras/stab229},
	doi = {10.1093/mnras/stab229},
	number = {1},
	urldate = {2025-05-20},
	journal = {Monthly Notices of the Royal Astronomical Society},
	author = {Croom, Scott M and Owers, Matt S and Scott, Nicholas and Poetrodjojo, Henry and Groves, Brent and van de Sande, Jesse and Barone, Tania M and Cortese, Luca and D’Eugenio, Francesco and Bland-Hawthorn, Joss and Bryant, Julia and Oh, Sree and Brough, Sarah and Agostino, James and Casura, Sarah and Catinella, Barbara and Colless, Matthew and Cecil, Gerald and Davies, Roger L and Drinkwater, Michael J and Driver, Simon P and Ferreras, Ignacio and Foster, Caroline and Fraser-McKelvie, Amelia and Lawrence, Jon and Leslie, Sarah K and Liske, Jochen and López-Sánchez, Ángel R and Lorente, Nuria P F and McElroy, Rebecca and Medling, Anne M and Obreschkow, Danail and Richards, Samuel N and Sharp, Rob and Sweet, Sarah M and Taranu, Dan S and Taylor, Edward N and Tescari, Edoardo and Thomas, Adam D and Tocknell, James and Vaughan, Sam P},
	month = jul,
	year = {2021},
	pages = {991--1016},
}

@article{demaio_lost_2018,
	title = {Lost but not forgotten: intracluster light in galaxy groups and clusters},
	volume = {474},
	issn = {0035-8711},
	shorttitle = {Lost but not forgotten},
	url = {https://doi.org/10.1093/mnras/stx2946},
	doi = {10.1093/mnras/stx2946},
	number = {3},
	urldate = {2023-10-19},
	journal = {Monthly Notices of the Royal Astronomical Society},
	author = {DeMaio, Tahlia and Gonzalez, Anthony H and Zabludoff, Ann and Zaritsky, Dennis and Connor, Thomas and Donahue, Megan and Mulchaey, John S},
	month = mar,
	year = {2018},
	pages = {3009--3031},
}

@article{vazdekis_uv-extended_2016,
	title = {{UV}-extended {E}-{MILES} stellar population models: young components in massive early-type galaxies},
	volume = {463},
	issn = {0035-8711, 1365-2966},
	shorttitle = {{UV}-extended {E}-{MILES} stellar population models},
	url = {https://academic.oup.com/mnras/article-lookup/doi/10.1093/mnras/stw2231},
	doi = {10.1093/mnras/stw2231},
	language = {en},
	number = {4},
	urldate = {2024-08-15},
	journal = {Monthly Notices of the Royal Astronomical Society},
	author = {Vazdekis, A. and Koleva, M. and Ricciardelli, E. and Röck, B. and Falcón-Barroso, J.},
	month = dec,
	year = {2016},
	pages = {3409--3436},
}

@article{montes_buildup_2021,
	title = {The {Buildup} of the {Intracluster} {Light} of {A85} as {Seen} by {Subaru}’s {Hyper} {Suprime}-{Cam}},
	volume = {910},
	issn = {0004-637X},
	url = {https://dx.doi.org/10.3847/1538-4357/abddb6},
	doi = {10.3847/1538-4357/abddb6},
	language = {en},
	number = {1},
	urldate = {2023-10-19},
	journal = {The Astrophysical Journal},
	author = {Montes, Mireia and Brough, Sarah and Owers, Matt S. and Santucci, Giulia},
	month = mar,
	year = {2021},
	note = {Publisher: The American Astronomical Society},
	pages = {45},
}

@article{gonzalez_galaxy_2013,
	title = {Galaxy {Cluster} {Baryon} {Fractions} {Revisited}},
	volume = {778},
	issn = {0004-637X},
	url = {https://ui.adsabs.harvard.edu/abs/2013ApJ...778...14G},
	doi = {10.1088/0004-637X/778/1/14},
	urldate = {2025-04-22},
	journal = {The Astrophysical Journal},
	author = {Gonzalez, Anthony H. and Sivanandam, Suresh and Zabludoff, Ann I. and Zaritsky, Dennis},
	month = nov,
	year = {2013},
	note = {Publisher: IOP
ADS Bibcode: 2013ApJ...778...14G},
	pages = {14},
}

@article{gonzalez_census_2007,
	title = {A {Census} of {Baryons} in {Galaxy} {Clusters} and {Groups}},
	volume = {666},
	issn = {0004-637X},
	url = {https://iopscience.iop.org/article/10.1086/519729/meta},
	doi = {10.1086/519729},
	language = {en},
	number = {1},
	urldate = {2023-10-19},
	journal = {The Astrophysical Journal},
	author = {Gonzalez, Anthony H. and Zaritsky, Dennis and Zabludoff, Ann I.},
	month = sep,
	year = {2007},
	note = {Publisher: IOP Publishing},
	pages = {147},
}

@article{bahe_disruption_2019,
	title = {Disruption of satellite galaxies in simulated groups and clusters: the roles of accretion time, baryons, and pre-processing},
	volume = {485},
	issn = {0035-8711},
	shorttitle = {Disruption of satellite galaxies in simulated groups and clusters},
	url = {https://doi.org/10.1093/mnras/stz361},
	doi = {10.1093/mnras/stz361},
	number = {2},
	urldate = {2025-05-23},
	journal = {Monthly Notices of the Royal Astronomical Society},
	author = {Bahé, Yannick M and Schaye, Joop and Barnes, David J and Dalla Vecchia, Claudio and Kay, Scott T and Bower, Richard G and Hoekstra, Henk and McGee, Sean L and Theuns, Tom},
	month = may,
	year = {2019},
	pages = {2287--2311},
}

@article{montenegro-taborda_growth_2023,
	title = {The growth of brightest cluster galaxies in the {TNG300} simulation: dissecting the contributions from mergers and in situ star formation},
	volume = {521},
	issn = {0035-8711},
	shorttitle = {The growth of brightest cluster galaxies in the {TNG300} simulation},
	url = {https://ui.adsabs.harvard.edu/abs/2023MNRAS.521..800M},
	doi = {10.1093/mnras/stad586},
	urldate = {2024-11-19},
	journal = {Monthly Notices of the Royal Astronomical Society},
	author = {Montenegro-Taborda, Daniel and Rodriguez-Gomez, Vicente and Pillepich, Annalisa and Avila-Reese, Vladimir and Sales, Laura V. and Rodríguez-Puebla, Aldo and Hernquist, Lars},
	month = may,
	year = {2023},
	note = {Publisher: OUP
ADS Bibcode: 2023MNRAS.521..800M},
	pages = {800--817},
}

@article{murante_importance_2007,
	title = {The importance of mergers for the origin of intracluster stars in cosmological simulations of galaxy clusters},
	volume = {377},
	issn = {0035-8711},
	url = {https://doi.org/10.1111/j.1365-2966.2007.11568.x},
	doi = {10.1111/j.1365-2966.2007.11568.x},
	number = {1},
	urldate = {2023-10-19},
	journal = {Monthly Notices of the Royal Astronomical Society},
	author = {Murante, Giuseppe and Giovalli, Martina and Gerhard, Ortwin and Arnaboldi, Magda and Borgani, Stefano and Dolag, Klaus},
	month = may,
	year = {2007},
	pages = {2--16},
}

@article{barfety_assessment_2022,
	title = {An {Assessment} of the {In} {Situ} {Growth} of the {Intracluster} {Light} in the {High}-redshift {Galaxy} {Cluster} {SpARCS1049}+56},
	volume = {930},
	issn = {0004-637X},
	url = {https://ui.adsabs.harvard.edu/abs/2022ApJ...930...25B},
	doi = {10.3847/1538-4357/ac61dd},
	urldate = {2024-11-19},
	journal = {The Astrophysical Journal},
	author = {Barfety, Capucine and Valin, Félix-Antoine and Webb, Tracy M. A. and Yun, Min and Shipley, Heath and Boone, Kyle and Hayden, Brian and Hlavacek-Larrondo, Julie and Muzzin, Adam and Noble, Allison G. and Perlmutter, Saul and Rhea, Carter and Wilson, Gillian and Yee, H. K. C.},
	month = may,
	year = {2022},
	note = {Publisher: IOP
ADS Bibcode: 2022ApJ...930...25B},
	pages = {25},
}

@article{desmons_galaxy_2023,
	title = {Galaxy and mass assembly ({GAMA}): comparing visually and spectroscopically identified galaxy merger samples},
	volume = {523},
	issn = {0035-8711},
	shorttitle = {Galaxy and mass assembly ({GAMA})},
	url = {https://doi.org/10.1093/mnras/stad1639},
	doi = {10.1093/mnras/stad1639},
	number = {3},
	urldate = {2025-06-06},
	journal = {Monthly Notices of the Royal Astronomical Society},
	author = {Desmons, Alice and Brough, Sarah and Martínez-Lombilla, Cristina and De Propris, Roberto and Holwerda, Benne and López-Sánchez, {\'A}ngel R},
	month = aug,
	year = {2023},
	pages = {4381--4393},
}

@book{sersic_atlas_1968,
	title = {Atlas de {Galaxias} {Australes}},
	url = {https://ui.adsabs.harvard.edu/abs/1968adga.book.....S},
	urldate = {2025-06-10},
	author = {Sersic, Jose Luis},
	month = jan,
	year = {1968},
	note = {Publication Title: Cordoba
ADS Bibcode: 1968adga.book.....S},
}

@article{pillepich_first_2018,
	title = {First results from the {IllustrisTNG} simulations: the stellar mass content of groups and clusters of galaxies},
	volume = {475},
	issn = {0035-8711},
	shorttitle = {First results from the {IllustrisTNG} simulations},
	url = {https://doi.org/10.1093/mnras/stx3112},
	doi = {10.1093/mnras/stx3112},
	number = {1},
	urldate = {2023-10-19},
	journal = {Monthly Notices of the Royal Astronomical Society},
	author = {Pillepich, Annalisa and Nelson, Dylan and Hernquist, Lars and Springel, Volker and Pakmor, Rüdiger and Torrey, Paul and Weinberger, Rainer and Genel, Shy and Naiman, Jill P and Marinacci, Federico and Vogelsberger, Mark},
	month = mar,
	year = {2018},
	pages = {648--675},
}




\appendix

\section{Examples of membership assignment}
\label{appendix:membership_examples}
\begin{figure*}
    \centering
    \includegraphics[width=\linewidth]{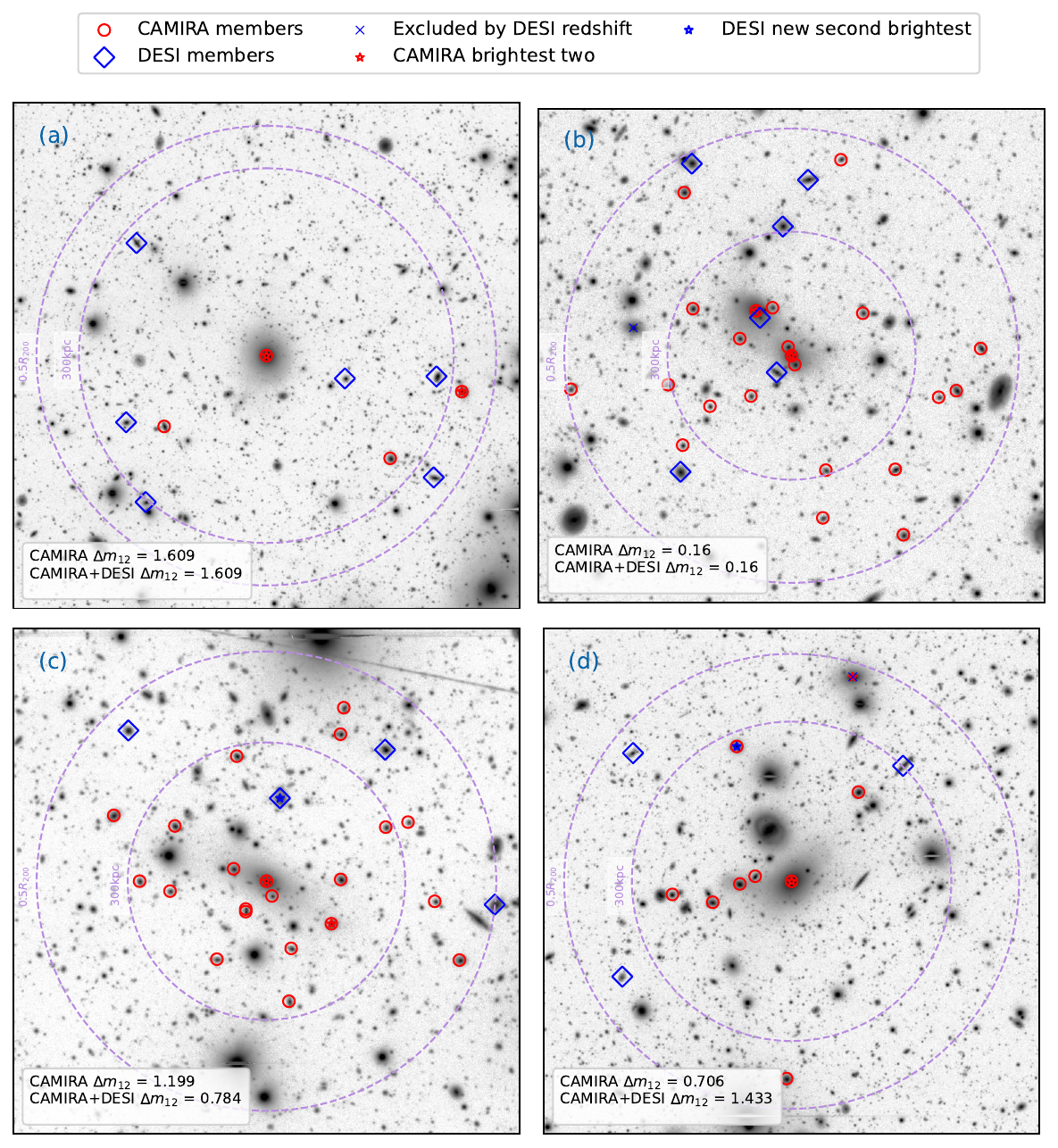}
    \caption{Four example clusters that have additional DESI spectroscopic information. We mark CAMIRA and DESI members that fall within $0.5R_{200}$ of the BCG, which is the radius within which we consider members for the magnitude gap measurement. Panels (a) and (b) show examples where the magnitude gap is unaffected, whereas panel (c) shows an example of DESI providing a new second brightest member that was not part of the CAMIRA catalogue, and panel (d) shows an example of a DESI redshift excluding the CAMIRA second brightest member. Also marked is the 300 kpc radius, which is the radius within which we measure the ICL fraction.}
    \label{fig:desi_membership}
\end{figure*}

This section gives examples of how DESI data is used in adding or excluding members from CAMIRA clusters when calculating the magnitude gap described in Section \ref{sec:group_and_cluster_parameters}. Figure \ref{fig:desi_membership} shows examples of how this additional membership information can affect (or not affect) our magnitude gap measurements. 

Panels (a) and (b) show one low magnitude gap and one high magnitude gap example where the additional membership does not affect the magnitude gap with just CAMIRA information (this makes up 23 of 49 CAMIRA clusters that have additional membership information). In these cases, there are additional DESI members, but because the CAMIRA members remain the brightest and second brightest, $\Delta m_{12}$ does not change. 

Panel (c) shows an example of a new DESI member becoming the second brightest member of the cluster. This results in a lower magnitude gap than using CAMIRA information alone. This is the case for 18 of our clusters.

Panel (d) shows an example of the previous second brightest member being excluded by its DESI redshift. This results in a higher magnitude gap than using CAMIRA information alone. This is the case of 8 of our clusters.

\bsp	
\label{lastpage}
\end{document}